\documentclass[a4paper,11pt,oneside]{article}

\usepackage[margin=1in]{geometry}
\usepackage{physics}
\usepackage{enumerate}
\usepackage[normalem]{ulem}
\usepackage{empheq}
\usepackage{booktabs}
\usepackage{amsmath,amssymb,amsbsy,amstext, amsthm, simplewick}
\usepackage{hyperref}
\usepackage{graphicx}
\usepackage{amsfonts}
\usepackage{color}
\usepackage{wasysym}
\usepackage{tikz, pict2e}
\usepackage{tcolorbox}

\usepackage{colortbl}
\definecolor{summersky}{cmyk}{0.71,0.33,0,0.5}
\definecolor{flamingo}{cmyk}{0,0.51,0.71,0.5}
\definecolor{rp}{cmyk}{0.2, 1, 0.6, 0}
\definecolor{pacificblue}{cmyk}{0.95,0.3,0, 0.5}
\definecolor{gray60}{cmyk}{0.4,0.4,0,0.8}

\hypersetup{
 pdftoolbar=true, % show Acrobat’s toolbar?
 pdfmenubar=true, % show Acrobat’s menu?
 pdffitwindow=true, % window fit to page when opened
 pdfstartview={FitH}, % fits the width of the page to the window
 pdfnewwindow=true, % links in new window
 colorlinks=true, % false: boxed links; true: colored links
 linkcolor=pacificblue, % colour of internal links (change box colour with linkbordercolour)
 citecolor=flamingo, % color of links to bibliography
 filecolor=magenta, % color of file links
 urlcolor=pacificblue % color of external links
}

\usepackage{titlesec}
\titleformat{\section}{\normalfont\fontsize{12}{16}\bfseries}{\thesection}{1em}{}

\numberwithin{equation}{section}

\newcommand{\ex}[1]{\langle #1 \rangle}
\newcommand{\be}{\begin{eqnarray} }
\newcommand{\ee}{\end{eqnarray} }
\newcommand{\bs}{\begin{split} }
\newcommand{\es}{\end{split} }

\renewcommand{\O}{\mathcal{O}}

\newcommand{\bfx}{{\mathbf{x}}}
\newcommand{\bfk}{{\mathbf{k}}}
\newcommand{\bfp}{{\mathbf{p}}}

\newcommand{\Hi}{H_{\text{int}}}
\newcommand{\ki}{\mathbf{k}_{\text{in}}}
\newcommand{\ko}{\mathbf{k}_{\text{out}}}
\newcommand{\ksi}{ k_{\text{in}}}
\newcommand{\Ei}{E_{\text{in}}}
\newcommand{\Eo}{E_{\text{out}}}

\newcommand{\e}{\epsilon}

\newcommand{\then}{\quad \Rightarrow\quad}
\newcommand{\bea}{\begin{eqnarray}}
\newcommand{\eea}{\end{eqnarray}}

\newcommand{\Gf}{G_F}
\newcommand{\Gp}{G^+}

\newcommand{\Gio}{B_{\text{in-out}}}
\newcommand{\Gii}{B_{\text{in-in}}}
\newcommand{\T}{T}

\renewcommand{\eqref}[1]{Eq.~(\ref{#1})}

\usetikzlibrary{calc,backgrounds,intersections}
\usetikzlibrary{arrows,shapes,positioning,shadows,trees}
\usetikzlibrary{mindmap}
\usetikzlibrary{decorations.pathreplacing}
\usetikzlibrary{decorations.pathmorphing}
\usetikzlibrary{decorations.markings}
\usetikzlibrary{matrix}
\usetikzlibrary{patterns}

\tikzstyle arrowstyle=[scale=1]
\tikzstyle directed=[postaction={decorate,decoration={markings,
 mark=at position .5 with {\arrow[arrowstyle]{stealth}}}}]
\tikzstyle reverse directed=[postaction={decorate,decoration={markings,
 mark=at position .5 with {\arrowreversed[arrowstyle]{stealth};}}}]

\begin{document}

\begin{titlepage}
\setcounter{page}{1} \baselineskip=15.5pt

\thispagestyle{empty}

\renewcommand*{\thefootnote}{\fnsymbol{footnote}}

\begin{center}

{\fontsize
{20}{20} \bf The In-Out Formalism for In-In Correlators} \\

\end{center}

\vskip 18pt
\begin{center}
\noindent
{\fontsize{12}{18}\selectfont Yaniv Donath\footnote{\tt yaniv@donath-zafir.com} and Enrico Pajer\footnote{\tt enrico.pajer@gmail.com}}
\end{center}

\begin{center}
\vskip 8pt
\textit{Department of Applied Mathematics and Theoretical Physics, University of Cambridge, Wilberforce Road, Cambridge, CB3 0WA, UK} 
\end{center}

%=========================================

\vskip 50pt
\noindent \textbf{Abstract} ~ \noindent
Cosmological correlators, the natural observables of the primordial universe, have been extensively studied in the past two decades using the in-in formalism pioneered by Schwinger and Keldysh for the study of dissipative open systems. Ironically, most applications in cosmology have focused on non-dissipative closed systems. We show that, for non-dissipative systems, correlators can be equivalently computed using the in-out formalism with the familiar Feynman rules. In particular, the myriad of in-in propagators is reduced to a single (Feynman) time-ordered propagator and no sum over the labelling of vertices is required. In de Sitter spacetime, this requires extending the expanding Poincar\'e patch with a contracting patch, which prepares the bra from the future. Our results are valid for fields of any mass and spin but assuming the absence of infrared divergences. 

We present three applications of the in-out formalism: a representation of correlators in terms of a sum over residues of Feynman propagators in the energy-momentum domain; an algebraic recursion relation that computes Minkowski correlators in terms of lower order ones; and the derivation of cutting rules from Veltman's largest time equation, which we explicitly develop and exemplify for two-vertex diagrams to all loop orders. 

The in-out formalism leads to a natural definition of a de Sitter scattering matrix, which we discuss in simple examples. Remarkably, we show that our scattering matrix satisfies the standard optical theorem and the positivity that follows from it in the forward limit.

%=========================================

\

\end{titlepage}

\setcounter{tocdepth}{2}
{
\hypersetup{linkcolor=black}
\tableofcontents
}

\renewcommand*{\thefootnote}{\arabic{footnote}}
\setcounter{footnote}{0} 

\newpage

\section{Introduction}

Conquering of the atomic scale required not only cutting-edge mathematical structures but also a major overhaul of our ideas of the natural world. The demise of determinism implied by the uncertainty principle and a new picture of physical reality were not, and still are not, easy to swallow. Nevertheless, repeated confrontation with data and mathematical consistency has left us no choice but to abandon classical realism and embrace the confounding beauty of the quantum world. Many of us suspect, hope and fear that the conquering of the Planck scale will similarly require abandoning cherished principles of physics such as locality and the fundamental nature of space and time. To be successful we will no doubt need new advanced mathematics and, at the same time, a tight web of experiments and observation that yet again leave us no choice but to abandon our old prejudices. Currently, our best hope for a confrontation with nature about the character of gravity and spacetime comes from the study of the primordial universe and the treasure trove of information that is stored in cosmological correlators. Meaningful observational data might not be collected next year, the next decade, or even during our life span. Nevertheless, such data will one day be collected and will constitute a milestone of human civilization. In this work, we put forward a small piece of technology that we hope will help us better compute and understand cosmological correlators in quantum field theory. \\

A correlator is the quantum expectation value of the product of a set of operators in a given state. When we apply quantum field theory in curved spacetime to the study of the primordial universe, we are interested in local quantum fields that we can later measure in cosmological data sets. Moreover, we focus on the unique de Sitter invariant quantum state that reduces to the Minkowski vacuum on short distances, namely the Bunch-Davies state \cite{Bunch:1978yq} (a.k.a. Hartle-Hawking or Euclidean state \cite{Hartle:1983ai}). In the past twenty years, these correlators have been studied using the so-called \textit{in-in formalism}, following suggestions in \cite{Maldacena:2002vr,Weinberg:2005vy} (see \cite{Chen:2017ryl} for a review). This formalism had been developed much earlier in the pioneering work of Schwiger \cite{Schwinger:1960qe}, Keldysh \cite{Keldysh:1964ud} and Feynman and Vernon \cite{feynman2000theory} for the study of out-of-equilibrium open quantum systems (see e.g. \cite{breuer2002theory,kamenev2023field} for modern textbooks). The main raison d'être of this formalism is to account for the exchange of energy and information between an open system and its environment, which leads to dissipation, fluctuations and non-unitary evolution. Ironically, the vast majority of applications of the in-in formalism to cosmology have been restricted to closed quantum systems undergoing unitary, non-dissipative time evolution. Here we point out that, for these non-dissipative systems, we actually have the alternative and equivalent option of using the \textit{in-out formalism}, which is more familiar to many from the study of scattering amplitudes.\\

Indeed, correlators in the Heisenberg picture have no allegiance to in-in or in-out: they are just correlators of operators on a given state, which is often taken to be the ``vacuum" $\ket{\Omega}$ of the interacting theory. The catch is that often we don't know non-perturbatively what $\ket{\Omega}$ is. Instead, we approximate it via an iteratively perturbative expansion. This is most apparent in the so-called interaction picture, where the easy evolution described by the quadratic Hamiltonian is accounted for by working with free fields and the difficult non-linear interactions appear in the preparation of the bra and ket of a correlator. It is here that we face a choice. We can prepare the bra and the ket by adiabatically turning on interactions in the infinite past and evolving forward or in the infinite future and evolving backwards. A natural choice may be to mimic the physical problem under investigation. For example in a scattering experiment we like to think of particles coming from the infinite past and wandering off to the infinite future and the in-out formalism fits this intuition. But we don't have to make this choice. We can perfectly well use the Lehmann–Symanzik–Zimmermann (LSZ) formula on the in-in correlators instead. A major advantage of the in-out formalism is that it minimizes bookkeeping: all operators are in the same time ordering, whether they are fields or interactions in the Hamiltonian. As a consequence, each Feynman diagram corresponds to a single product of propagators and vertices and most importantly all propagators are time-ordered Feynman propagators. \\

\begin{figure}
 \centering
 \includegraphics[width=\textwidth]{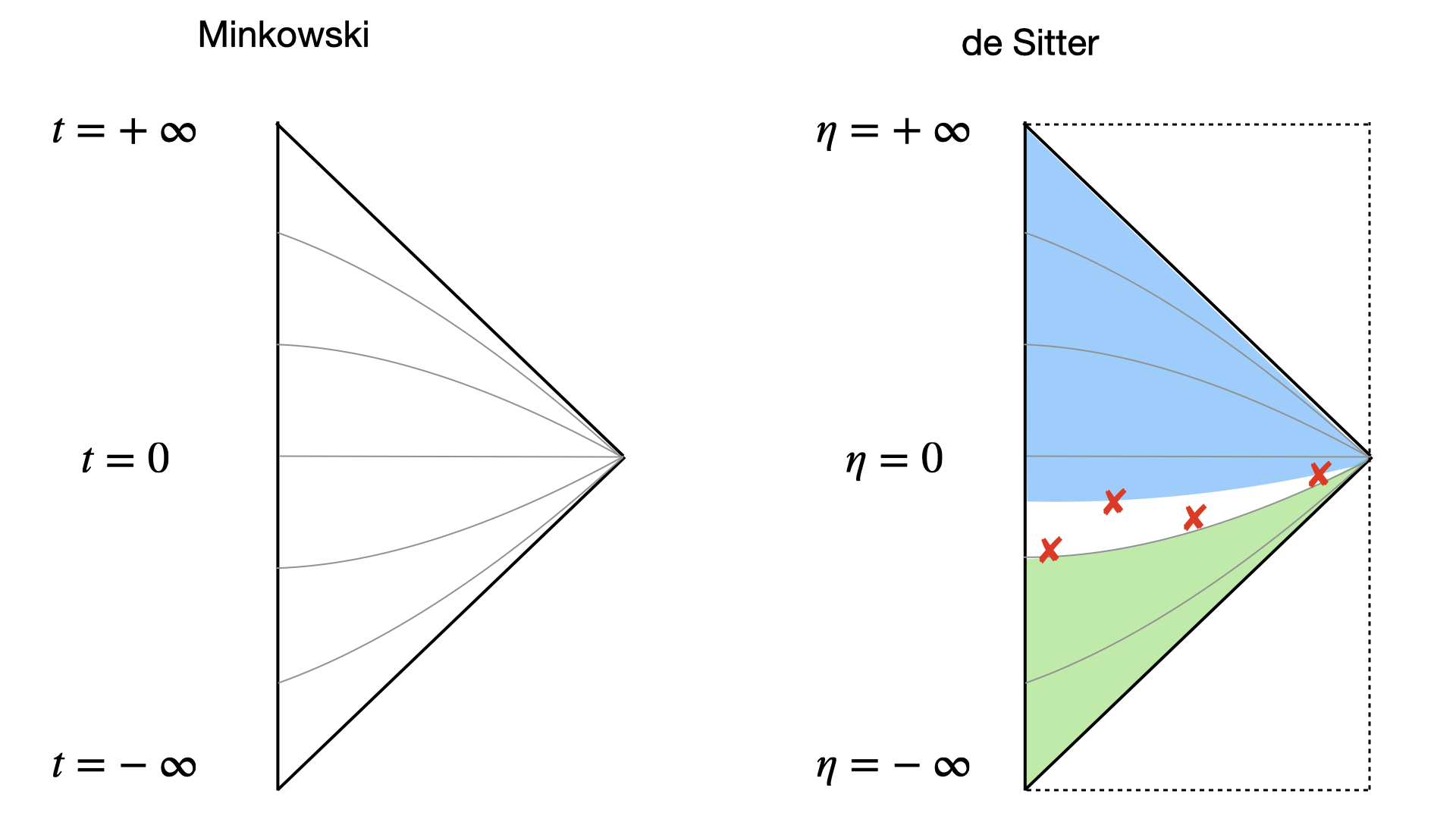}
 \caption{The Penrose diagrams of Minkowski (left) and de Sitter spacetime (right). The de Sitter diagram has been extended with an additional copy of the contracting Poincar\'e patch that allows one to prepare the bra (blue-shaded region) using the in-out formalism. The green shaded region is the preparation of the ket and the red crosses represent insertions of local operators.}
 \label{figextdS}
\end{figure}

When working with a given physical system, there may be limits in which it is consistent to adiabatically turn on and off interactions and limits in which this is not possible. For example, for correlators in de Sitter spacetime in the expanding Poincar\'e patch, as relevant for cosmology, we have the spacelike future conformal boundary and the null past cosmological horizon (which can be reached in finite proper time but always with an infinite proper volume \cite{Marolf:2012kh}). The condition of starting with the Bunch-Davies state tells that we can prepare both the bra and the ket by evolving from the Fock vacuum on the past cosmological horizon and this is why this has been the prominent choice in the literature so far. Conversely, it seems more complicated to turn off interactions towards the future conformal boundary because of the phenomenon of particle production and the decay instability of particles of any mass. It turns out that these are not insurmountable obstacles and concrete constructions have been devised to obtain a well-defined set of amplitudes \cite{Marolf:2012kh,Melville:2023kgd}. Here instead we explore a different possibility: we prepare the ket from the past null cosmological horizon as usual, but instead \textit{we prepare the bra from the future null conformal horizon of an auxiliary contracting Poincar\'e patch}, as shown in the right panel of Fig.~\ref{figextdS}. The similarity to Minkowski spacetime is evident. With this extended spacetime in mind, we set up the in-out formalism and we prove it's equivalent to the traditional in-in formalism, in the context of QFT in curved spacetime.\\

Our main motivation to develop an in-out formalism was to find a non-perturbative optical theorem that can be used in cosmology to constrain low-energy theories that admit a standard UV completion. At the perturbative level, some consequences of unitary time evolution in FLRW spacetime have been understood in the form of the cosmological optical theorem \cite{COT,Cespedes:2020xqq,sCOTt,Goodhew:2021oqg,Bonifacio:2021azc,Pimentel:2022fsc,Jazayeri:2022kjy}. This has been useful to bootstrap \cite{Baumann:2022jpr} perturbative correlators \cite{MLT,Bonifacio:2022vwa,Cabass:2022rhr,Baumann:2021fxj,DuasoPueyo:2023viy}, but it is insufficient to derive general positivity bounds, where one constrains an unknown and not-necessarily perturbative UV-completion. Beautiful progress on the non-perturbative side has been obtained in \cite{Hogervorst:2021uvp,DiPietro:2021sjt} using the K\"ahllen-Lehman representation (see also \cite{Bros:1990cu,Hollands:2011we,Loparco:2023rug}) and related ideas that leverage group theory and harmonic analysis. In this work, a non-perturbative constraint from unitarity can be obtained because the textbook derivation of the optical theorem in Minkowski also applies to our de Sitter scattering matrix (see Sec.~\ref{sec:Smatrix}), which is naturally defined in our in-out formalism. We only preview some results about this topic, which will be discussed in a separate publication. 

%%%%%%%%%%%%%%%%%%%%%%%%%%%%%%%%%%%%%%%%%%%%%%%%%%%%

\paragraph{Shopping advice} So you have to compute a few correlators explicitly, but don't know what formalism to choose? Here are some handy top tips to consider before you start scribbling away on your tablet:
\begin{itemize}
 \item In-in formalism.
 \begin{itemize}
 \item[Pros:] It can handle dissipation, fluctuations and non-unitary evolution. Indeed this was the main reason why this formalism was developed (see e.g. \cite{kamenev2023field} and \cite{Liu:2018kfw}). Even if the evolution of a ``closed" system such as the universe is expected to be unitary (but see \cite{Cotler:2022weg,Cotler:2023eza} for a different point of view), if we only observe part of the systems we are working with an \textit{open system} and at the quantum level we need to use words such as density operator, Liouvillian, pure-to-mix state evolution and Markovian approximation.
 \item[Cons:] The large number of propagators (four bulk-bulk and two bulk boundary) and the exponential proliferation of labellings of diagrams ($2^{V-1}$ for a diagram with $V$ vertices) are a considerable nuisance. Moreover, we miss out on a lot of the intuition coming from the extensive study of scattering amplitudes.
 \end{itemize}
 \item In-out formalism.
 \begin{itemize}
 \item[Pros:] There is a single propagator! And it's everyone's favourite: the time-ordered Feynman propagator. The $2^{V-1}$ contributions of the in-in calculation are nicely repackaged into a ``single" (nested) integral expression\footnote{To be fair we should point out that in the time domain, the Feynman propagator has two time orderings. However in the frequency/energy domain in Minkowski both are captured by a single term, courtesy of the Feynman $i\e$ prescription, $1/(p^2+i\e)$.}. 
 \item[Cons:] It cannot handle dissipation. For scattering experiments, this is not a problem because of the excruciating care that experimentalists put into shielding particle collisions from the rest of the world. Conversely, in less artificial systems such as many condensed matter systems and cosmology, this limitation prevents us from accessing many beautiful phenomena. 
 \end{itemize} 
\end{itemize}

%%%%%%%%%%%%%%%%%%%%%%%%%%%%%%%%%%%%%%%%%%%%%%%%%%%%

\begin{figure}
 \centering
 \includegraphics[width=\textwidth]{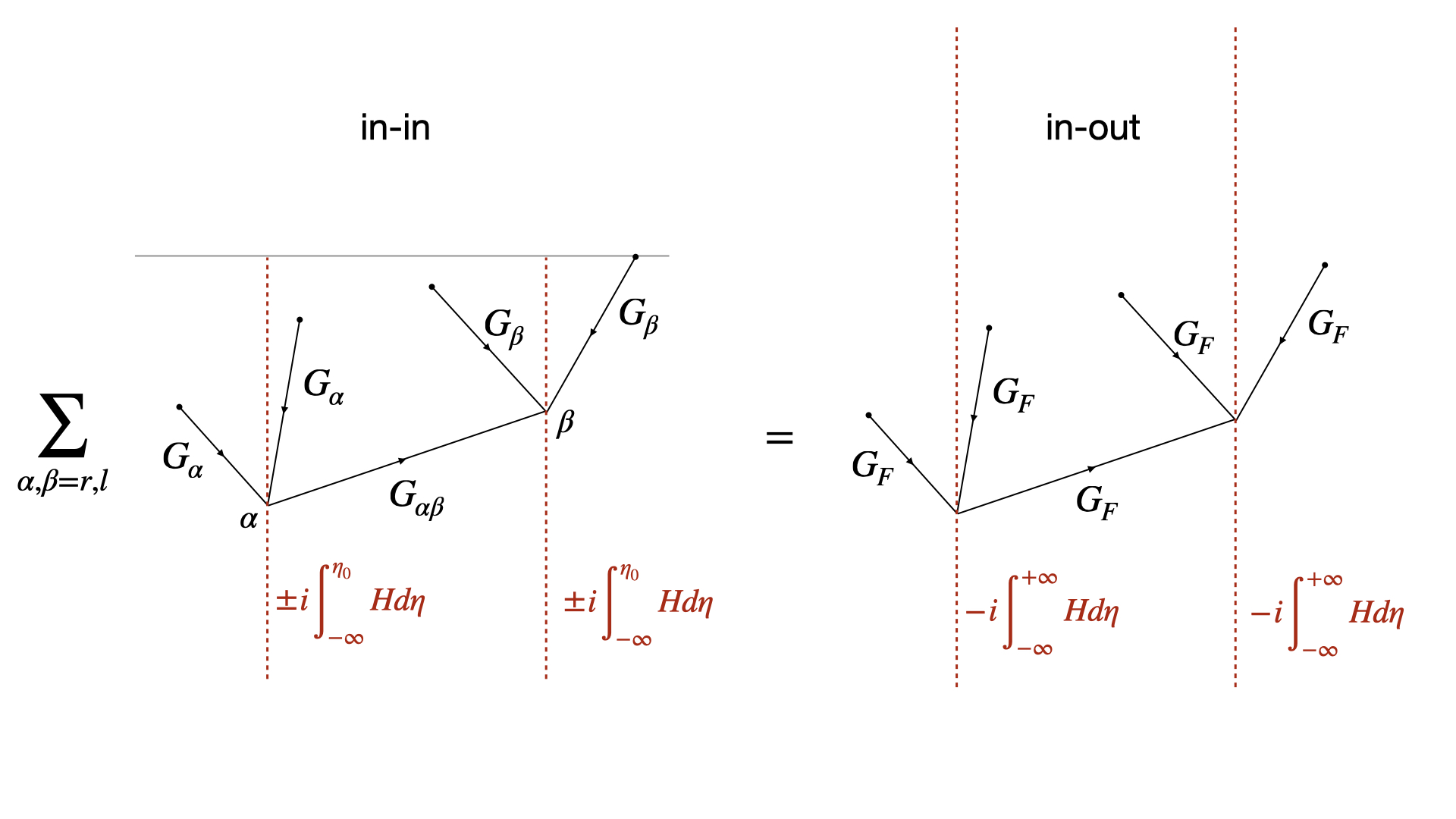}
 \caption{An example of the equivalence of the in-in and in-out formalism. Left: the in-in formalism requires summing over all possibilities to label vertices as ``$l$" (left) or ``$r$" (right); the bulk-boundary propagators $G_{r}$ and $G_{l}$ and bulk-bulk propagators $G_{ll}$, $G_{lr}$, $G_{rl}$ and $G_{r}$ are then chosen accordingly. Right: the in-out formalism is a single expression where all propagators are Feynman propagators.}
 \label{fig2}
\end{figure}

\paragraph{Summary of the results} For the convenience of the reader we summarize here our main results:
\begin{itemize}
 \item We developed an in-out formalism to compute unequal time correlators in Minkowski, de Sitter and more general cosmological spacetimes. The formalism crucially assumes the absence of late-time IR divergences, which in practice means that the divergence of the flat-slicing volume at future infinity, $\eta \to 0$, has to be offset by the decay of fields and their derivatives (see discussion around~\eqref{Delta}). The formalism applies to fields of any mass and spin. The Feynman rules, outlined around~\eqref{inoutrules}, are the same as for the standard in-out Minkowski correlators (also the same as for amplitudes except one does not amputate external legs). In particular, all lines, both internal and external, correspond to a time-order Feynman propagator. As compared to the in-in formalism, this removes the need to sum over all the possible ways to label each vertex as ``left" and ``right". We presented a formal argument for the equivalence of in-in and in-out formalism and some explicit checks in perturbation theory. The equivalence is depicted in Fig.~\ref{fig2}.
 \item We used the in-out formalism to provide two new procedures to compute correlators, and for concreteness we focus on equal time products of scalars. These results are obtained with similar manipulations as in a parallel study of the wavefunction \cite{Arkani-Hamed:2017fdk}. The first procedure, which we dub ``pole bagging", leverages the simplicity of the Feynman propagator in the energy-momentum domain in Minkowski to write a (loop integrand of a) correlator as a sum of residues of products of propagators. This can also be extended to massless and conformally coupled scalars in de Sitter. The second procedure consists of an algebraic recursion relation that iteratively removes internal lines of a diagram reducing it to linear combinations of simpler diagrams. The main novelty of our results is that we work directly at the level of the observable correlators, rather than the somewhat more primitive wavefunction.
 \item Time-ordered and anti-time-ordered products of operators are related by an operator identity\footnote{This was used in a related context in \cite{Gillioz:2016jnn,Meltzer:2020qbr}, where it was called ``CFT optical theorem". },~\eqref{CFTopt}, which is sometimes equivalently stated as the ``largest time equation" \cite{Veltman:1994wz} and leads to the amplitude cutting rules. We use this identity to derive an infinite number of cutting rules for correlators in Minkowski and cosmological spacetimes including de Sitter, where we restrict to massless and conformally coupled scalars. The number of correlator cutting rules grows quickly with the complexity of the diagram and we provide a systematic study of one- and two-vertex diagrams (see Sec.~\ref{sec:cuts} and the summary~\eqref{finaleq}) to all loop orders and discuss three-vertex diagrams in App.~\ref{app:C}.
 \item The in-out formalism suggests a straightforward definition of a de Sitter S-matrix describing scattering from the past to the future null horizon (see also \cite{Melville:2023kgd} for the discussion of a similar but not identical object). We show some simple examples for a number of conformally coupled scalars. A main advantage of our definition is that amplitudes satisfy the standard optical theorem. Moreover, the symmetry between the initial and final states ensures that the imaginary part of the forward scattering amplitudes, which in Minkowski becomes a discontinuity under appropriate analyticity assumptions, is positive because of unitarity (see Sec.~\ref{sec:Smatrix}). A dedicated analysis will appear in a separate publication.
\end{itemize}

%%%%%%%%%%%%%%%%%%%%%%%%%%%%%%%%%%%%%%%%%%%%%%%%%%%%%%%%%%%%%%%%%%%%%%%%%

\paragraph{Relation to previous work} Some aspects of our discussion are closely related to previous work (see e.g. \cite{Landsman:1986uw} for a review of in-in, in-out and Euclidean formalism in the finite temperature context). In particular, in a series of nice papers initiated with \cite{Marolf:2010zp,Marolf:2012kh}, Marolf and Morrison studied perturbation theory in de Sitter. Particularly relevant for us is their construction of an S-matrix for \textit{global} de Sitter spacetime by glueing together a contracting and an expanding Poincar\'e patch \textit{along the common cosmological horizon}. An interesting aspect of their setup is that the cosmological horizon of the Poincar\'e patch can be reached from anywhere in global dS in a \textit{finite} proper time. Hence it is natural to extend the path integral contour right through this surface. Here we take a similar but complementary route by glueing two Poincar\'e patches along their future/past conformal boundaries (see Fig.~\ref{figextdS}). This has the advantage that our path integral contour runs straight, just like in Minkowski. In particular, our path integral does not bend on itself creating so-called ``timefolds" and hence we avoid the associated proliferation of propagators and labelling of interaction vertices. Conversely, since it takes an infinite amount of proper time to reach the future conformal boundary, in our construction perturbations can move from one Poincar\'e patch to the other only in a conformal sense, i.e. only after jettisoning a divergent conformal factor. While we concede that this makes the physical interpretation less intuitive, we don't think this is a problem because all operators are inserted in a single patch and the second patch is just invoked as an equivalent preparation of the bra of the correlator. Moreover, we postpone to future work a discussion of how to apply an LSZ-like reduction to correlators in our in-out formalism to derive an S-matrix. 

The ``extended" spacetime in Fig.~\ref{figextdS} is reminiscent of the suggestion of so-called ``conformal cyclic cosmology" \cite{penrose2010cycles}. Here however the (upper) contracting Poincar\'e patch is just auxiliary and all operators are inserted in the (lower) expanding patch. In passing we do notice that correlators of a Lorentzian conformal field theory naturally live on an infinite cylinder, rather than on a single copy of the conformal Minkowski diagram \cite{Luscher:1974ez}. This ensures that finite special conformal transformations don't violate causality (see e.g. \cite{Gillioz:2022yze}). Since de Sitter is conformally flat it is equally natural to consider an infinite conformal extension, of which we are only making partial use here.

The in-out formalism suggests a natural definition of a dS scattering matrix, with asymptotic states in the past and future null infinities of Fig.~\ref{figextdS}. This S-matrix turns out to be quite similar to the ``Bunch-Davies S-matrix" recently discussed in \cite{Melville:2023kgd}. We comment on this in Sec.~\ref{sec:Smatrix}.\\

\begin{figure}[h]
 \centering
 \includegraphics[width=0.9\textwidth]{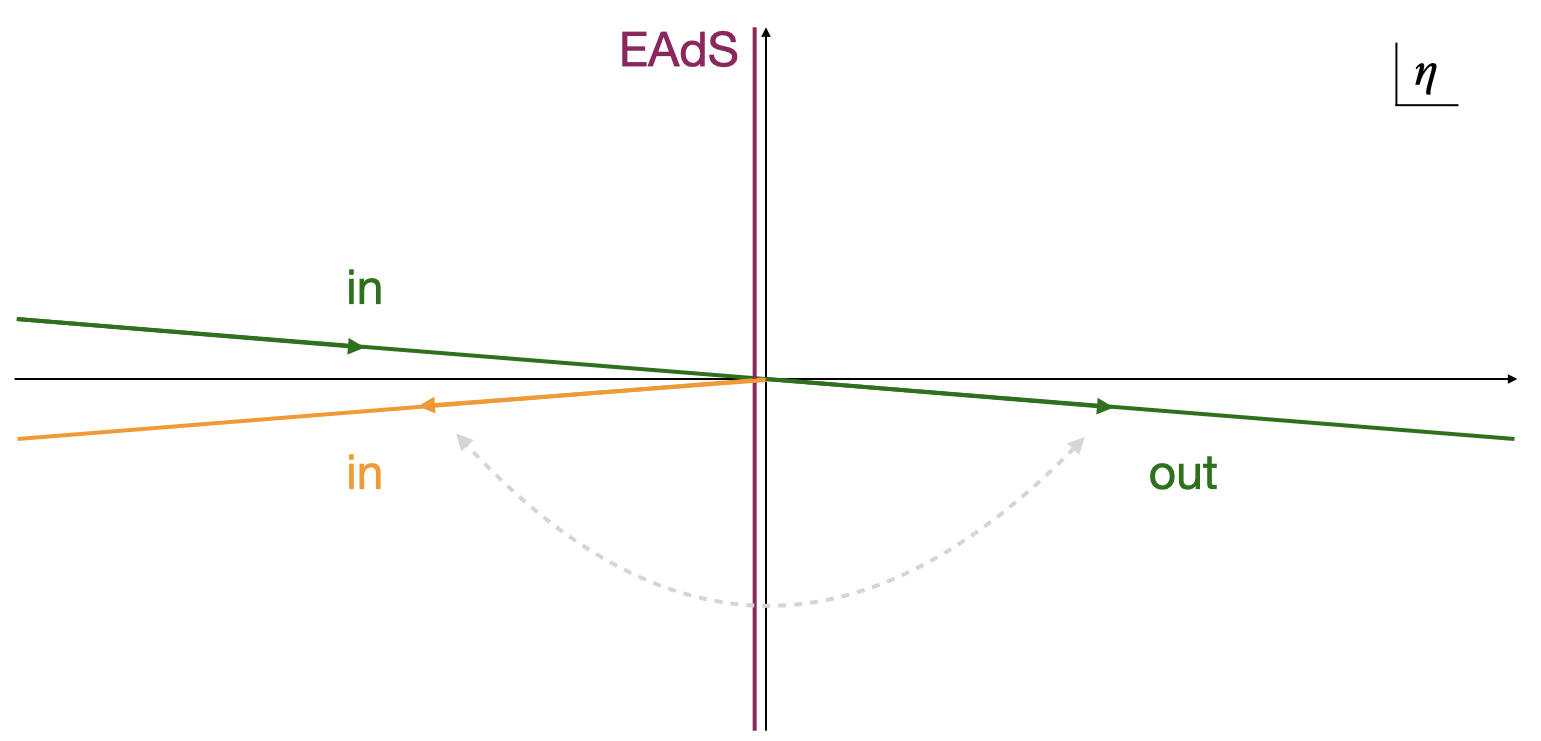}
 \caption{The figure shows various equivalent contours that have been proposed for the calculation of cosmological correlators. In the traditional Schwinger-Keldysh formalism, the path integral runs forwards over the ``in" green line and then backwards over the "in" orange line. Rotating both contours to the imaginary axis connects to the Euclidean AdS calculation. The "in-out" contour discussed in this paper (the green line) rotates the backwards "in" contour to the forwards "out" contour.}
 \label{fig:contours}
\end{figure}
We are not the first to propose a rotation of the in-in contour that leads to a simplification of the calculation. One proposal, going back to \cite{Maldacena:2002vr} and then fully developed in \cite{Sleight:2021plv} and 
\cite{DiPietro:2021sjt}, is to straighten out the closed-time contour by rotating the time-ordered and anti-time-ordered branches by 90 degrees counterclockwise and clockwise respectively, so that the contour coincides with the purely imaginary axis of the complex $\eta$ plane (see Fig.~\ref{fig:contours}). The result is then precisely related to a perturbative calculation in Euclidean AdS, where the imaginary part of $\eta$ is interpreted as the radial coordinate. A second and related proposal was put forward in \cite{Behbahani:2012be} and used again in \cite{Green:2013rd}. It consists of the same contour rotation where one recognizes that the time and anti-time ordering of the in-in contour combine into a single anti-time ordered Euclidean Green's function. Our proposal in this work shares with previous proposals the idea of straightening the in-in contour so that the multiple propagators reduce to a single one. In contrast to previous proposals our rotation of the anti-time ordered in-in contour goes all the way to the positive real axis, so that the calculation remains firmly within the realm of Lorentzian time. All these proposals are summarised and compared in Fig.~\ref{fig:contours}. Finally, we should mention that the relation between in-in and in-out formalisms in Minkowski has also been investigated recently in \cite{Chaykov:2022pwd,Chaykov:2022zro}, where it was shown, among other things, that imposing initial conditions in the infinite past is a necessary requirement.\\

%%%%%%%%%%%%%%%%%%%%%%%%%%%%%%%%%%%%%%%%%%%%%%%%%%%%%%%%%%%%%%%%%%%%%%%%%%

The rest of this paper is organized as follows. In Sec.~\ref{sec:inininout} we define the in-out formalism and prove that it gives the same result for time-ordered unequal time correlators as the in-in formalism. To this end, we review a formal non-perturbative argument in Minkowski and adapt it to de Sitter and then provide explicit checks of the equivalence to various orders in perturbation theory at tree level. In Sec.~\ref{sec:mink} we consider two applications of the in-out formalism to correlators. The first, in Secs.~\ref{poleb} and \ref{polebdS} is a representation of correlators as a sum of residues of the product of Feynman propagators in the energy-momentum domain. The second, in Sec.~\ref{ss:recursions}, is a set of purely algebraic set of recursion relations for equal-time Minkowski correlators, which computes all tree-level diagrams and a large class of ``melonic" loop diagrams. Next, in Sec.~\ref{sec:cuts}, we use Veltman's largest time equation to derive an infinite set of propagator identities, which can in turn be expressed in terms of correlator identities. We present explicit formulae for all two-vertex diagrams to any number of loops and external legs. In Sec.~\ref{sec:Smatrix} we give a preview of how the in-out formalism leads to a natural definition of a de Sitter scattering matrix, which moreover obeys the standard optical theorem. We present some simple examples and consistency checks. Finally, we conclude in Sec.~\ref{concl}.

%%%%%%%%%%%%%%%%%%%%%%%%%%%%%%%%%%%%%%%%%%%%%%%%%%%%%%%%%%%%%%%%%%%%%%%%

\paragraph{Notation and conventions} We use a prime to remove the ubiquitous momentum-conserving Dirac delta,
\begin{align}
 \ex{\prod_a^n \phi(\eta,\bfk_a)}\equiv (2\pi)^3 \delta\left( \sum_a^n \bfk_a \right) \ex{\prod_a^n \phi(\eta,\bfk_a)}'\,.
\end{align}
We denote time-ordered correlators by
\begin{align}
 \ex{T\prod_a^n \phi(\eta,\bfk_a)}'\equiv B_n(\{\eta_a,\bfk_a\})\,,
\end{align}
where $\{\eta_a,\bfk_a\}$ collectively refers to the spacetime positions of the operators. We define the following two-point functions or propagators as
\begin{align}
 G^+(\eta_1,\eta_2,p)\equiv &\bra{0} \phi(\eta_{1},\bfp)\phi(\eta_{2},\bfp') \ket{0}'=f_{p}(\eta_{1})f_{p}^{\ast}(\eta_{2})\,, \\ \nonumber
 \Gf(\eta_1,\eta_2,p)\equiv &\bra{0} T\phi(\eta_{1},\bfp)\phi(\eta_{2},\bfp') \ket{0}'=f_{p}(\eta_{1})f_{p}^{\ast}(\eta_{2})\theta(\eta_{1}-\eta_{2})+f_{p}^{\ast}(\eta_{1})f_{p}(\eta_{2})\theta(\eta_{2}-\eta_{1})\,,
\end{align}
where the mode functions in de Sitter and Minkowski are\footnote{Here we restrict our discussion to positive masses, but see \cite{McCulloch:2024hiz} for an extensive discussion of tachyonic fields and their phenomenology.}
\begin{align}\label{modefs}
f_{k}(\eta)&=-i\frac{\sqrt{\pi}H}{2}e^{i\frac{\pi}{4}(1+2\nu)}(-\eta)^{3/2}H^{(1)}_{\nu}(-k\eta)\,,\quad \nu\equiv\sqrt{\frac{9}{4}-\frac{m^{2}}{H^{2}}}\,, \\
f_{k}(\eta)&=i\eta\frac{H}{\sqrt{2k}}e^{-ik\eta}\,. \quad \text{(conformally coupled)}\,, \label{ccmode} \\ \label{masslessmode}
f_{k}(\eta)&=\frac{H}{\sqrt{2k^{3}}}(1+ik\eta)e^{-ik\eta} \quad \text{(massless, dS)}\,, \\
f_{E}(t)&=\frac{e^{-iEt}}{\sqrt{2E}}\,. \label{Minkmodef} 
\end{align}
Many of the integrals we encounter lead to distributions rather than functions and should be understood as acting on appropriate test functions. As commonly done in the physics literature, we will often represent these distributions as the limit of functions using a small parameter that is taken to zero at the end of the calculation as for example in the Sokhotski–Plemelj theorem or the Feynman propagator. More in detail, we specify that a certain integral to $t,\eta = \pm \infty$ should be computed assuming a positive or negative imaginary part to guarantee convergence even in the absence of a test function. To convey this in a compact way we use the shorthand notation
\begin{align}
 \pm \infty_+ &\equiv \pm \infty (1+i\e)\,, & \pm \infty_- &\equiv \pm \infty (1-i\e)\,,
\end{align}
where $\e>0$ is a real and positive parameter that should be taken to zero at the end of the derivation.

%%%%%%%%%%%%%%%%%%%%%%%%%%%%%%%%%%%%%%%%%%%%%%%%%%%%%%%%%%%%%%%%%%%%%%%%%%%%%%%%%%%%%%%%%%%%

\section{In-in equals in-out}\label{sec:inininout}

The main motivation behind the construction of the in-in formalism by Keldysh \cite{Keldysh:1964ud}, foreshadowed by the work of Schwinger \cite{schwinger1961brownian}, was to describe the non-unitary and dissipative evolution of an open system in contact with an environment. However, it can also be used to study the unitary evolution of isolated systems and indeed almost all applications of the in-in formalism in early universe cosmology have considered a situation of this type. As we discussed, for the unitary evolution of an isolated system, there is an equivalent in-out description that computes correlators. In the following we review a general non-perturbative argument for the equality between in-in and in-out, we specify the conditions under which this holds and finally, we present some explicit checks.\\

First of all, let's clarify some nomenclature. In the \textit{interaction picture}, where states evolve according to the interaction Hamiltonian and operators according to the free Hamiltonian, we say that a certain matrix element is an \textit{in-out correlator} if it contains a single time\footnote{Here we focus on time-ordering as opposed to path ordering to set up our nomenclature. When re-writing these correlators as path integrals one can sometimes combine different time orderings into a single path ordering, as famously done with the closed time contour of the in-in path integral.} ordering. This corresponds to a single time evolution operator, possibly with some insertion of local operators. Conversely, \textit{in-in correlators} contain two separate time orderings, one going forward in time, which evolves the ket, and one going backwards in time, which evolves the bra. With this nomenclature in place, we move on to define our main objects of study. \\

We define an \textit{in-out correlator} of the product $\O(\{t,\bfx\})$ of local operators at positions $\{t_a,\bfx_a\}$ for $a=1,\dots,n$, as the following expectation value in the interaction picture
\begin{tcolorbox}[colback=gray!10!white, colframe=white]
\vspace{-1em}
\begin{align}\label{gio}
\Gio \equiv \frac{\bra{0} T\left[ \O(\{t,\bfx\})e^{-i\int_{-\infty (1- i\e)}^{+\infty (1- i\e)} \Hi dt } \right] \ket{0}' }{ \bra{0} T\left[ e^{-i\int_{-\infty (1- i\e)}^{+\infty (1- i\e)} \Hi dt } \right] \ket{0}' }\,,
\end{align}
\end{tcolorbox}
where a prime removes the Dirac delta of momentum conservation. Here $\Hi $ is the interaction Hamiltonian that generates time evolution of states, $T$ denotes time ordering (early to the right, late to the left) and $\ket{0}$ is the Fock vacuum, a.k.a. the ``vacuum" of the free theory. Notice that the factor in the denominator does not depend on where the fields are inserted: it is the familiar sum over so-called \textit{vacuum bubbles}. Heuristically one can simply justify it by demanding that $\Gio$ becomes unity when we don't insert any operators. In diagrammatic language, the denominator in $\Gio$ simply tells us that we should disregard all diagrams that contain a subdiagram that is not connected to any inserted operator, i.e. a vacuum bubble. More precisely this factor arises when we re-write the vacuum of the interacting theory $\ket{\Omega}$ as that of the free theory $\ket{0}$ with interactions turned on adiabatically. The adiabatic turning on and off of interactions in the asymptotic past and future is implemented by the $i\e$ rotations of the boundary of integration in~\eqref{gio}. The correct sign of this imaginary part, namely $\pm\infty (1- i\e)$ can be simply determined by demanding convergence of the integral in perturbation theory.\\

Typically we will consider operators of the form
\begin{align}
 \O(\{t,\bfx\})=\prod_a^n \phi(t_a,\bfx_a)\,.
\end{align}
Notice that $\Gio$ is a very familiar object in QFT: it is precisely the object appearing on the right-hand side of the Lehmann–Symanzik–Zimmermann (LSZ) reduction formula, which gets amputated and delivers scattering amplitudes in Minkowski. Since all the operators in~\eqref{gio} appear inside the same time ordering, $\Gio$ is computed in perturbation theory from the product of time-ordered two-point functions, a.k.a. Feynman propagators, stitched together by non-linear interactions. While we will say more about this later, for the moment the reader should have in mind that the time foliation of spacetime has been chosen such that $t=\pm \infty$ represents a null ``initial" surface and a null ``final" surface, as it is familiar from Minkowski. \\

Next, let's define \textit{in-in correlators} of the product $\O(\{t,\bfx\})$ of local operators at positions $\{t_a,\bfx_a\}$ as the following expectation value in the interaction picture
\begin{align}\label{gii}
\Gii \equiv \bra{0} \bar{T} \left[e^{+i\int_{-\infty (1 + i\e)}^{t_{0}} \Hi dt } \right]T\left[ \O(\{t,\bfx\}) e^{-i\int_{-\infty (1- i\e)}^{t_{0}} \Hi dt }\right] \ket{0}'\,.
\end{align}
Here $\bar\T$ indicates anti-time ordering (early to the left, late to the right) and $ t_{0}$ is an arbitrary time that is \textit{later} than any of the $ t_{a}$ appearing in the inserted fields, $t_0>t_a$. Notice the absence of the vacuum bubble factor appearing in $\Gio$. Heuristically this is easily understood noticing that if we insert the identity operator, namely $\O=1$, then $\Gii$ is already equal to 1, just like $\Gio$. The imaginary part in the boundary of integration, namely $-\infty(1-i\e)$ in the time-ordered part and $-\infty (1 +i\e) $ in the anti-time-ordered part, ensure convergence and implement mathematically the adiabatically switching on of interactions in the infinite past.

Note that our definition of $\Gii$ is slightly more general than what is usually considered in the cosmological literature because we allow for unequal time operators. An even more general possibility would be to insert some operators in the time ordering, some in the anti-time-ordering and some in between\footnote{The case where all operators are in the anti-time-ordering is trivially related to the case we consider in this paper}. In this case, we don't expect it to be possible to find a simple relation to the ``straight" in-out formalism. However these situations are examples of out-of-time-order contours or ``time-folds" and have also been studied extensively (see e.g. \cite{Heemskerk:2012mn,Haehl:2017qfl}).

As long as $ t_{0}>\bar t$ with $\bar t = \text{max}_a(t_a)$, $ \Gii$ does not depend on $ t_{0}$ because the time evolution after the latest $ t_{a}$ cancels out,
\begin{align}
 \Gii &= \ex{ \bar{T} \left[ e^{+i\int_{-\infty}^{\bar t} \Hi \,dt} \right] U^{-1}(t_0,\bar t) U(t_0,\bar t) T\left[ \O(\{t,\bfx\})e^{-i\int_{-\infty}^{\bar t} \Hi \,dt} \right]}' \\
 &= \ex{ \bar{T} \left[ e^{+i\int_{-\infty}^{\bar t} \Hi \,dt} \right] T\left[ \O(\{t,\bfx\}) e^{-i\int_{-\infty}^{\bar t} \Hi \,dt} \right] }'\,,
\end{align}
where we used the following notation for the time evolution operator $U$ in the interaction picture
\begin{align}
\bar T\exp \left[ +i\int_{t_{1}}^{t_{2}} \Hi dt \right] = U^{\dagger}(t_{2},t_{1})=U^{-1}(t_{2},t_{1})\quad \text{ for } t_{2}>t_{1}\,.
\end{align}
Notice that our definition of $ \Gii $ is slightly more general than what is usually encountered in the cosmology literature. Usually one considers the product of operators at the same time. In that case, it does not matter if the operators are inside the time ordering, the anti-time-ordering or just in between,
\begin{align}
 \ex{ \bar{T} \left[ e^{+i\int_{-\infty}^{\bar t} \Hi \,dt} \right] T\left[ \O(\bar t)e^{-i\int_{-\infty}^{\bar t} \Hi \,dt} \right]}
 &= \ex{ \bar{T} \left[ e^{+i\int_{-\infty}^{\bar t} \Hi \,dt} \right] \O(\bar t)T\left[ e^{-i\int_{-\infty}^{\bar t} \Hi \,dt} \right] }\\
 &= \ex{ \bar{T} \left[ e^{+i\int_{-\infty}^{\bar t} \Hi \,dt} \O(\bar t) \right] T\left[ e^{-i\int_{-\infty}^{\bar t} \Hi \,dt} \right] } \,.
\end{align}
Also, we usually don't write the time ordering of operators when computing in-in correlators because we compute them at the same time. When we extend to unequal time of course we can choose whether to time order or not. The object that nicely relates to $ \Gio$ is the time-ordered in-in correlator.

%%%%%%%%%%%%%%%%%%%%%%%%%%%%%%%%%%%%%%%%%%%%%%%%%%%%%%%%%%%%%%%%%%%%%%%%%%%%%%%%%%%%%%%%%%%%%%%%%%%%

\subsection{A formal proof} 

Here is a formal proof that $ \Gio=\Gii $ following the intro of \cite{kamenev2023field}. First we notice that if we start with the Fock vacuum $ \ket{0}$, which is annihilated by the lowering ladder operators of all fields for all momenta, and turn on and off interactions adiabatically, we expect to go back to $ \ket{0}$ up to a multiplicative factor corresponding to the sum over vacuum bubbles\footnote{Here we are in the interaction picture and $U$ is the associated time-evolution operator, which depends on the interaction Hamiltonian $H$ evaluated on free fields (i.e. fields evolved with the free Hamiltonian $H_0$).}
\begin{align}
U(+\infty,-\infty)\ket{0}=\ket{0} \bra{0} U(+\infty,-\infty)\ket{0}\,, \label{belief}
\end{align}
where we are assuming the normalization $\bra{0}\ket{0}=1$. This expectation can be justified in a few ways. 
Here we simply remark that the standard $i\epsilon$ rotation of the time integral contour induces a suppression $e^{-2\e T (E_n-E_0)}$ for time evolution $U(T,-T)$ on excited energy eigenstates with eigenvalues $E_n$, compared to the lowest energy state. In the limit $T\to +\infty$ all excited states are projected out and only the Fock vacuum survives. To see this more explicitly, we compute in perturbation theory the projection of the left-hand side of~\eqref{belief} onto an n-particle state $\ket{n}$, with total energy $k_T$. To lowest order in the coupling constants, the result\footnote{The exact result is given in \eqref{contactA} where we also discuss that this matrix element can be interpreted as a 0 to $n$ scattering in de Sitter.} of the time integral is proportional to derivatives of a Dirac delta of energy conservation
\begin{align}
\bra{n} U(+\infty,-\infty)\ket{0} \propto \partial^n_{k_T} \delta(k_T)\,.    
\end{align}
This distribution does not have support on physical states and hence $U(+\infty,-\infty)\ket{0}$ does not have support on any excited state\footnote{This argument might be subtle due to IR divergences in the initial and final state, $|\eta|\to \infty$ (not at $\eta\to 0$), where the analogue of soft and collinear divergences might arise. We hope to come back to this issue in the future.}.

Since unitary evolution preserves the norm, the multiplicative factor on the right-hand side of~\eqref{belief} is a pure phase
\begin{align}
 \bra{0} U(+\infty,-\infty)\ket{0}\bra{0} U(+\infty,-\infty)\ket{0}^* =1\,.
\end{align}
This is precisely the vacuum bubble factor, by which we divided out in the definition of in-out correlators $\Gio$,~\eqref{gio}. Taking the dagger of~\eqref{belief} (more precisely using the Riesz representation theorem) gives us
\begin{align}\label{adiabatic}
\bra{0}U^{\dagger}(+\infty,-\infty)=\bra{0} \bra{0} U(+\infty,-\infty)\ket{0}^*\,.
\end{align} 
The imaginary rotation of the contour at infinity is quite important in the derivation. To keep our expression compact we introduce the useful shorthand notation
\begin{align}
 \pm \infty_+ &\equiv \pm \infty (1+i\e)\,, & \pm \infty_- &\equiv \pm \infty (1-i\e)\,,
\end{align}
Then the derivation proceeds as follows
\begin{align}
\Gio&=\frac{\ex{ T \left[ \prod^{n}_{a}\phi(t_{a})e^{-i\int_{-\infty_-}^{+\infty_-} \Hi dt } \right]}'}{\ex{ T\left[ e^{-i\int_{-\infty_-}^{+\infty_-} \Hi dt } \right]}' } \\
&= \ex{ U^{\dagger}(+\infty_+,-\infty_+) T\left[ \prod^{n}_{a}\phi(t_{a})e^{-i\int_{-\infty_-}^{+\infty_-} \Hi dt } \right]}'\\
&=\ex{ \bar T \left[ e^{ +i\int_{-\infty_+}^{+\infty_+} \Hi dt } \right] \T\left[ e^{-i\int^{+\infty_-}_{t_{0}} \Hi dt } \right] \T\left[\prod^{n}_{a}\phi(t_{a})e^{-i\int_{-\infty_-}^{t_{0}} \Hi dt } \right]}'\\
&= \ex{ \bar T \left[ e^{+i\int^{t_{0}}_{-\infty_+} \Hi dt } \right] U^{\dagger}(+\infty,t_{0})U(+\infty,t_{0}) \T\left[\prod^{n}_{a}\phi(t_{a})e^{-i\int^{t_{0}}_{-\infty_-} \Hi dt } \right]}'\\
&=\Gii\,,
\end{align}
where in the second line we used~\eqref{adiabatic} and in the third line the fact that $ t_{0}>t_{a}$. \\

This equality can be stated with more evocative language. The only difference between the in-in and in-out formalism is how the bra is prepared. In both cases, we would like to prepare it by acting on the Fock vacuum with an evolution operator, which we can readily expand in perturbation theory for practical calculations. The in-in formalism prepares this bra by slowly turning on interactions from past (null) infinity. The in-out formalism instead slowly turns on interactions from future (null) infinity and evolves ``backwards" in time. 

\paragraph{An extended de Sitter spacetime} The equality of in-in and in-out correlators is not surprising in Minkowski and is already well-known (see e.g. \cite{kamenev2023field}). Here we claim that the same result also applies to de Sitter spacetime. In particular, we propose to extend the expanding Poincar\'e patch of dS by glueing on top of it a copy of the contracting Poincar\'e patch, as in Fig.~\ref{figextdS}. This means that the standard conformal time now can run over all real values, $-\infty < \eta< \infty$, with positive values representing the expanding patch, namely $a=e^{+Ht}$ in cosmological time, and negative values representing the contracting patch, $a=e^{-Ht}$. The two patches are glued together at $\eta=0$, which represents the future and past conformal boundary of the two Poincar\'e patches. To be absolutely explicit, we can hence define in-out correlators in de Sitter by
\begin{align}\label{giodS}
\Gio \equiv \frac{\bra{0} T\left[ \O(\{\eta,\bfx\})e^{-i\int_{-\infty (1-i\e)}^{+\infty(1-i\e)} \Hi (\eta) d\eta } \right] \ket{0}'}{ \bra{0} T\left[ e^{-i\int_{-\infty(1-i\e)}^{+\infty(1-i\e)} \Hi (\eta) d\eta } \right] \ket{0}'}\,.
\end{align}
This glueing at $\eta=0$ is consistent as long as no divergences take place at $\eta=0$. In turn, this means that the divergent volume factor $\sqrt{-g}=(H\eta)^{-4}$ coming from the measure of time integration must be more than offset by powers of $\eta$ coming from the inverse metric contracting space and time derivatives and by the decay of massive fields towards $\eta \to 0$. To be more specific, let the conformal dimension $\Delta$ of a scalar field of mass $m$ be\footnote{For an extension to $d$ spacial dimensions set $3\to d$. For the scaling of spinning fields see e.g. \cite{Lee:2016vti}.}
\begin{align}\label{Delta}
 \Delta\equiv \frac{3}{2}-\sqrt{\frac{3}{2}-m^2H^2}\,.
\end{align}
For an interaction involving $n$ fields of dimensions $\Delta_a$ with $a=1,\dots, n$ and a number $n_{\partial_i}$ of spacial derivatives the condition for tree-level IR finiteness is
\begin{align}
\sum_a^n \Delta_a+n_{\partial_i} > 3\,.
\end{align}
For massless fields, we have $\Delta=0$ and then the condition becomes
\begin{align}
2n_{\partial_\eta}+n_{\partial_i} > 3\,.
\end{align}
where $n_{\partial_\eta}$ is the number of time derivatives.

In the rest of the paper, we will assume that \textit{there are no IR divergences at $\eta \to 0$}. It is possible that our result can be extended also to the case of IR divergent interactions (recently discussed in \cite{Bzowski:2023nef}). This would require introducing an IR regulator that respects the relation in~\eqref{belief}. Our preliminary investigation suggests that such a regulator exists but we postpone a thorough discussion to future work.

%%%%%%%%%%%%%%%%%%%%%%%%%%%%%%%%%%%%%%%%%%%%%%%%%%%%%%%%%%%%

\subsection{In-in and in-out Feynman rules}

The Feynman rules are familiar both for the in-in and in-out formalism. Here we briefly review them for completeness. In both cases, we consider diagrams with $ V $ vertices, $ I $ internal lines, each connecting two vertices, and $ n $ external lines for an $n$-point correlator (time runs vertically with the past below the future). For concreteness, we use $\eta$ for time, but the same expressions apply to Minkowski changing $\eta$ to $t$. 

%%%%%%%%%%%%%%%%%%%%%%%%%%%%%%%%

\paragraph{In-in Feynman rules} Our definition of the un-equal time in-in correlators in~\eqref{gii} generalized what is typically used in cosmology, where fields are inserted at the same time. Hence the Feynman rules below are also slightly different in the case of un-equal time correlators but ``backward compatible" with the equal-time case. Then, for in-in correlators:
\begin{itemize}
\item A vertex can be either a ``right'' vertex, labelled by ``$ r $'' or a left vertex, labelled by ``$ l $''. Hence one needs to sum over $ 2^{V} $ possible labellings.
\item External lines are associated to a momentum $ \bfk_{a} $ with $ a=1,\dots,n $. Each vertex comes with a momentum-conserving Dirac delta. The $L=I-V+1$ internal ``loop" momenta are not fixed by these Dirac deltas and must be integrated over.
\item Internal lines are called bulk-to-bulk propagators and come in four types
\begin{align}
\bullet - \bullet = G_{rr}(\eta_{1},\eta_{2},p) &= \bra{0}T \phi(\eta_{1},\bfp)\phi(\eta_{2},\bfp') \ket{0}'=\Gf(\eta_1,\eta_2,\bfp)\\
&=f_{p}(\eta_{1})f_{p}^{\ast}(\eta_{2})\theta(\eta_{1}-\eta_{2})+f_{p}^{\ast}(\eta_{1})f_{p}(\eta_{2})\theta(\eta_{2}-\eta_{1}) \label{Grr} \\
\circ - \bullet= G_{lr}(\eta_{1},\eta_{2},p) &=\bra{0} \phi(\eta_{1},\bfp)\phi(\eta_{2},\bfp') \ket{0}'=f_{p}(\eta_{1})f_{p}^{\ast}(\eta_{2})\\
 \bullet-\circ=G_{rl}(\eta_{1},\eta_{2},p)&=\bra{0} \phi(\eta_{2},\bfp')\phi(\eta_{1},\bfp) \ket{0}'=G_{lr}^{\ast}(\eta_{1},\eta_{2},p) \\
\circ - \circ = G_{ll}(\eta_{1},\eta_{2},p)&=\bra{0} \bar T \phi(\eta_{1},\bfp)\phi(\eta_{2},\bfp') \ket{0}'=G_{rr}^{\ast}(\eta_{1},\eta_{2},p)\\
&=f_{p}^{\ast}(\eta_{1})f_{p}(\eta_{2})\theta(\eta_{1}-\eta_{2})+f_{p}(\eta_{1})f_{p}^{\ast}(\eta_{2})\theta(\eta_{2}-\eta_{1})\,,
\end{align}
where $f_p(\eta)$ are the mode functions, namely the solutions of the linearized classical equations of motion with momentum $\bfp$. For later use, we denote the non-time-ordered two-point function as 
\begin{align}
 G^+(\eta_1,\eta_2,p)\equiv &\bra{0} \phi(\eta_{1},\bfp)\phi(\eta_{2},\bfp') \ket{0}'=f_{p}(\eta_{1})f_{p}^{\ast}(\eta_{2})\,,
\end{align}
which is not symmetric under the exchange of the time variables. Notice that
\begin{align}
 G_{lr}&=G^+ = G_{rl}^* \,.
\end{align}
For spinning fields, it is convenient to strip off all polarization tensors from the propagators and move them to the vertices, where they are contracted with each other and with derivatives. Here $ G_{rr} $ is the Feynman propagator, and $ G_{ll} $ is its complex conjugate. 
\item External lines connected to external fields at times $\eta_a$ with $a=1,\dots,n$ are called bulk-to-boundary propagators and come in two types\footnote{Actually the bulk-boundary propagator $G_r$ coincides with the time-ordered bulk-bulk propagator $G_{rr}$ and the other bulk-boundary propagator $G_l$ coincides with non-time ordered bulk-bulk propagator $G_{lr}$.}
\begin{align}
\bullet -&=G_{r}(\eta,\eta_a,p)=\Gf(\eta,\eta_a,p)\,, \\ 
\circ- &=G_{l}(\eta,\eta_a,p)=G^+(\eta,\eta_a,p)\,.
\end{align}
Notice that $G_{r}$ is symmetric in its time arguments while $G_l$ is not. This is where our definition of in-in correlators generalizes the traditional one for equal-time correlators where $G_r$ is not time-ordered because one assumes that $\eta<\eta_a$.
\item The time $ \eta $ of a vertex must be integrated over $ d\eta $ with a measure $ \sqrt{-g}=(\eta H)^{-4} $ and boundaries 
\begin{align}
\text{right vertex: } -\infty(1-i\e)<\eta\leq 0 \,,\\
\text{left vertex: } -\infty(1+i\e)<\eta\leq 0 \,.
\end{align}
where we are assuming IR finiteness. As can be seen from~\eqref{gii}, right vertices come with a $-i$ times the appropriate interaction, while for left vertices we have a $+i$.
\item A discussion of the combinatorial factors can be found in \cite{Goodhew:2023bcu}.
\end{itemize}
It turns out that half of the above diagrams are related to the other half simply by complex conjugation,
\begin{align}
D[\sigma]=D[\bar \sigma]^{\ast} (-)^{n_{i}}\,,
\end{align}
where $ n_{i}$ is the total number of spatial derivatives. Here $D$ represents a diagram, $\sigma$ represents the collection $\{ r,l\}$ labels of its vertices and $\bar \sigma$ refers to the opposite labelling, $r\leftrightarrow l$.

%%%%%%%%%%%%%%%%%%%%%%%%%%%%%%%%%%%%%%%%%%%%%%%%%%%%%%%%%%%%%%%%%%%%%%%%%%%%%%%%%%%%%%%%%%%%%%%%%%%%%%%%

\paragraph{In-out Feynman rules} We now move on to the discussion of in-out rules. These are exactly the well-known Feynman rules we all learn in introductory courses on QFT. Because of this we will be very concise and focus only on what's different with respect to the in-in rules:
\begin{itemize}
 \item All interaction vertices and external fields are inside the same time ordering, they are on the same footing. Hence, there is no need to label them in different ways or to distinguish between bulk and boundary, as we did in the in-in formalism. All vertices receive a factor of $-i$ as appropriate for expanding the time evolution operator $e^{-i\int \Hi }$.
 \item Internal lines connecting two interaction vertices and external lines connecting a vertex to an external field correspond to one and the same propagator, the time-ordered Feynman propagator:
 \begin{align}\label{inoutrules}
 \bullet - \bullet = \Gf(\eta_{1},\eta_{2},p) &= \bra{0}T \phi(\eta_{1},\bfp)\phi(\eta_{2},\bfp') \ket{0}'\\
&=f_{p}(\eta_{1})f_{p}^{\ast}(\eta_{2})\theta(\eta_{1}-\eta_{2})+f_{p}^{\ast}(\eta_{1})f_{p}(\eta_{2})\theta(\eta_{2}-\eta_{1}) \label{Grr}\,.
 \end{align}
 \item The time $ \eta $ of a vertex must be integrated over the whole real line, with a measure $ \sqrt{-g}=(\eta H)^{-4} $ and boundaries 
 \begin{align}\label{inouteps}
 -\infty(1-i\e)<\eta < +\infty (1-i\e) \,,
 \end{align}
 where we are again assuming IR finiteness. This contour ensures the convergence of all time integrals for (interaction picture) correlators in the Fock vacuum where fields are inserted at arbitrary but finite and negative times.
\end{itemize}
Two comments are in order. First, let us stress that these in-out diagrams should not be amputated, as we do when computing amplitudes. They include all the propagators relevant for connecting to external fields. Second, as it is well known, one should only include diagrams where each propagator is eventually connected to an external field. This excludes all the vacuum bubble contributions which are exactly cancelled by the denomination in $\Gio$.

%%%%%%%%%%%%%%%%%%%%%%%%%%%%%%%%%%%%%%%%%%%%%%%%%%%%%%%%%%%%%%%%%%%%%%%%%%%%%%%%%%%%%%%%%%%%%%%%%%%%

\subsection{Explicit checks}\label{checks}

Here we will present some very simple calculations to see how the equality between in-in and in-out pans out in practice. Since the calculations are very similar between Minkowski and de Sitter, with the simple identification of $t$ with $\eta$, we will discuss both spacetimes jointly in each example. 

\paragraph{Tree-level contact diagrams} Now consider a simple theory of a scalar with a cubic interaction\footnote{This notation is meant to account for both time and space derivatives. A more explicit notation would specify on which field the time derivative acts, as e.g. in \cite{Goodhew:2021oqg}. Since at the end, the proof proceeds unchanged with or without time derivatives, we prefer to adopt a sloppier but more streamlined notation.}
\begin{align}
 \Hi(\eta) =\int_\bfx \frac{\lambda}{(n!)} \,F(\eta \partial_i,\eta \partial_\eta) \phi^{n}(\eta)
\end{align}
where $F$ represents a generic set of space and time derivatives that can act on any of the fields. Moreover, $\phi$ can be any set of fields of any mass and spin, but to simplify the presentation we focus on a single massive scalar. We will now compute the $n$-point function to $ \O(\lambda)$ with fields inserted at time $\eta_a\leq 0$ for $a=1,\dots,n$. For the in-in formalism, we find (setting for simplicity $H=1$)
\begin{align}
\Gii&=\Gii^{r}+\Gii^{l}\\
&=-i\lambda \int_{-\infty(1-i\e)}^{0} \frac{d\eta}{\eta^{4}}\, F \,\prod_{a}^{n} G_{F}(\eta,\eta_{a};k_{a})+i\lambda \int_{-\infty(1+i\e)}^{0} \frac{d\eta}{\eta^{4}} \,F \,\prod_{a}^{n} G^{+}(\eta,\eta_{a};k_{a})^*\,,
\end{align}
As long as there are more than four factors of $\eta$ coming from $F$ and the propagators this is convergent at $ \eta=0$. This is for example the case of any local interaction with more than three conformally coupled scalars or an interaction of massless scalars with $2n_{\partial_\eta}+n_{\partial_i}\geq 4$. 
Using the in-out formalism we find
\begin{align}\label{dsio}
\Gio&=-i\lambda \int_{-\infty(1-i\e)}^{+\infty(1-i\e)} \frac{d\eta}{\eta^{4}}\, F\, \prod_{a}^{n} G_{F}(\eta,\eta_{a};k_{a})\,.
\end{align}
To check that this is equivalent to the in-in expression we compute the difference. The part of the in-out time integral form $-\infty$ to $0$ exactly cancels out the ``right" contribution of $\Gii$ and we are left with 
\begin{align}
\Gio-\Gii&=-i\lambda \int_{0}^{+\infty_-} \frac{d\eta}{\eta^4} \, F\, \left[ \prod_{a}^{n} f_a(\eta) f_a^{*}(\eta_{a}) \right]-i\lambda \int_{-\infty_+}^{0} \frac{d\eta}{\eta^4} \, F\, \left[ \prod_{a}^{n} f_a(\eta) f_a^{*}(\eta_{a}) \right] \\
&=-i\lambda \prod_{a}^{n} f_a^{*}(\eta_{a}) \int_{-\infty (1+i\e)}^{+\infty(1-i\e)} \frac{d\eta}{\eta^4} \, F f_a(\eta)\, ,\label{argument}
\end{align}
where the label $a$ on the mode function refers to the different momenta $\bfk_a$, but may also indicate different fields with different masses. As a warm-up, let's see what happens for conformally coupled or massless fields, in which case the mode functions are just an exponential multiplied by a polynomial in $k\eta$ (see~\eqref{ccmode}). We find 
\begin{align}
\Gio-\Gii &\propto \sum_p \int_{-\infty(1+i\e)}^{+\infty(1-i\e)} d\eta \, \eta^{p} e^{-ik_{T}\eta}\,, \label{integral}
\end{align}
where the polynomial in $\eta$ makes explicit the factors present in the vertex $F$ and in the mode functions that make the $\eta \to 0$ limit convergent. The key to computing this integral is to notice the ``mixed" $i\e$ deformations at the two different boundaries of the integral. One comes from the in-in and one from the in-out. They are such that the integral is exponentially converging in both limits, as it should be. These boundaries invite us to close the contour in the lower-half complex plane, where we can drop the circle at infinity\footnote{Here we are assuming $k_T \neq 0$. To be more precise we should add a Dirac delta $\delta(k_T)$. These distributional terms were discussed in \cite{Salcedo2022} and in greater detail in \cite{Albayrak:2023hie}.}. Since the integrand is analytic in $\eta$ in the lower-half complex plane, the integral vanishes and $\Gii$ coincides with $\Gio$. 

More generally, we observe that the equivalence of in-in and in-out for contact diagrams relies on two properties: the mode functions and interaction vertices are analytic in the lower-half complex plane, and their product vanishes for $\Im \eta <0$ and $|\eta|\to \infty$. For fields of mass $m$ the mode functions are Hankel functions of $-k\eta$ times appropriate factors of $\eta$,
\begin{align} \label{modefct}
f_{k}(\eta)&=-i\frac{\sqrt{\pi}H}{2}e^{i\frac{\pi}{4}(1+2\nu)}(-\eta)^{3/2}H^{(1)}_{\nu}(-k\eta)\,,\quad \nu\equiv\sqrt{\frac{9}{4}-\frac{m^{2}}{H^{2}}}\,.
\end{align}
We can choose $f(\eta)$ to have a single branch cut running along the negative real axis and to be analytic everywhere else. Moreover, $f_k(\eta)$ vanishes for $\Im \eta <0$ and $|\eta|\to \infty$. This can be seen by expanding it in this limit or from the integral representation
\begin{align}
 H^{(1)}_\nu(z)=\frac{e^{-i\pi\nu/2}}{i\pi}\int_{-\infty}^{+\infty} e^{iz\cosh t-\nu t}dt\,, \quad \text{for}\quad \pi< \text{Arg}z<0\,.
\end{align}
These two properties combined tell us that we can close the integral in~\eqref{argument} in the lower-half complex plane where it is analytic and hence the equivalence between in-in and in-out is established. Finally, we note that the above calculation can be easily adapted to Minkowski: switch $\eta$ to $t$ and drop the time-dependent factors in the measure of integration $\eta^{-4}$ and in the interactions. The conclusion is hence unchanged.

%%%%%%%%%%%%%%%%%%%%%%%%%%%%%%%%%%%%%%%%%%%%%%%%%%%%%%%%%%%%%%%%%%%%%%%%%%%%%%%%%%%%%%%%%%%%%%%%%%%%

\paragraph{Tree-level exchange diagram} For in-in diagrams, we generally need to consider $2^V$ ($2^{V-1}$ if we use that half of the diagrams are conjugate) diagrams. We here give the explicit matching of diagrams from the in-in to the in-out formalism, for the two-to-two exchange diagram. The correspondence is then readily generalized to more complicated diagrams. We consider two interactions of the form
\begin{align}\label{Hamiltonian}
 \Hi =\int_\bfx \frac{\lambda_1}{(n+1)!} \,F_1(\eta \partial_i,\eta \partial_\eta) \phi^{n+1} + \int_\bfx \frac{\lambda_2}{(m+1)!} \,F_2(\eta \partial_i,\eta \partial_\eta) \phi^{m+1},
\end{align}
where $F_1$ and $F_2$ again capture derivatives. Furthermore, we here focus on a particular channel with $k_1,...,k_n$ attaching to the $\lambda_1$ vertex and $k_{m+1},...,k_{n+m}$ attaching the $\lambda_2$ vertex. The in-out correlator is then given by 
\begin{align}
\Gio&=-\lambda_{1} \lambda_2 \int_{-\infty_{-}}^{\infty_{-}} \int_{-\infty_{-}}^{\infty_{-}} \frac{d\eta}{\eta^{4}} \frac{d\eta'}{\eta'^{4}}F_1 F_2\, \left[G_{F}(\eta,\eta';s)\prod_{a=1}^{n} G_{F}(\eta,\eta_{a};k_{a})\prod_{b=n+1}^{n+m} G_{F}(\eta',\eta_{b};k_{b})\right]\,,
\end{align}
where $s$ is the energy of the internal leg. As we know, the in-in correlator comes in four parts: 
\begin{align}
\Gii&=\Gii^{rr}+\Gii^{rl}+\Gii^{lr}+\Gii^{ll}\,,
\end{align}
each given by
\begin{align}
\Gii^{rr}&=-\lambda_1 \lambda_2 \int_{-\infty_{-}}^0\int_{-\infty_{-}}^0 \frac{d\eta}{\eta^{4}} \frac{d\eta'}{\eta'^{4}}F_1 F_2\, \left[G_{F}(\eta,\eta';s)\prod_{a=1}^{n} G_{F}(\eta,\eta_{a};k_{a})\prod_{b=n+1}^{n+m} G_{F}(\eta',\eta_{b};k_{b})\right]\,,\nonumber \\ \nonumber
\Gii^{ll}&=-\lambda_1 \lambda_2 \int_{-\infty_{+}}^0\int_{-\infty_{+}}^0 \frac{d\eta}{\eta^{4}} \frac{d\eta'}{\eta'^{4}}F_1 F_2\, \left[G^*_{F}(\eta,\eta';s)\prod_{a=1}^{n} G_{l}(\eta,\eta_{a};k_{a})\prod_{b=n+1}^{n+m} G_{l}(\eta',\eta_{b};k_{b})\right]\,, \\ \nonumber
\Gii^{lr}&=\lambda_1 \lambda_2 \int_{-\infty_{+}}^0\int_{-\infty_{-}}^0 \frac{d\eta}{\eta^{4}} \frac{d\eta'}{\eta'^{4}}F_1 F_2\, \left[G_{l}(\eta,\eta';s)\prod_{a=1}^{n} G_{l}(\eta,\eta_{a};k_{a})\prod_{b=n+1}^{n+m} G_{F}(\eta',\eta_{b};k_{b})\right]\,, \\ \nonumber
\Gii^{rl}&=\lambda_1 \lambda_2 \int_{-\infty_{-}}^0\int_{-\infty_{+}}^0 \frac{d\eta}{\eta^{4}} \frac{d\eta'}{\eta'^{4}}F_1 F_2\, \left[G^*_{l}(\eta,\eta';s)\prod_{a=1}^{n} G_{F}(\eta,\eta_{a};k_{a})\prod_{b=n+1}^{n+m} G_{l}(\eta',\eta_{b};k_{b})\right]\,, 
\end{align}
where we have used that $G_r=G_{rr}=\Gf$, and $G_{lr}=G_{l}=\Gp$, as well as the fact that a lot of the propagators are conjugates of each other. 
The trick to match the two diagrams is now to split the in-out integrals in a similar way
\begin{align}
\Gio&=\Gio^{rr}+\Gio^{rl}+\Gio^{lr}+\Gio^{ll},
\end{align}
where
\begin{align}
\Gio^{rr}&:=-\lambda_1 \lambda_2\int_{-\infty_{-}}^{0} \frac{d\eta}{\eta^{4}} \int_{-\infty_{-}}^{0} \frac{d\eta'}{\eta'^{4}}F_1 F_2\, \left[G_{F}(\eta,\eta';s)\prod_{a=1}^{n} G_{F}(\eta,\eta_{a};k_{a})\prod_{b=n+1}^{n+m} G_{F}(\eta',\eta_{b};k_{b})\right]\,,\nonumber \\ \nonumber
\Gio^{ll}&:=-\lambda_1 \lambda_2 \int_{0}^{\infty_{-}} \frac{d\eta}{\eta^{4}}\int_{0}^{\infty_{-}}\frac{d\eta'}{\eta'^{4}}F_1 F_2\, \left[G_{F}(\eta,\eta';s)\prod_{a=1}^{n} G_{F}(\eta,\eta_{a};k_{a})\prod_{b=n+1}^{n+m} G_{F}(\eta',\eta_{b};k_{b})\right]\,, \\ \nonumber
\Gio^{lr}&:=-\lambda_1 \lambda_2\int_{0}^{\infty_{-}} \frac{d\eta}{\eta^{4}} \int_{-\infty_{-}}^{0}\frac{d\eta'}{\eta'^{4}}F_1 F_2\, \left[G_{F}(\eta,\eta';s)\prod_{a=1}^{n} G_{F}(\eta,\eta_{a};k_{a})\prod_{b=n+1}^{n+m} G_{F}(\eta',\eta_{b};k_{b})\right]\,, \\ \nonumber
\Gio^{rl}&:=-\lambda_1 \lambda_2 \int_{-\infty_{-}}^{0} \frac{d\eta}{\eta^{4}}\int_{0}^{\infty_{-}}\frac{d\eta'}{\eta'^{4}}F_1 F_2\, \left[G_{F}(\eta,\eta';s)\prod_{a=1}^{n} G_{F}(\eta,\eta_{a};k_{a})\prod_{b=n+1}^{n+m} G_{F}(\eta',\eta_{b};k_{b})\right]\,.\end{align}
We can now show that each of these four terms exactly cancel each other. Trivially we can see that $\Gio^{rr}=\Gii^{rr}$. Then for the mixed $rl$ case, we have
\bea\label{excharg}
&&\Gii^{rl}-\Gio^{rl}=\\ \nonumber
&&=\lambda_1 \lambda_2 \int_{-\infty_{-}}^{0} \frac{d\eta}{\eta^{4}}\int_{-\infty _{+}}^{\infty_{-}}\frac{d\eta'}{\eta'^{4}}F_1 F_2\, \left[ G_{l}(\eta',\eta;s)\prod_{a=1}^{n} G_{F}(\eta,\eta_{a};k_{a}) \prod_{b=n+1}^{n+m} G_{l}(\eta',\eta_{b};k_{b})\right] \\ \nonumber
&&\propto \int_{-\infty_{-}}^{0} \frac{d\eta}{\eta^{4}}F_1 \left[\prod_{a=1}^{n} G_{F}(\eta,\eta_{a};k_{a}) f_s^*(\eta)\right]\int_{-\infty _{+}}^{\infty_{-}}\frac{d\eta'}{\eta'^{4}} F_2\, \left[f_s(\eta')\prod_{b=n+1}^{n+m} f_{k_{b}}(\eta')\right],
\eea
which vanishes by the same argument given for contact diagrams, since the last integral is exactly the one found in~\eqref{argument}. Note that, from their definition in the in-out case, some of the Feynman propagators collapsed to unordered propagators since in the $rl$ case $\eta_b<0$ for all $b$, and in the in-out definition, $\eta<\eta'$ by the integral structure. The $lr$ case follows a similar story. The $ll$ case is the one that requires the most work. First let's rewrite the in-out case, using that $\eta_a<0$ and $\eta_b<0$. Moreover, we expand the Feynman propagator into two parts and write everything in terms of mode functions
\bea
\Gio^{ll}&=&-\lambda_1 \lambda_2 \prod_{a=1}^{n+m} f^*_{k_a} (\eta_a)\int_{0}^{\infty_{-}} \frac{d\eta}{\eta^{4}}F_1\left[ f_{s}(\eta)\prod_{a=1}^{n} f_{k_a}(\eta)\right]\int_{0}^{\eta}\frac{d\eta'}{\eta'^{4}}F_2 \left[f^*_{s}(\eta')\prod_{b=n+1}^{n+m} f_{k_b}(\eta')\right]\,\nonumber \\ \nonumber
&&-\lambda_1 \lambda_2 \prod_{a=1}^{n+m} f^*_{k_a} (\eta_a)\int_{0}^{\infty_{-}} \frac{d\eta'}{\eta'^{4}}F_2\left[ f_{s}(\eta')\prod_{b=n+1}^{n+m} f_{k_b}(\eta')\right]\int_{0}^{\eta'}\frac{d\eta}{\eta^{4}}F_1 \left[f^*_{s}(\eta)\prod_{a=1}^{n} f_{k_a}(\eta)\right]\,,
\eea
where we assume that the differential operators in $F_1$ and $F_2$ only act on the mode functions, not the Heaviside functions in the Feynman propagator. When they act on the Heaviside functions we go back to the contact diagram we discussed in the previous example. For the equivalent expressions of the in-in correlator, we get
\bea
\Gii^{ll}&=&\lambda_1 \lambda_2 \prod_{a=1}^{n+m} f^*_{k_a} (\eta_a) \int_{-\infty_{+}}^0 \frac{d\eta}{\eta^{4}}F_1\left[ f_{s}(\eta)\prod_{a=1}^{n} f_{k_a}(\eta)\right]\int_{0}^{\eta}\frac{d\eta'}{\eta'^{4}}F_2 \left[f^*_{s}(\eta')\prod_{b=n+1}^{n+m} f_{k_b}(\eta')\right]\,\nonumber \\ \nonumber
&&+\lambda_1 \lambda_2 \prod_{a=1}^{n+m} f^*_{k_a} (\eta_a) \int_{-\infty_{+}}^0 \frac{d\eta'}{\eta'^{4}}F_2\left[ f_{s}(\eta')\prod_{b=n+1}^{n+m} f_{k_b}(\eta')\right]\int_{0}^{\eta'}\frac{d\eta}{\eta^{4}}F_1 \left[f^*_{s}(\eta)\prod_{a=1}^{n} f_{k_a}(\eta)\right]\,,
\eea
and therefore
\bea
\Gii^{ll}-\Gio^{ll}&\propto& \int_{-\infty_{+}}^{\infty_{-}} \frac{d\eta}{\eta^{4}}F_1\left[ f_{s}(\eta)\prod_{a=1}^{n} f_{k_a}(\eta)\right]\int_{0}^{\eta}\frac{d\eta'}{\eta'^{4}}F_2 \left[f^*_{s}(\eta')\prod_{b=n+1}^{n+m} f_{k_b}(\eta')\right]\,\nonumber \\ \nonumber
&&+ \int_{-\infty_{+}}^{\infty_{-}} \frac{d\eta'}{\eta'^{4}}F_2\left[ f_{s}(\eta')\prod_{b=n+1}^{n+m} f_{k_b}(\eta')\right]\int_{0}^{\eta'}\frac{d\eta}{\eta^{4}}F_1 \left[f^*_{s}(\eta)\prod_{a=1}^{n} f_{k_a}(\eta)\right]\,.
\nonumber
\eea
The general proof that this expression vanishes for fields of arbitrary mass is a bit lengthy and therefore we postpone it to App.~\ref{app:A}. Here instead we focus on massless and conformally couples fields, for which the calculation is straightforward. Let's focus on the first line of the above expression since the story is analogous for the second line. The $d\eta'$ integrand takes the form
\begin{align}
 \int_{0}^{\eta}\frac{d\eta'}{\eta'^{4}}F_2 \left[f^*_{s}(\eta')\prod_{b=n+1}^{n+m} f_{k_b}(\eta')\right] &= \int_{0}^{\eta} d\eta'\, \text{Poly}(k,s,\eta')e^{-i(k_R-s)\eta'}\,,
\end{align}
where $k_R=\sum_{i=n+1}^{n+m}k_i$ and the polynomial depends on the details of the interaction and on whether we have massless or conformally coupled scalars. Then the integral in $d\eta'$ can be easily performed and leads to some other polynomial in $k_a$, $s$ and $\eta$ times $e^{-i(k_R-s)\eta}$. The final $d\eta$ integral can then be easily seen to vanish because the integrand is analytic and vanishes on the arc at infinity in the lower-half complex plane,
\begin{align}
 \Gii^{ll}-\Gio^{ll}&\propto \int_{-\infty_{+}}^{\infty_{-}} \frac{d\eta}{\eta^{4}}F_1\left[ f_{s}(\eta)\prod_{a=1}^{n} f_{k_a}(\eta)\right] \text{Poly}(k_a,s,\eta)e^{-i(k_R-s)\eta} \\
 &=\int_{-\infty_{+}}^{\infty_{-}} \text{Poly}'(k_a,s,\eta)e^{-i(k_R-s+s+k_L)\eta}=0\,,
\end{align}
where Poly' is some other polynomial and $k_L=\sum_{a=1}^m k_a$. Just like for the contact case, the two key steps of the proof are (i) that the integrand of the $d\eta$ integral from $+\infty(1-i\e)$ to $-\infty(1+i\e)$ is analytic in the lower-half complex plane and (ii) that it vanishes on the arc at infinity $\eta\sim \infty e^{i\theta}$ with $-\pi<\theta<0$. This is exactly what we show for general masses in App~\ref{app:A}.

%%%%%%%%%%%%%%%%%%%%%%%%%%%%%%%%%%%%%%%%%%%%%%%%%%%%%%%%%%%%%%%%%%%%%%%%%%%%%%%%%%%%%%%%%%%%

\section{Pole bagging: in-in correlators from Feynman propagators}\label{sec:mink}

In this section, we show a first applications of the in-out formalism to the calculation of correlators that leverages the simplicity of the Feynman propagator in energy-momentum space and is dubbed ``pole bagging" because it boils down to summing over poles. We discuss explicitly the case of Minkowski and comment on how the analysis can be extended to massless and conformally coupled scalars in de Sitter. 

%%%%%%%%%%%%%%%%%%%%%%%%%%%%%%%%%%%%%%%%%%%%%%%%%%%%%%%%%%%%%%%%%%%%%%%%%%%%%%%%%%%%%%%%%%%%

\subsection{Flat space}\label{poleb}

We have argued that in-in correlators equal in-out correlators. In Minkowski, in-out correlators take a very simple form in energy-momentum space, with both time orderings combined in a single term by Feynman's $i\e$ prescription. Indeed, tree-level Feynman diagrams in the energy-momentum domain can be computed by purely algebraic manipulations, no integral needed. Here we want to show how to use this simplicity to compute correlators in the time domain. The only work we need to do is to transform energies of the external fields back into time. This is easily done using the residue theorem because the integrands are simple rational functions and the $i\e$ prescription instructs us to pick up only half of the poles (those in the upper-half complex plane in our conventions). The result is a very different way to compute tree-level correlators or integrands for loop correlators simply by evaluating the products of Feynman diagrams on the relevant poles. Our manipulations below are analogous to similar manipulations presented in \cite{Arkani-Hamed:2017fdk}. The main difference is that (i) we work directly with correlators instead of wavefunction coefficients, and (ii) the Feynman propagators we encounter are simpler than the wavefunction propagators in that they don't have the extra boundary term, which often leads to a considerable algebraic simplification. \\

\paragraph{Contact diagrams} Consider the equal-time three-point function of a scalar in the time-momentum domain. Using the in-out formalism, this is simply the Fourier transform from frequency to time of the product of Feynman propagators
\begin{align}
B_{n,\text{in-out}} = (2\pi)^{4}\delta^{(4)}(\sum_{a} p_{a}^{\mu})\times (-iF) \times \prod_{b}^n \frac{i}{p_{b}^{2}+i\e} \quad \text{(contact)}\,,
\end{align}
where $F$ is some vertex accounting for derivatives. To see how the calculation progresses, let's focus on the simplest case of a cubic polynomial interaction $\lambda \phi^3/(3!)$ so that $F=\lambda$ and $n=3$. The Fourier transform to \textit{equal-time} correlators then gives
\begin{align}
B_3^{\text{flat}}&=\int_{-\infty}^{\infty} \frac{d\omega_{1}d\omega_{2}d\omega_{3} }{(2\pi)^{3}} (2\pi)\delta^{}\left(\sum_{a}^3 \omega_{a} \right) e^{it \sum_{a} \omega_{a} } \lambda \prod_{b}^3 \frac{1}{p_{b}^{2}+i\e}\\
&= \int_{-\infty}^{\infty} \frac{d\omega_{1}d\omega_{2}}{(2\pi)^2} \frac{\lambda}{(\omega_{1}^{2}-\Omega_{1}^{2}+i\e)(\omega_{2}^{2}-\Omega_{2}^{2}+i\e)((\omega_{1}+\omega_{2})^{2}-\Omega_{3}^{2}+i\e)}\,,
\end{align}
where the prime on the correlator means that we have dropped $(2\pi)^3\delta^{(3)}\left( \sum \bfk_a \right)$ and 
\begin{align}
\Omega_{a}^{2}\equiv |\bfk_{a}|^{2}+M_{a}^{2}\,.
\end{align}
The integrals can be computed using the residue theorem (we close the contour in the upper-half plane). The first integral has poles at $ \omega_{1}=-\Omega_{1}$ and $ \omega_{1}=-\omega_{2}-\Omega_{3}$. This gives
\begin{align}
-i \lambda\int \frac{d\omega_{2}}{2\pi} \frac{\Omega_{13}}{2\Omega_{1}\Omega_{3}(\omega_{2}-\Omega_{2} )(\omega_{2}+\Omega_{2} ) (\omega_{2}+\Omega_{13})(\omega_{2}-\Omega_{13})}\,,
\end{align}
where 
\begin{align}
 \Omega_{ij}=\Omega_i + \Omega_j\,,
\end{align} 
and we left the $i\e$'s implicit. Notice that there is a cancellation between the two residues which removes two of the zeros in the denominator ($\omega_{2}=\pm(\Omega_{1}-\Omega_{3} $)), leaving only the zeros at $ \omega_{2}=\pm \Omega_{2}$ and $ \omega_{2}=\pm(\Omega_{1}+\Omega_{3} $). The $ i\e$ prescription\footnote{A way to keep clarity in this calculation is to remove the $ i\e$s in the integrand and introduce a small counterclockwise rotation of the integration contours so that in practice one picks up only the residues from poles on the negative real axis.} instructs us to pick up only the two poles on the negative real axis, $ -\Omega_{2}$ and $ -\Omega_{13}$, which gives
\begin{align}
B_3^{\text{flat}} = - \frac{\lambda}{4\Omega_{1}\Omega_{2}\Omega_{3}(\Omega_{1}+\Omega_{2}+\Omega_{3})}\,.
\end{align}
This is indeed the expected results from the bulk time integral with the simple $ E_{T}$ pole and the correct normalization factor for each $ 1/\Omega_{a}$. The same calculation actually works for all contact correlators in just the same way. To see this one can make repeated use of the master formula
\begin{align}
 \int_{-\infty}^{\infty} \frac{d\omega_i} {2\pi} \frac{1}{(\omega_i^{2}-\Omega_{i}^{2} )((\omega_{i}+\omega_{X})^{2}-\Omega_{j}^{2} )} =-i \frac{\Omega_{ij}}{2\Omega_{i}\Omega_{j}(\omega_{X}^2-\Omega_{ij}^2)}\,.
\end{align}
This can be understood either as a brute force sum over two residues or as first separating the integrand into four partial fractions, corresponding to the four poles and then picking up the two partial fractions with poles on the negative real axis. At each iteration, one finds an additional factor of the pole $\Omega$ and the numerator is cancelled and substituted by itself plus the new pole. In the last integral one has $\omega_X=0$ and the numerator cancels the denominator leaving a simple $\Omega_T$ pole. 

\paragraph{Exchange diagram} Here we see explicitly how the same derivation also goes through for the tree-level exchange diagram. The idea is just the same, namely sum over the residues along the negative real axis for each of the integrals in $ d\omega_{a}$. The only difference is that there are three residues for two of the integrals because of the extra factor of the exchange propagator. We focus on a single cubic polynomial interaction and on the kinematics of the $s$-channel. More in detail
\begin{align}\label{springboard}
&B_4^{ex.}=\int \frac{d^{4}\omega_{a}}{(2\pi)^4} \frac{e^{it\sum_{a}\omega_{a}}\delta\left( \sum_{a}\omega_{a}\right)}{ \left(\omega _s^{2}-\Omega _s^{2}+i\e\right)}\times (-i)^2 \times \prod_{a}^{4}\frac{i}{\left(\omega _a^{2}-\Omega _a^{2}+i\e\right) }\\
=&\int_{\omega_{1,2}} \frac{\Omega _{34}}{2 \Omega _3 \Omega _4 \left(\omega _1^{2}-\Omega
 _1^{2}\right)\left(\omega _2^{2}-\Omega
 _2^{2}\right) \left((\omega _1+\omega
 _2)^{2}-\Omega _{34}^{2}\right) \left((\omega _1+\omega _2)^{2}-\Omega _s^{2}\right)
 }\\
 =&\int_{\omega_{2}} \frac{\omega
 _2^2 \Omega _1- \Omega _{134} \left(\Omega _s+\Omega
 _1\right) \left(\Omega _s+\Omega _{134}\right)}{4 \Omega _1 \Omega _3 \Omega _4 \Omega _s
 \left(\Omega _s+\Omega _{34}\right) \left(\omega _2^{2}-\Omega
 _2^{2}\right) \left(\omega _2^{2}-\Omega
 _{134}^{2}\right) \left(\omega _2^2-(\Omega _s+\Omega _1)^2\right) }\\
 =&-\frac{\Omega _s+\Omega _T}{8 \Omega _1
 \Omega _2 \Omega _3 \Omega _4 \Omega _T \Omega _s \left(\Omega _s+\Omega _{12}\right) \left(\Omega _s+\Omega _{34}\right)}\,,
\end{align}
where
\begin{align}
\Omega_{s}^{2}&=|\bfk_{1}+\bfk_{2}|^{2}+M^{2}\,, & \omega_{s}=\omega_{1}+\omega_{2}\,.
\end{align}
The result exactly agrees with the time-integral calculation and displays the characteristic $ E_{L}$, $ E_{R}$ and $ E_{T}=\sum \Omega_{a}$ poles, which are present in the quartic wavefunction coefficient, as well as the $ E_{s}$ pole which appears only in the correlator.\\

The same procedure works for arbitrary tree-level diagrams and for the integrands of loop diagrams. However, the bookkeeping becomes a hindrance for more complicated diagrams. One should develop a more streamlined graphical notation. In Sec.~\ref{sec:cuts} we will see that this can be achieved from a different point of view and so we refrain from developing this further here. \\

In passing, we would like to notice that~\eqref{springboard} is a springboard for the discussion of the nature of effective field theory for equal-time ``bulk" correlators. This is quite different from the usual expectation of effective field theories for amplitudes. For example, for amplitudes, we expect that a heavy field with mass $M$ can be removed at tree-level by collapsing its propagators and inserting an infinite sum over higher derivative operators, which are organised in powers of $\Box/M^2$. Conversely, by expanding~\eqref{springboard} in large $M$ one finds terms that are odd in $1/M $ and are hence not captured by this expansion (as also noticed in \cite{Salcedo2022}). This is where dissipation rears its head and we come to appreciate the real power of the in-in formalism. We will discuss this elsewhere. 

%%%%%%%%%%%%%%%%%%%%%%%%%%%%%%%%%%%%%%%%%%%%%%%%%%%%%%%%%%%%%%%%%%%

\subsection{Massless and conformally coupled scalars in de Sitter}\label{polebdS}
For massless and conformally coupled scalars we can generalise the pole bagging procedure from flat space to de Sitter. The key ingredient is that, when $\nu$ a half-integer, we can re-write the Feynman propagator as
\bea\label{key}
G_F^{dS}(\eta,\eta', k) &=& \left(\frac{1}{2 \pi i}\right)\int_{-\infty}^{\infty} d\omega \frac{2 \omega f_{\omega}(\eta)f^*_{\omega}(\eta')}{\omega^2 -k^2+i \epsilon}. \\ \nonumber
\eea
This only works for half-integer values of $\nu$ because the residue at $\omega=0$ is only zero for these masses. Nevertheless, it allows us to compute cosmological correlators through an interesting integral. For simplicity, we here focus on conformally coupled fields, but the procedure is exactly the same for massless scalars. We have
\bea\label{key}
G_F^{cc}(\eta,\eta', k) &=& \left(\frac{1}{2 \pi i}\right)\int_{-\infty}^{\infty} d\omega \frac{H^2 \eta \eta' e^{-i \omega (\eta-\eta')}}{\omega^2 -k^2+i \epsilon}. 
\eea
The simplest example is the contact four-point function with a $\phi^4$ interaction. Since this interaction is conformally invariant at this order, the calculation is identical to the Minkowski calculation (technically the $\eta$ in the four mode functions cancels the $\eta{-4}$ in the volume measure). A slightly more interesting computation shows the general strategy of how we can perform a pole bagging calculation in de Sitter and rephrase it as derivatives of the flat-space result. Consider now the five-point contact diagram with a $\phi^5/(5!)$ interaction. This interaction is not conformal invariant and so now the peculiarities of de Sitter will show up. Using the shorthand notation $\omega_T= \sum_{i=1}^5 \omega_i$ we write
\bea
B_5^{cc}&=& -i H^6\left(\frac{ \eta_0}{2 \pi i}\right)^5 \int_{-\infty}^{\infty}\left(\prod_{i=1}^5 \frac{d\omega_i}{\omega_i^2 -k^2+i \epsilon}\right)e^{i \omega_T\eta_0}\int_{-\infty}^{\infty}d\eta \eta e^{-i \omega_T\eta} \\ \nonumber
&=& 2\pi H^6 \left(\frac{ \eta_0}{2 \pi i}\right)^5 \int_{-\infty}^{\infty}\left(\prod_{i=1}^5 \frac{d\omega_i}{\omega_i^2 -k_i^2+i \epsilon}\right)e^{i \omega_T\eta_0}\delta'(\omega_T) \\ \nonumber
&=& -2\pi i H^6 \left(\frac{ \eta_0}{2 \pi i}\right)^5 \int_{-\infty}^{\infty} \frac{d\omega_1d\omega_2d\omega_3d\omega_4(\eta_0((\omega_T-\omega_5)^2 -k_5^2))-2i(\omega_T-\omega_5))}{\left(\prod_{i=1}^4\left(\omega_i^2 -k_i^2+i \epsilon\right)\right)\left((\omega_T-\omega_5)^2 -k_5^2+i \epsilon\right)^2}\\ \nonumber
&=& \frac{ (H \eta_0)^6}{(k_1+k_2+k_3+k_4+k_5)(16 k_1 k_2 k_3 k_4 k_5)}\eea.

Therefore the general strategy is to transfer the derivatives acting on the Dirac delta to the rest of the integral, which effectively is the flat space analogue of the integrals above. The only thing one has to keep track of is the number of these derivatives. The integrals pick up the poles in the upper half plane and can be computed in an algorithmic manner. So algorithmic in fact, that we can quite easily derive the general formula for the conformally coupled contact diagrams for any $n$.

\paragraph{Conformally coupled n-point correlators}
With pole bagging, it is quite simple in fact to do any n-point correlator with a $\phi^n/(n!)$ interaction. The outline is solving the integral
\bea\label{ccgeneral}
B_n^{cc}&=& -i H^{2n-4}\left(\frac{ \eta_0}{2 \pi i}\right)^n \int_{-\infty}^{\infty}\left(\prod_{i=1}^n \frac{d\omega_i}{\omega_i^2 -k_i^2+i \epsilon}\right)e^{i \omega_T\eta_0}\int_{-\infty}^{\infty}d\eta \eta^{n-4} e^{-i \omega_T\eta} \\ \nonumber
&=& -2 \pi i (i)^{n-3} H^{2n-4}\left(\frac{ \eta_0}{2 \pi i}\right)^n \int_{-\infty}^{\infty}\left(\prod_{i=1}^n \frac{d\omega_i}{\omega_i^2 -k_i^2+i \epsilon}\right)e^{i \omega_T\eta_0}\delta^{(n-4)}(\omega_T)\\ \nonumber
&=& -(-i\eta_0)^{n} H^{2n-4}\left(\frac{ 1}{2 \pi i}\right)^{n-1} \int_{-\infty}^{\infty}\partial_{\omega_n}^{n-4}\left(\left(\prod_{i=1}^n \frac{d\omega_i}{\omega_i^2 -k_i^2+i \epsilon}\right)e^{i \omega_T\eta_0}\right)\delta(\omega_T). \eea

While solving this integral is not particularly insightful, we here simply quote the result, because of its simplicity, and leave the derivation to App.~\ref{app:PoledS}. In particular, there we show the general method, how we can write this diagram in terms of the derivatives of its flat space analogue. Finally, we get a simple closed-form expression for contact diagrams of conformally coupled scalars in de Sitter, given by
\bea\label{nicecc}
B_{n+4}^{cc}&=&\frac{n! H^{n}(-H\eta_0)^{n+4}}{\left(\prod_{i=1}^{n+4}2 k_i\right)k_T^{n+1}}2\text{Re}\left[i^n e^{i \eta_0 k_T}\bigg|_{n}\right]
\eea
where the bar indicates the Taylor series in $\eta_0$ up to that order.

This technique can be further developed, for example for exchange diagrams. In particular, one can show that we can always write diagrams for conformally coupled scalars as sums of derivatives of the flat space one, as in~\eqref{flatder}, but we will not do this in this work. Other examples can be obtained using integration by parts, perhaps along the lines of \cite{Chen:2023iix}.

%%%%%%%%%%%%%%%%%%%%%%%%%%%%%%%%%%%%%%%%%%%%%%%%%%%%%%%%%%%%%

\section{Correlator recursion relations}
We here leverage that with the in-out formalism, there is no longer a distinction between bulk-to-bulk and bulk-to-boundary propagators. This allows us to find an algebraic recursion relation valid at all loop orders that computes correlators in Minkowski, somewhat analogous to the recursion relation for wavefunction coefficients derived in \cite{Arkani-Hamed:2017fdk}.
%%%%%%%%%%%%%%%%%%%%%%%%%%%%%%%%%%%%%%%%%%%%%%%%%%%%%%%%%%%

\subsection{From correlators to chains in flat space}

In Minkowski, bulk-to-boundary propagators $K_E(t)$ for the calculation of the wave function have the nice property that
\bea
K_{E_1}(t)K_{E_2}(t)=K_{E_1+E_2}(t).
\eea
This allows us to describe a diagram with any number of external legs, by simply considering the total energy flowing into a vertex \cite{Arkani-Hamed:2017fdk}. Therefore, it is natural to summarize a large family of Feynman-Witten diagrams by so-called ``chains", namely diagrams that have exactly one external leg per vertex. Since all vertices have a single external leg, one often omits to draw them and simply adds an ``external" energy $x$ to each vertex.

For correlators, the situation can in fact be simplified in a similar way. Given the in-out formalism described in Sec.~\ref{sec:inininout}, the analogue to a bulk-to-boundary correlator is a Feynman propagator. In flat space, Feynman propagators in the time-momentum domain obey
\bea\label{proprule}
\Gf(t,t',E_1)\Gf(t,t',E_2)=\frac{E_1+E_2}{2 E_1 E_2}\Gf(t,t',E_1+E_2).
\eea
Notice that it is crucial that the propagators have the same time variables, which ensures that there are only two possible time orderings, which match on the left- and the right-hand side. More generally, the product of $n$ Feynman propagators obeys
\bea\label{proprulegen}
\Gf(t,t',E_1)...\Gf(t,t',E_n)=\frac{2 E_T}{\prod_{i=1}^n 2 E_i}\Gf(t,t',E_T).
\eea
 We stress that the Feynman rules for the ``correlator-chains" we discuss here are different to the ``wavefunction-chains" discussed in \cite{Arkani-Hamed:2017fdk}. \\

The above discussion leads us to the following set of Feynman-rules for what we call a chain in the context of correlators:
\begin{itemize}
 \item Every vertex with energy $x_i$ and time $t_i$ leads to a Feynman-propagator $\Gf(t_i,t_0,x_1)$. This corresponds to the omnipresent single external line that we omit to draw.
 \item Every internal line connecting a vertex $\{x_i,t_i\}$ to $\{x_j,t_j\}$ leads to a factor $\Gf(t_i,t_j,y_{ij})$, with $y_{ij}$ being the exchanged momenta (which could be loops).
\end{itemize}

Therefore, if we label the energies at vertices of the chain with $\{x_1,...,x_m\}$ and the external legs of the correlator carry energies $\{E_1,...,E_n\}$ (note $n\geq m$) we can write. 

\begin{tcolorbox}[colback=gray!10!white, colframe=white] 
\begin{equation}
\text{correlator} = \frac{\prod_{i=1}^m 2 x_i}{\prod_{i=1}^n 2 E_i} \times\text{chain}
\end{equation}
\end{tcolorbox}

For example, with $x_1=E_1+E_2$, $x_2=E_3$ and $x_3=E_4+E_5$, we have

\begin{equation}\label{onex}
 \begin{tikzpicture}[ball/.style = {circle, draw, align=center, anchor=north, inner sep=0}]
 \node[color=black] at (-1,-0.5)(ppp1) {\large $\prod_{i=1}^5 2 E_i$};

 \node[ball,text width=.18cm,fill,color=black] at (1,-1) (xx1) {}; 
 \node[ball,text width=.18cm,fill,color=black, right=1cm of xx1] (xx2) {};
 \node[ball,text width=.18cm,fill,color=black, right=1cm of xx2] (xx3) {};

 \draw[-,thick,color=black] (xx1) -- (xx2);
 \draw[-,thick,color=black] (xx2) -- (xx3);

 \node[ball,text width=.1cm,fill,color=black,label=above:{\scriptsize $E_1$}] at (.75,0) (pp1) {};
 \node[ball,text width=.1cm,fill,color=black,label=above:{\scriptsize $E_2$}] at (1.25,0) (pp2) {};
 \node[ball,text width=.1cm,fill,color=black,label=above:{\scriptsize $E_3$}] at (2.2,0) (pp3) {};
 \node[ball,text width=.1cm,fill,color=black,label=above:{\scriptsize $E_4$}] at (3.15,0) (pp4) {};
 \node[ball,text width=.1cm,fill,color=black,label=above:{\scriptsize $E_5$}] at (3.65,0) (pp5) {};

 \draw[-,color=black] (pp1) -- (xx1) -- (pp2);
 \draw[-,color=black] (pp3) -- (xx2) ;
 \draw[-,color=black] (pp4) -- (xx3) -- (pp5);
 
 \draw[-,thick,color=black] (xx1.east) edge node [text width=.18cm,above=-0.05cm,midway] {{\scriptsize $y_{12}$}} (xx2.west);
 \draw[-,thick,color=black] (xx2.east) edge node [text width=.18cm,above=-0.05cm,midway] {{\scriptsize $y_{23}$}} (xx3.west);

 \node[ball,text width=.1cm,fill,color=white] at (0,0) (d1) {};
 \node[ball,text width=.1cm,fill,color=white] at (4.4,0) (d2) {};
 \draw[thick,color=black] (d1) -- (d2); 
 \node[color=black] at (5.5,-0.5) (x0) {\large $=\prod_{i=1}^3 2 x_i$};
 \node[ball,text width=.18cm,fill,right=0.4cm of x0.east,color=black,label=below:{\scriptsize $x_1$}] (x1) {};
 \node[ball,text width=.18cm,fill,color=black,right=1.5cm of x1.east,label=below:{\scriptsize $x_2$}] (x2) {};
 \node[ball,text width=.18cm,fill,color=black,right=1.5cm of x2.east,label=below:{\scriptsize $x_3$}] (x3) {}; 
 \draw[-,thick,color=black] (x1.east) edge node [text width=.18cm,above=-0.05cm,midway] {{\scriptsize $y_{12}$}} (x2.west);
 \draw[-,thick,color=black] (x2.east) edge node [text width=.18cm,above=-0.05cm,midway] {{\scriptsize $y_{23}$}} (x3.west);
 \node[ right = 0.05cm of x3] {.}; 
 \end{tikzpicture}
\end{equation}

With the rules above, we can comfortably consider chains exclusively and eventually relate them easily back to correlators. Let us make two simple examples. The contact diagram would simply be
\begin{equation}\label{twoex}
 \begin{tikzpicture}[ball/.style = {circle, draw, align=center, anchor=north, inner sep=0}]

 \node[ball,text width=.18cm,fill,right=0.4cm of x0.east,color=black,label=below:{\scriptsize $x$}]at (-4,-0.5) (x) {};
 \node[color=black,right=0.4cm of x] {\Large $=$};
 
 \node[color=black] at (-1,-0.5)(ppp1) {\Large \hspace{-2em}$\frac{\prod_{i=1}^n 2 E_i}{2 x}$};
\hspace{-1em}
 \node[ball,text width=.18cm,fill,color=black] at (1,-1) (xx1) {}; 
 
 \node[ball,text width=.1cm,fill,color=black,label=above:{\scriptsize $E_1 $}] at (0.15,0) (pp1) {};
 \node[color=black] at (1.2,-0.3)(ppp1) {\small $\ldots$};
 \node[ball,text width=.1cm,fill,color=black,label=above:{\scriptsize\quad $E_2 \quad \ldots$}] at (0.7,0) (pp2) {};
 \node[ball,text width=.1cm,fill,color=black,label=above:{\scriptsize $E_n$}] at (1.8,0) (pp3) {};

 \draw[-,color=black] (pp1) -- (xx1) -- (pp2);
 \draw[-,color=black] (pp3) -- (xx1);

 \node[ball,text width=.1cm,fill,color=white] at (0,0) (d1) {};
 \node[ball,text width=.1cm,fill,color=white] at (2,0) (d2) {};
 \draw[thick,color=black] (d1) -- (d2); 
 \node[color=black] at (2.8,-0.5) {\Large $=-\frac{1}{x^2}$.}; \end{tikzpicture}
\end{equation}
And for a single exchange, we have the identification
\begin{equation}\label{eq:wvn}
 \begin{tikzpicture}[ball/.style = {circle, draw, align=center, anchor=north, inner sep=0}]

 \node[ball,text width=.18cm,fill,color=black,label=below:{\scriptsize $x_1$}] at (-5,-0.5) (x1) {};
 \node[ball,text width=.18cm,fill,color=black,right=1.5cm of x1.east,label=below:{\scriptsize $x_2$}] (x2) {};
 \draw[-,thick,color=black] (x1.east) edge node [text width=.18cm,above=-0.05cm,midway] {{\scriptsize $y$}} (x2.west);
 \node[color=black,right=0.4cm of x2] {\Large $=$};
 
 \node[color=black] at (-1,-0.5)(ppp1) {\Large $\frac{\prod_{i=1}^n 2 E_i}{4 x_1 x_2}$};

 \node[ball,text width=.18cm,fill,color=black] at (1,-1) (xx1) {};

 \node[ball,text width=.18cm,fill,color=black] at (1,-1) (xx1) {}; 
 \node[ball,text width=.18cm,fill,color=black, right=1.5cm of xx1] (xx2) {};

 \draw[-,thick,color=black] (xx1) -- (xx2);

 \node[ball,text width=.1cm,fill,color=black,label=above:{\scriptsize $E_1 \ldots$}] at (.75,0) (pp1) {};
 \node[ball,text width=.1cm,fill,color=black,label=above:{\scriptsize $E_m$}] at (1.35,0) (pp2) {};

 \node[ball,text width=.1cm,fill,color=black,label=above:{\scriptsize $E_{m+1}\quad$}] at (2.4,0) (pp3) {};
 \node[ball,text width=.1cm,fill,color=black,label=above:{\scriptsize $\ldots E_n$}] at (3.0,0) (pp4) {};

 \draw[-,color=black] (pp1) -- (xx1) -- (pp2);
 
 \draw[-,color=black] (pp3) -- (xx2) -- (pp4);
 
 \draw[-,thick,color=black] (xx1.east) edge node [text width=.18cm,above=-0.05cm,midway] {{\scriptsize $y$}} (xx2.west);

 \node[ball,text width=.1cm,fill,color=white] at (0,0) (d1) {};
 \node[ball,text width=.1cm,fill,color=white] at (3.75,0) (d2) {};
 \draw[thick,color=black] (d1) -- (d2);

 \node[color=black] at (7,-0.5) {\Large $=\frac{x_1+x_2+y}{2 x_1 x_2 y (x_1+x_2)(x_1+y)(x_2+y)}$.}; \end{tikzpicture}
\end{equation}
With these definitions, we can move on to derive recursion relations following \cite{Arkani-Hamed:2017fdk}.

%%%%%%%%%%%%%%%%%%%%%%%%%%%%%%%%%%%%%%%%%%%%%%%%%%%%%%%%%%%%%%%%%%%%%%%%%%%%%%%

\subsection{Recursion relations}\label{ss:recursions}

In \cite{Arkani-Hamed:2017fdk}, two sets of recursion relations for wavefunction coefficients were derived. We did not find a counterpart of the ``primary" recursion relations obtained by inserting the time-translation operator and integrating by part. We will comment on this at the end of the section. Instead here we focus on the ``secondary" recursion relations derived by integrating one site of a tree-level chain. In that context, the generalization to loop diagrams was found to be possible but cumbersome because of the proliferation of diagrams induced by the boundary term in the wavefunction calculation\footnote{Here we refer to the fact that, since the wave function answers a boundary value question rather than computing an average like a correlator, its propagator contains both the time-order Feynman propagator and a homogeneous solution to the equations of motion that enforces the vanishing of the propagator at the time surface where the wavefunction is evaluated.}. Here we follow the same logic for the correlators computed in the in-out formalism and find a remarkable simplification. Since now all the propagators are Feynman propagators, without any boundary terms, we are able to derive recursion relations both for tree-level diagrams and for a class of ``melonic" loop diagrams. 

\paragraph{Tree-level relations} For the tree-level case, we similarly start with a single tree-level edge, and an arbitrary remainig tree-level chain. That is

\bea\label{eq:TreeInt}
 \raisebox{-3ex}{\begin{tikzpicture}[ball/.style = {circle, draw, align=center, anchor=north, inner sep=0}]
 \node[ball,text width=.18cm,fill,color=black, label=below:$\mbox{\scriptsize $x_1$}$] at (0,0) (x1) {};
 \node[ball,text width=1cm,shade,right=1.1cm of x1.east] (S1) {$\mathcal{B}$};
 \node[ball,text width=.18cm,fill,color=black,right=1cm of x1.east, label=below:$\mbox{\scriptsize $\hspace{-.3cm}x_2$}$] (x2) {};
 \draw[-,thick,color=black] (x1.east) edge node [text width=.18cm,above=-0.05cm,midway] {{\scriptsize $y_{12}$}} (x2.west);

 \draw[-,thick,color=black] (x1.east) -- (x2.west);
 \end{tikzpicture}}
&&\equiv\:\int_{-\infty}^\infty\prod_{v\in\mathcal{B}\setminus\{2\}}i\, dt_v \,\Gf(t_v,t_0,x_t)\times\\ \nonumber 
&&\qquad\qquad\times\int_{-\infty}^\infty i\, dt_2 \Gf(t_2,t_0,x_2) \prod_{e\in\mathcal{B}}\Gf(t_{v_e},t_{v'_e},y_{v_e})I_1(y_{12},t_2),
\eea
where
\begin{equation}\label{eq:I1}
 I_1(y_{12},t_2)\:\equiv\:\int_{-\infty}^{\infty}i\, dt_1\, \Gf(t_1,t_0,x_1)\Gf(t_1,t_2,y_{12}).
\end{equation}
We can fully perform this integral
\bea
 I_1(y_{12},t_2)&=&\:\left(\frac{1}{2\pi i}\right)^2\int_{-\infty}^{+\infty}d\omega_1\int_{-\infty}^{+\infty}d\omega_2\int_{-\infty}^{\infty} i\, dt_1\ \frac{e^{i\omega_1(t_0-t_1)}}{\omega_1^2-x^{2}_{1}+i\varepsilon}\frac{e^{i\omega_2(t_1-t_2)}}{\omega_2^2-y^{2}_{12}+i\varepsilon} \\ \nonumber
 &&= 2\pi i \:\left(\frac{1}{2\pi i}\right)^2\int_{-\infty}^{+\infty}d\omega_1\int_{-\infty}^{+\infty}d\omega_2\delta(\omega_1-\omega_2)\ \frac{e^{i\omega_1 t_0}}{\omega_1^2-x^{2}_{1}+i\varepsilon}\frac{e^{-i\omega_2 t_2}}{\omega_2^2-y^{2}_{12}+i\varepsilon} \\ \nonumber
 &&=\left(\frac{1}{2\pi i}\right)\int_{-\infty}^{+\infty}d\omega\ \frac{e^{i\omega(t_0-t_2)}}{\omega^2-x^{2}_{1}+i\varepsilon}\frac{1}{\omega^2-y^{2}_{12}+i\varepsilon}\\ \nonumber
 &&=\left(\frac{1}{2\pi i}\right)\int_{-\infty}^{+\infty}d\omega\ \frac{1}{x_1^2-y_{12}^2}\left[\frac{e^{i\omega(t_0-t_2)}}{\omega^2-x^{2}_{1}+i\varepsilon}-\frac{e^{i\omega(t_0-t_2)}}{\omega^2-y^{2}_{12}+i\varepsilon}\right]\\ \nonumber
 &&= \frac{1}{x_1^2-y_{12}^2}\left[\Gf(t_0,t_2,x_1)-\Gf(t_0,t_2,y_{12})\right].
\eea
Finally, these Feynman propagators can be viewed as an additional external leg connecting to the $x_2$ vertex. Therefore, absorbing the propagators with the appropriate rules in~\eqref{proprule}, we can write the recursion relation
\begin{equation}\label{eq:TreeRRf}
 \begin{tikzpicture}[ball/.style = {circle, draw, align=center, anchor=north, inner sep=0}]
 \node[ball,text width=.18cm,fill,color=black, label=below:$\mbox{\scriptsize $x_1$}$] at (0,0) (x1) {};
 \node[ball,text width=1cm,shade,right=.8cm of x1.east] (S1) {$\mathcal{B}$};
 \node[ball,text width=.18cm,fill,color=black,right=.7cm of x1.east, label=below:$\mbox{\scriptsize $\hspace{-.3cm}x_2$}$] (x2) {};
 \draw[-,thick,color=black] (x1.east) edge node [text width=.18cm,above=-0.05cm,midway] {{\scriptsize $y_{12}$}} (x2.west);
 \draw[-,thick,color=black] (x1.east) -- (x2.west);
 \node[right=.2cm of S1.east] (for1) {$\displaystyle\:=\:\frac{1}{x_1^2-y_{12}^2}\left[\frac{x_1+x_2}{ 2 x_1 x_2}\quad\right.$};
 \node[ball,text width=1cm,shade,right=.2cm of for1.east] (S2) {$\mathcal{B}$};
 \node[ball,text width=.18cm,fill,color=black,right=.1cm of for1.east, label=below:$\mbox{\scriptsize $\hspace{-.4cm}x_1+x_2$}$] (x2a) {};
 \node[right=.2cm of S2.east] (for2) {$\displaystyle\:-\:\frac{y_{12}+x_2}{2 y_{12} x_2}\quad$}; 
 \node[ball,text width=1cm,shade,right=.2cm of for2.east] (S3) {$\mathcal{B}$};
 \node[ball,text width=.18cm,fill,color=black,right=.1cm of for2.east, label=below:$\mbox{\scriptsize $\hspace{-.4cm}y_{12}+x_2$}$] (x2b) {};
 \node[right=.2cm of S3.east] (for3) {$\displaystyle\hspace{-.3cm}{\left.\phantom{\frac{1}{y}}\right].}$}; 
 \end{tikzpicture}
\end{equation}
As an example consider the two chain and one chain in~\eqref{onex} and~\eqref{twoex}, we have
\begin{equation}\label{eq:wvn}
 \begin{tikzpicture}[ball/.style = {circle, draw, align=center, anchor=north, inner sep=0}]

 \node[ball,text width=.18cm,fill,color=black,label=below:{\scriptsize $x_1$}] at (-5,-0.5) (x1) {};
 \node[ball,text width=.18cm,fill,color=black,right=1.5cm of x1.east,label=below:{\scriptsize $x_2$}] (x2) {};
 \draw[-,thick,color=black] (x1.east) edge node [text width=.18cm,above=-0.05cm,midway] {{\scriptsize $y$}} (x2.west);
\hspace{1em}
 \node[color=black] at (2,-0.5) {\Large $\hspace{-2em}=\frac{1}{x_1^2-y^2}\left[\frac{x_1+x_2}{2 x_1 x_2}\left(-\frac{1}{(x_1+x_2)^2}\right)-\frac{y+x_2}{ 2 y x_2}\left(-\frac{1}{(y+x_2)^2}\right)\right]$.}; \end{tikzpicture}
\end{equation}
The way this recursion relation can be used repeatedly was shown in \cite{Arkani-Hamed:2017fdk}, and it can also iteratively be used for instance to relate the double exchange to a number of contact diagrams, which works similarly here. The interesting thing in the case here, however, is that we can generalise this to loops.

%%%%%%%%%%%%%%%%%%%%%%%%%%%%%%%%%%%%%%%%%%%%%%%%%%%%%%%%%%%%%%%%%%%%%%%%%%%%%%%%%

\paragraph{Recursion relations for melonic loop integrands}

The derivation for the recursion relation at loop level does in principle require not many more steps. We simply need to utilise~\eqref{proprulegen}. The recursion relations we write here are purely for the integrand, and all results in this section should be interpreted to be integrated over internal momenta. Furthermore, we focus on any number of loops, but we keep the number of vertices fixed, that is, we for instance do not consider box diagrams. The methods we present here do however generalize to these kinds of diagrams. In conclusion, we focus on diagrams of the type
 \bea\label{eq:TreeInt}
 \raisebox{-2.3em}{
 \begin{tikzpicture}[ball/.style = {circle, draw, align=center, anchor=north, inner sep=0}]
 \node[ball,text width=.18cm,fill,color=black, label=below:$\mbox{\scriptsize $x_1$}$] at (0,0) (x1) {};
 \node[ball,text width=1cm,shade,right=2.1cm of x1.east] (S1) {$\mathcal{B}$};
 \node[ball,text width=.18cm,fill,color=black,right=2cm of x1.east, label=below:$\mbox{\scriptsize $\hspace{-.2cm}x_2$}$] (x2) {};
\draw[thick] (x1) to[bend right=50] node[above, yshift=9.5 mm] {\scriptsize $y_1$} (x2);
 \draw[thick] (x1) to[bend right=25] node[above, yshift=5mm] {\scriptsize $y_2$} (x2);
 \draw[thick] (x1) to[bend left=50] node[ below, yshift=-7.2mm] {\scriptsize $y_{n-1}$} (x2);
 \draw[thick] (x1) to[bend left=25] node[ below, yshift=-7.5mm] {\scriptsize $y_n$} (x2);
 \node[rotate=90] at (1.1,-0.1) {\scriptsize ...};
 \end{tikzpicture}}
 &&\hspace{-1em}\displaystyle\:\equiv\:\int_{-\infty}^\infty \hspace{-0.25em}\prod_{v\in\mathcal{B}\setminus\{2\}} \hspace{-0.6em} i\, dt_v \,G_F(t_v,t_0,x_t) \times \\ \nonumber
 && \qquad  \times \int_{-\infty}^\infty \hspace{-0.7em}i\, dt_2 G_{F}(t_2,t_0,x_2)
 \prod_{e\in\mathcal{B}}G_{F}(t_{v_e},t_{v'_e},y_{v_e})I_n(\{y_i\},t_2),
\eea
where now we have
\begin{equation}\label{eq:I1}
I_n(\{y_i\},t_2)\:\equiv\:\int_{-\infty}^{\infty}i\, dt_1\, \Gf(t_1,t_0,x_1)\prod_{i=1}^n\Gf (t_1,t_2,y_{i})=\frac{2 y_T}{\prod_{i=1}^n 2 y_i}I_1(y_T,t_2),
\end{equation}
where in the last step we related $I_n$ to $I_1$ using~\eqref{proprulegen}.
On the level of the diagrams we, therefore, have
 \begin{equation}\label{eq:TreeInt}
 \begin{tikzpicture}[ball/.style = {circle, draw, align=center, anchor=north, inner sep=0}]
 \node[ball,text width=.18cm,fill,color=black, label=below:$\mbox{\scriptsize $x_1$}$] at (0,0) (x1) {};
 \node[ball,text width=1cm,shade,right=2.1cm of x1.east] (S1) {$\mathcal{B}$};
 \node[ball,text width=.18cm,fill,color=black,right=2cm of x1.east, label=below:$\mbox{\scriptsize $\hspace{-.2cm}x_2$}$] (x2) {};
\draw[thick] (x1) to[bend right=50] node[above, yshift=9.5 mm] {\scriptsize $y_1$} (x2);
 \draw[thick] (x1) to[bend right=25] node[above, yshift=5mm] {\scriptsize $y_2$} (x2);
 \draw[thick] (x1) to[bend left=50] node[ below, yshift=-7.2mm] {\scriptsize $y_{n-1}$} (x2);
 \draw[thick] (x1) to[bend left=25] node[ below, yshift=-7.5mm] {\scriptsize $y_n$} (x2);
 \node[rotate=90] at (1.1,-0.1) {\scriptsize ...};
 \node[right=.1cm of S1.east] (for1) {\Large$=\frac{2 y_T}{\prod_{i=1}^n 2 y_i}$} (S3);
 \node[ball,text width=.18cm,fill,color=black,left=5cm of S3, label=below:$\mbox{\scriptsize $x_1$}$] (xx1) {};
 \node[ball,text width=1cm,shade,right=1.6cm of xx1.east] (SS1) {$\mathcal{B}$};
 \node[ball,text width=.18cm,fill,color=black,right=1.5cm of xx1.east, label=below:$\mbox{\scriptsize $\hspace{-.2cm}x_2$}$] (xx2) {};
\draw[thick] (xx1) to[bend right=0] node[above] {\scriptsize $y_T$} (xx2);
\end{tikzpicture}\raisebox{1em}{  .}
\end{equation}
Of course, as in the previous section we can also fully find $I_n$ to get a generalisation of~\eqref{eq:TreeRRf} by writing
\bea
I_n(\{y_i\},t_2)&=&\frac{2 y_T}{\prod_{i=1}^n 2 y_i} \frac{1}{x_1^2-y_{T}^2}\left[\Gf(t_0,t_2,x_1)-\Gf(t_0,t_2,y_{T})\right],
\eea
which written as diagrams, means
\begin{equation}\label{eq:LoopRRf}
 \begin{tikzpicture}[ball/.style = {circle, draw, align=center, anchor=north, inner sep=0}]
 \node[ball,text width=.18cm,fill,color=black, label=below:$\mbox{\scriptsize $x_1$}$] at (0,0) (x1) {};
 \node[ball,text width=1cm,shade,right=2.1cm of x1.east] (S1) {$\mathcal{B}$};
 \node[ball,text width=.18cm,fill,color=black,right=2cm of x1.east, label=below:$\mbox{\scriptsize $\hspace{-.2cm}x_2$}$] (x2) {};
\draw[thick] (x1) to[bend right=50] node[above, yshift=9.5 mm] {\scriptsize $y_1$} (x2);
 \draw[thick] (x1) to[bend right=25] node[above, yshift=5mm] {\scriptsize $y_2$} (x2);
 \draw[thick] (x1) to[bend left=50] node[ below, yshift=-7.2mm] {\scriptsize $y_{n-1}$} (x2);
 \draw[thick] (x1) to[bend left=25] node[ below, yshift=-7.5mm] {\scriptsize $y_n$} (x2);
 \node[rotate=90] at (1.1,-0.1) {\scriptsize ...};
 \node[right=.2cm of S1.east] (for1) {$\displaystyle\:=\:\frac{2 y_T}{\prod_{i=1}^n 2 y_i}\frac{1}{x_1^2-y_{T}^2}\left[\frac{x_1+x_2}{ 2 x_1 x_2}\quad\right.$};
 \node[ball,text width=1cm,shade,right=.2cm of for1.east] (S2) {$\mathcal{B}$};
 \node[ball,text width=.18cm,fill,color=black,right=.1cm of for1.east, label=below:$\mbox{\scriptsize $\hspace{-.4cm}x_1+x_2$}$] (x2a) {};
 \node[right=.2cm of S2.east] (for2) {$\displaystyle\:-\:\frac{y_{T}+x_2}{ 2 y_{T} x_2}\quad$}; 
 \node[ball,text width=1cm,shade,right=.2cm of for2.east] (S3) {$\mathcal{B}$};
 \node[ball,text width=.18cm,fill,color=black,right=.1cm of for2.east, label=below:$\mbox{\scriptsize $\hspace{-.4cm}y_{T}+x_2$}$] (x2b) {};
 \node[right=.2cm of S3.east] (for3) {$\displaystyle\hspace{-.3cm}{\left.\phantom{\frac{1}{y}}\right].}$}; 
 \end{tikzpicture}
\end{equation}
For a simple application, consider a 1-loop exchange diagram. Then we have
\begin{equation}\label{eq:2s1l}
 \begin{tikzpicture}[ball/.style = {circle, draw, align=center, anchor=north, inner sep=0}]
 \node[ball,text width=.18cm,fill,color=black] at (0,0) (x1) {};
 \node[left=.0cm of x1.west,label=below:$x_1$] (x1l) {};
 \node[ball,text width=.18cm,fill,color=black,right=1.5cm of x1.east] (x2) {};
 \node[right=.0cm of x2.east,label=below:$x_2$] (x2l) {};
 \draw[thick] (.85,0) circle (.85);
 \node at (.85,1) {$\displaystyle y_a$};
 \node at (.85,-1) {$\displaystyle y_b$}; 
 \node[right=.2cm of x2.east] (f1) {\Large$=\frac{y_a+y_b}{ 2 y_a y_b}\frac{1}{x_1^2-(y_a+y_b)^2}\left[\frac{x_1+x_2}{ 2 x_1 x_2}\left(-\frac{1}{(x_1+x_2)^2}\right)-\frac{y_a+y_b+x_2}{ 2(y_a+y_b) x_2}\left(-\frac{1}{(y_a+y_b+x_2)^2}\right)\right]$};
 \node[yshift=-1.5cm, left= -7.5 cm of f1] (f2) {\Large$=\frac{x_1+x_2+y_a+y_b}{4 x_1 x_2 y_a y_b (x_1+x_2)(x_1+y_a+y_b)(x_2+y_a+y_b)},$};
 \end{tikzpicture} \nonumber
\end{equation}
and the result for the n-loop exchange can be derived very similarly.\\

We would like to conclude with a final comment. In \cite{Arkani-Hamed:2017fdk} the main recursion relation was derived by using the fact that a time-translation of the Minkowski wavefunction is simple in terms of the external energies. Upon integration by parts and by virtue of the properties of the wavefunction propagator, a diagram can written as an appropriate sum of single cuts. Unfortunately, we were not able to export this type of recursion relation to the correlators. There are two reasons for this. First, for wave function coefficients the time translation operator acts non-trivially on bulk-bulk propagators, since the boundary term is not time translation invariant, even in Minkowski. In our case, we simply have Feynman propagators and get zero. The second reason is that after integrating by parts, in the case of wave function coefficients, the derivative of the bulk-boundary propagators is proportional to the bulk-boundary propagator. However, the derivative of the Bulk-boundary Feynman propagator simply gives us a derivative interaction, that we cannot be related to the original expression. For the exchange, for example, the results we get are
\bea
\int dt dt' G^{\phi \dot{\phi}}_{F}(t_0,t,x_1)G_{F}(t,t',y_{12})G_{F}(t_0,t',x_2)+G_{F}(t_0,t,x_1)G_{F}(t,t',y_{12})G^{\phi \dot{\phi}}_{F}(t,t',x_2)=0, \nonumber
\eea
of which we could not make practical use. 

%%%%%%%%%%%%%%%%%%%%%%%%%%%%%%%%%%%%%%%%%%%%%%%%%%%%%%%%%%%%%%%%%%%%%%%%%%%%%%%%%%%%%%%%%%%%

\section{Cutkosky cutting rules for correlators}\label{sec:cuts}

Let us now move on to a different application of the in-out formalism: cutting rules for correlators. Given that the in-out formalism features only one type of propagator, one might expect that a version of Cutkosky's cutting rules might apply \cite{Cutkosky:1960sp}. In this section, we confirm this expectation and derive explicit results for all diagrams with one or two interaction vertices, to all loops. As long as we consider IR-finite interactions, our results apply to Minkowski as well as to de Sitter space-time.\\

The ``primum mobile" of Cutkosky's cutting rules is Veltman's largest time equation \cite{Veltman:1994wz}, which in turn can be traced back to the following operator identity:
\begin{align}\label{CFTopt}
\sum_{r=0}^n (-1)^r \sum_{\sigma\in\Pi(r,n-r)} \bar T \left[ \mathcal{O}_{\sigma(1)}(t_{\sigma(1)})...\mathcal{O}_{\sigma(r)}(t_{\sigma(r)}) \right] T\left[ (\mathcal{O}_{\sigma(r+1)}(t_{\sigma(r+1)})...\mathcal{O}_{\sigma(n)}(t_{\sigma(n)}) \right]=0\,.
\end{align}
Here $\Pi(r,n-r)$ is the set of partitions of $\{1,..,n\}$ into two subsets of lengths $r$ and $n-r$, so the sum involves $2^n$ terms. The fields $\mathcal{O}_i$ are arbitrary products of operators at the same time. We will mostly focus on cases where these operators are monomials in the fields of the theory and their derivatives. The identity above can be proven by induction. To lowest non-trivial order, $n=2$,~\eqref{CFTopt} simply restates the well-known fact that the two non-time-ordered propagators and the two time-ordered and anti-time ordered propagators are related by\footnote{Since~\eqref{CFTopt} is an operator identity, this propagator identity is valid in any state of the theory.}
\begin{align}\label{simpler}
 \Gf(t,t')+\Gf^*(t,t')-\Gp(t,t')-\Gp(t,t')^*=0\,.
\end{align}
In this simple case, the connection to the largest time equation becomes apparent: whatever $t$ and $t'$ may be, one of the two must be larger\footnote{The case $t=t'$ is trivial because all propagators are equal in that case and the identity is trivially satisfied.}, say $t>t'$. Then the Feynman propagator $\Gf$ reduces to $\Gp$ and the anti-time-ordered propagator $\Gf^*$ reduces to $(\Gp)^*$, hence proving the validity of~\eqref{simpler}.

When one furthermore assumes that all the operators in~\eqref{CFTopt} are Hermitian the equation becomes a real equation, even though this is not apparent in that form. Indeed, we can combine terms pairwise to re-write it as 
\begin{align}\label{even}
 \sum_{r=0}^{n/2-1}& \sum_{\sigma\in\Pi(r,n-r)} (-1)^r 2\Re \ex{ \bar T \left[ \prod_{a=1}^r\mathcal{O}_{\sigma(a)} \right] T\left[ \prod_{b=r+1}^n \mathcal{O}_{\sigma(b)} \right] }\\ &+\sum_{\sigma\in\Pi(n/2,n/2)} (-1)^{n/2} \Re \ex{ \bar T \left[ \prod_{a=1}^{n/2} \mathcal{O}_{\sigma(a)} \right] T\left[ \prod_{b=n/2+1}^n \mathcal{O}_{\sigma(b)} \right]} =0 \,, \nonumber
\end{align}
for $n$ even and as
\begin{align}\label{odd}
 \sum_{r=0}^{(n-1)/2} \sum_{\sigma\in\Pi(r,n-r)} (-1)^r \Im \ex{ \bar T\left[ \prod_{a=1}^r \mathcal{O}_{\sigma(a)} \right] T\left[ \prod_{b=r+1}^n \mathcal{O}_{\sigma(b)} \right]}=0\,,
\end{align}
for $n$ odd, where we left the time arguments implicit. \\

We now follow a similar route to Veltman's derivation of cutting rules \cite{Veltman:1994wz}. We note, however, that the set of rules we establish here are slightly different form their amplitude counterparts because we are not amputating the diagrams, and so we still have to deal with the time ordering acting on the operator insertions. This is not an issue for amplitudes because the LSZ formula effectively pushes all the field insertions to future or past infinity for outgoing and incoming particles, respectively. Before getting into the details, let's summarize our general two-step strategy:
\begin{enumerate}
 \item \textbf{From the largest time equation to propagator identities:} We use the largest time equation on the operators appearing inside a correlator to a given order in perturbation theory. This includes the operators appearing explicitly in a correlator and any number of powers of the interaction Hamiltonian. The outcome is a set of identities relating different products of time-ordered and non-time-ordered propagators. We employ a nifty diagrammatic notation to write these identities in terms of cut diagrams.
 \item \textbf{From propagator identities to cutting rules for correlators:} Using properties of the propagators, we re-write the above propagator identities as relations among correlators for which some of the energies have been analytically continued to negative real values. 
\end{enumerate} 
We will derive explicitly cutting rules for diagrams with at most two interaction vertices, with any number of external legs and any number of loops. For three or more vertices, we can still write propagator identities, but to transform them into cutting rules for correlators one would need to appropriately generalise our derivation. Let us now proceed to the derivation of propagator identities. 

%%%%%%%%%%%%%%%%%%%%%%%%%%%%%%%%%%%%%%%%%%%%%%%%%%%%%%%%%%%%%%%%%%%%%%%%%%%%

\subsection{Propagator identities}

Here we use the largest time equation to find propagator identities. The starting point is to expand the time-evolution operator in an in-out correlator to some order in perturbation theory. Then we want to think of the various powers of $\phi$ and $\Hi$ as the different operators appearing in~\eqref{CFTopt}. This gives us a set of identities. To see how this works, let's start with 1-vertex diagrams and work our way up to two vertices. The case of three vertices is discussed in App.~\ref{app:C}.

%%%%%%%%%%%%%%%%%%%%%%%%%%%%%%%%%%%%%%%%%%%%%

\paragraph{1-vertex diagrams} A one vertex diagram has a single power of $\Hi$ and $n$ copies of the field $\phi$. As a simple example, consider the choice
\begin{align}
 \mathcal{O}_1&=\phi(\bfx_1,t_0)^m\,, &\mathcal{O}_2&=\phi(\bfx_2,t_0)^{n-m}\,, & \mathcal{O}_3&=\Hi(t)\,,
\end{align}
where $t_0$ is an arbitrary time in Minkowski or in de Sitter. Since we restrict our derivation to equal-time correlators we will omit this time dependence in the following. Inserting the above choice of operators into~\eqref{odd} and sandwiching between two ground states we get (in position space)
\bea\label{long1}
 \Im \left\{ \langle T [\phi^n \Hi]\rangle-\langle \phi^m T[\phi^{n-m} \Hi]\rangle -\langle \phi^{n-m} T[\phi^{m} \Hi]\rangle \right\} \simeq 0,
\eea
where we omitted to write the time-ordering or anti-time-ordering when a single time is present. Here we wrote $\simeq 0$ to indicate that this identity is only valid after time integration because we dropped the term $\Hi T[\phi^n]$. This is allowed because every term where a Hamiltonian interaction is not in the same time ordering as a field, it integrates to zero. Indeed, if the interaction has no relative time ordering to the boundary, we obtain an integral over the mode functions from minus infinity to infinity. As discussed previously, the integral over the mode functions in the in-out formalism is regularized in such a way that it goes to zero in the infinite past and future, so that the full integral evaluates to zero. We have hence omitted those terms in this expression and we used the symbol $\simeq 0$ as a reminder of this simplification.

To find the equivalent of~\eqref{long1} in momentum space we have to deal with the fact that $\phi(\bfk)$ is not Hermitian because $\phi(\bfk)^*=\phi(-\bfk)$ by the reality of $\phi(\bfx)$. To remove the extra minus sign we have to separate the discussion between interactions that are even (+) or odd (-) under spatial parity (point inversion). For example, we have 
\begin{align}
\langle T [\phi(\bfk)^n \Hi(t)]\rangle_{\text{PE}}=+ \langle \bar{T} [ \phi(\bfk)^{n} \Hi(t)]\rangle^*_{\text{PE}}\,, \\
\langle T [\phi(\bfk)^n \Hi(t)]\rangle_{\text{PO}}=- \langle \bar{T} [ \phi(\bfk)^{n} \Hi(t)]\rangle^*_{\text{PO}}\,, 
\end{align}
where the labels parity even (PE) and parity odd (PO) mean 
\begin{align}
 \ex{F(\bfk)}_{\text{PE}} &\equiv \frac12 \left[ \ex{F(\bfk)}+\ex{F(-\bfk)} \right]\,, & \ex{F(\bfk)}_{\text{PO}} &\equiv \frac12 \left[ \ex{F(\bfk)}-\ex{F(-\bfk)} \right]\,.
\end{align}
Hence in momentum space, the parity-even component satisfies the same equation as in~\eqref{long1}, while for the parity-odd component, the imaginary part is substituted by the real part.

We notice that the cutting rule is really a statement about the imaginary (parity even) or real (parity odd) part of~\eqref{long1}. As we will see, these are exactly the parts of the diagrams we are interested in, so this will lead to useful relations between diagrams. But first, let's give another example.

\paragraph{2-vertex diagrams} The procedure for two-vertex diagrams is very similar. We simply take a different group of operators in an in-out correlator and use the largest time equation. Here we will focus on a particular channel, but all other channels can be discussed in the same way. For example, after expanding the evolution operator to second order, let's insert the following identifications in the largest time equation,
\begin{align}
 \mathcal{O}_1&=\phi(t_0)^m\,, &\mathcal{O}_2&=\phi(t_0)^{n-m}\,, &\mathcal{O}_3&=\Hi^{(1)}(t)\,, &\mathcal{O}_4&=\Hi^{(2)}(t')\,.
\end{align} 
We get in real space
\bea\label{long2}
% 0&\simeq &\langle T [\phi^n \Hi^{(1)}\Hi^{(2)}]\rangle+\langle \bar{T}[\phi^{m}\Hi^{(1)}] T[ \phi^{n-m}\Hi^{(2)}]\rangle\\ \nonumber
% &-&\left(\langle \phi^{m} T[ \phi^{n-m}\Hi^{(1)}\Hi^{(2)}]\rangle+\langle \bar{T}[\phi^m\Hi^{(1)}\Hi^{(2)}]\phi ^{n-m}\rangle\right) \\
0&\simeq & \Re \left\{ \langle T [\phi^n \Hi^{(1)}\Hi^{(2)}]\rangle - \langle \phi^{m} T[ \phi^{n-m}\Hi^{(1)}\Hi^{(2)}]\rangle - \langle \phi^{n-m} T[ \phi^{m}\Hi^{(1)}\Hi^{(2)}]\rangle \right.\\ 
 \nonumber &+& \left. \langle \bar{T}[\phi^{m}\Hi^{(1)}] T[ \phi^{n-m}\Hi^{(2)}]\rangle + \langle \bar{T}[\phi^{m}\Hi^{(2)}] T[ \phi^{n-m}\Hi^{(1)}]\rangle \right\}\,,
\eea
where $\simeq 0$ again indicates that the identity is valid after integrating over the time of the interactions because we omitted a number of terms that integrate to zero. This is again due to the fact that the Hamiltonian interactions do not have a time ordering with respect to the insertions of the operators. In momentum space, the above expression is unchanged for the parity-even part, while for the parity-odd part, one needs to change $\Re \to \Im$.
The above procedure is straightforward but leads to lengthy expressions. Here we show how to streamline it using the diagrammatic notation of ``cut diagrams".

%%%%%%%%%%%%%%%%%%%%%%%%%%%%%%%%%%%%%%%%%%%%%%%%%%%

\paragraph{A diagrammatic representation: cut diagrams} We would like to represent expressions such as~\eqref{long1} and~\eqref{long2} in terms of diagrams. To this end, consider a Feynman diagram and imagine separating it into two subsets of vertices\footnote{This is different from deciding to cut or not cut \textit{each} line. For example, for a one-loop two-vertex diagram we cannot cut only one of the two loop propagators because cutting only one does not create two separated subsets of vertices.} by a ``cut". The meaning of a cut is that (i) all operators on one side of the cut are time-ordered with respect to each other, (ii) those on the other side are anti-time ordered with respect to each other and (iii) operators on different sides of the cut have no relative time ordering. As an intermediate step, let's introduce Feynman rules for shaded cuts, where the shading refers to the side that is anti-time-ordered, while the un-shaded side is time-ordered:
\begin{align}
 \text{Shaded}&\sim \text{anti-time ordered}\,, & \text{Un-shaded}&\sim \text{time ordered}\,.
\end{align}
The Feynman rules for a shaded cut diagram are as follows~\cite{Veltman:1994wz}:
\begin{enumerate}
\item A vertex at $t$ on the shaded side of the cut connected to a vertex at $t'$ on the un-shaded side of the cut leads to a factor $\Gp(t,t')=f(t)f^{\ast}(t')$.
\item Two vertices $t$ and $t'$ on the un-shaded side of the cut that are connected to each other lead to a time-ordered propagator $\Gf(t,t')$.
\item Two vertices $t$ and $t'$ on the shaded side of the cut that are connected to each other lead to an anti-time ordered propagator $\Gf^*(t,t')$.
\item A vertex on the un-shaded (shaded) side gets a $-i$ ($+i$) factor times the appropriate coupling constant. This choice corresponds to the process of expanding the forward time evolution operator $T e^{-iH}$ on the un-shaded side, and reverse time evolution operator $\bar T e^{+iH}$ on the shaded side. 
\end{enumerate}
Vertex factors and integration over the vertices are as usual. The only difference from the standard Feynman rules are the propagators. The rules for a right-shaded cut, are the same just under the exchange of the word 'left' and 'right'. Finally, an unshaded cut is the average of the two shadings, that is: 

\begin{equation}\label{eq:wvn}
 \begin{tikzpicture}[ball/.style = {circle, draw, align=center, anchor=north, inner sep=0}]
 \node[ball,text width=.18cm,fill,color=white,label=below:] at (0,0) (x1) {};
 \node[ball,text width=.18cm,fill,color=white,label=below:] at (2.5,0)(x2) {};
 \coordinate (A) at (1.25,-0.75);
 \coordinate (B) at (1.25,0.65);
 \draw[-,thick,color=black] (A) -- (B); 
 \draw[-,thick,color=black] (x1) -- (x2);
 \node[color=black,right=0.1cm of x2.east] (y1) {\Large $:=\frac{1}{2}\bigg($}; 
 \node[ball,text width=.18cm,fill,color=white,label=below:] at (4,0) (x1) {};
 \node[ball,text width=.18cm,fill,color=white,label=below:] at (6.5,0)(x2) {};
 \draw[-,thick,color=black] (x1) -- (x2);
 \coordinate (A) at (5.2,-0.75);
 \coordinate (B) at (5.2,0.65);
 \coordinate (C) at (5.3,0.65);  
 \coordinate (D) at (5.3,-0.75);  
 \fill[pattern=north east lines, pattern color=black] (A) -- (B) -- (C) -- (D) ;
 \draw[white, line width=0.1pt] (A) -- (B);
 \draw[white, line width=0.1pt] (B) -- (C);
 \draw[black, line width=1pt] (C) -- (D);
 \draw[white, line width=0.1pt] (D) -- (A);
 
; 

 \node[color=black,right=0.1cm of x2.east] (y1) {\large $+$};

 \node[ball,text width=.18cm,fill,color=white,label=below:] at (7.5,0) (x1) {};
 \node[ball,text width=.18cm,fill,color=white,label=below:] at (10,0)(x2) {};
 \coordinate (A) at (8.7,-0.75);
 \coordinate (B) at (8.7,0.65);
 \coordinate (C) at (8.8,0.65);  
 \coordinate (D) at (8.8,-0.75);  
 \fill[pattern=north west lines, pattern color=black] (A) -- (B) -- (C) -- (D) ;
 \draw[black, line width=1pt] (A) -- (B);
 \draw[white, line width=0.1pt] (B) -- (C);
 \draw[white, line width=0.1pt] (C) -- (D);
 \draw[white, line width=0.1pt] (D) -- (A);
 \draw[-,thick,color=black] (x1) -- (x2);
 \node[color=black,right=-0.3cm of x2.east] (y1) {\large $\bigg).$}; 
\end{tikzpicture}
\end{equation}

%%%%%%%%%%%%%%%%%%%%%%%%%%%%%%%%%%%%%%%%

\paragraph{Comments} Three comments are in order. The first is that cut diagrams have a meaning that is independent of the largest time equation. For example, we can consider the $s$-channel exchange for the four-point function in Minkowski with two insertions of a $\phi^3/(3!)$ interaction and directly write
\bea
\raisebox{-5ex}{\begin{tikzpicture}
\draw (-0.8,0) -- (2.3,0);
\foreach \x \y in {-.25/1.25 ,0.25/1.75} {
 \draw (\x,0) -- (0,-1.5);
 \draw (\y,0) -- (1.5,-1.5);
 \filldraw (\x,0) circle (2pt);
 \filldraw (\y,0) circle (2pt);
 \filldraw (0,-1.5) circle (2pt);
 \filldraw (1.5,-1.5) circle (2pt);
}
 \draw (0,-1.5) to (1.5,-1.5);
 
 \coordinate (A) at (-0.68,-1.5);
 \coordinate (B) at (0.77,0.4);
 \coordinate (C) at (0.9,0.4);  
 \coordinate (D) at (-0.55,-1.5);  
 \fill[pattern=north west lines, pattern color=black] (A) -- (B) -- (C) -- (D) ;
 \draw[white, line width=0.1pt] (A) -- (B);
 \draw[white, line width=0.1pt] (B) -- (C);
 \draw[black, line width=1pt] (C) -- (D);
 \draw[white, line width=0.1pt] (D) -- (A);
 
 \end{tikzpicture}} = -\int_{-\infty}^\infty \int_{-\infty}^\infty dt dt' \left(\prod_{i=1}^2 G^+(t_0,t,E_i)\right)G_F(t,t',s) \left(\prod_{i=3}^4 G_F(t_0,t',E_i)\right),
\eea
where $s$ is the energy of the internal line. This emphasises that a cut diagram is just a diagrammatic representation of the integral of the product of propagators. \\

Second, we want to discuss how to diagrammatically identify those terms in the largest time equation that vanish upon integration over time, namely cases where a Hamiltonian interaction does not appear in the same time ordering as an external operator. This observation leads to the diagrammatic rule that \textit{a diagram integrates to zero if it contains an interaction vertex that is separated by the cut from all (fixed) operator insertions}. As an example of this rule, consider a contact diagram in flat space with a single power of the interaction Hamiltonian, which we take to be a simple $\phi^3/(3!)$ interaction. This results in a correlator of the form 
\begin{align}\label{above}
\left\langle T \left(\prod_a^n\phi_{E_a}(t_0) \Hi(t)\right)\right\rangle\,. 
\end{align}
Applying the largest time~\eqref{CFTopt} to this diagram (viewing $\phi(t_0)^n$ as a single operator and $\Hi(t)$ as a second operator), we obtain a sum over terms, one of which takes the form
\begin{align}
 \int_{-\infty}^{\infty}dt \left\langle \Hi(t) T \left(\phi_{E_1}(t_0)\phi_{E_2}(t_0)\phi_{E_3}(t_0)\right)\right\rangle & \sim \int_{-\infty}^{\infty}dt \left\langle \phi(t)^3 T \left(\phi_{E_1}(t_0)\phi_{E_2}(t_0)\phi_{E_3}(t_0)\right)\right\rangle 
\end{align}
This integrates to zero because the mode functions do
\begin{align}
 & \propto \int_{-\infty}^{\infty}dt \prod_{i=1}^3G^+(t,t_0,E_i) =0\,.
\end{align}
Graphically this corresponds to
\bea
\raisebox{-5ex}{\begin{tikzpicture}
\draw (-1,0) -- (1,0);
\foreach \x in {-0.5,0,0.5} {
 \draw (\x,0) -- (0,-1.5);
 \filldraw (\x,0) circle (2pt);
 \filldraw (0,-1.5) circle (2pt);
}

\coordinate (A) at (-0.6,-0.63);
 \coordinate (B) at (-0.6,-0.5);
 \coordinate (C) at (0.8,-1.2);  
 \coordinate (D) at (0.8,-1.33);  
 \fill[pattern=north east lines, pattern color=black] (A) -- (B) -- (C) -- (D) ;
 \draw[white, line width=0.1pt] (A) -- (B);
 \draw[black, line width=1pt] (B) -- (C);
 \draw[white, line width=0.1pt] (C) -- (D);
 \draw[white, line width=0.1pt] (D) -- (A);

 \end{tikzpicture}} = 0.
\eea
Therefore, the actual number of cut diagrams that we have to consider when rephrasing the largest time equation as a propagator identity is less, and often much less, than the number of parting into two sets appearing in~\eqref{CFTopt}.\\

The third and final comment is that the results derived in this section are valid for fields of any mass in de Sitter or Minkowski. It is only in the next section that we will restrict to massless and conformally coupled fields in de Sitter.

%%%%%%%%%%%%%%%%%%%%%%%%%%%%%%%%%%%%%%%%%%%%%%%%%%%%%%%%%%%%%%%

\paragraph{$V=1$ cut diagrams} Let's see now how the one- and two-vertex examples we considered before look like in terms of cut diagrams. For contact operators~\eqref{long1} becomes
\bea\label{contactfin}
 \hspace{-4em}\raisebox{-10ex}{\begin{tikzpicture}[scale=0.8,ball/.style = {circle, draw, align=center, anchor=north, inner sep=0}]
 \draw (-0.2,0) -- (2.4,0);
 \foreach \x/\label in {0.1/E_1,0.4/ , 0.7/\hspace{1.7em}\ldots,1.3/,
 1.7/,2.1/\hspace{1em}E_n} {
 \draw (\x,0) -- (1.1,-1.7);
 \filldraw (\x,0) circle (2pt) node[above] {\footnotesize $\label$};
 }
 \filldraw (1.1,-1.7) circle (2pt) ;
 
 \foreach \x in {0.5, 0.62,0.8} {
 \draw (1.1,-1.7-\x) circle (\x);
 }
 \end{tikzpicture}} -
 \raisebox{-10ex}{\begin{tikzpicture}[scale=0.8,ball/.style = {circle, draw, align=center, anchor=north, inner sep=0}]
 \draw (-1.2,0) -- (3.4,0);
 \foreach \x/\label in {-0.6/\hspace{1em}E_1\ldots,-0.3/ , 0.6/\hspace{-0.5em}E_m, 1.6/\hspace{1.5em}E_{m+1}\,\ldots, 2.5/,2.8/\hspace{1em}E_n} {
 \draw (\x,0) -- (1.1,-1.7);
 \filldraw (\x,0) circle (2pt) node[above] {\footnotesize $\label$};
 }
 \filldraw (1.1,-1.7) circle (2pt) ;
 
 \foreach \x in {0.5, 0.62,0.8} {
 \draw (1.1,-1.7-\x) circle (\x);
 }
 \draw[black, line width=1pt] (1.2,0.7) -- (-0.1,-1.7);
 
 \end{tikzpicture}}-
 \raisebox{-10ex}{\begin{tikzpicture}[scale=0.8,ball/.style = {circle, draw, align=center, anchor=north, inner sep=0}]
 \draw (-1.2,0) -- (3.4,0);
 \foreach \x/\label in {-0.6/\hspace{1em}E_1\ldots,-0.3/ , 0.6/\hspace{-0.5em}E_m, 1.6/\hspace{1.5em}E_{m+1}\,\ldots, 2.5/,2.8/\hspace{1em}E_n} {
 \draw (\x,0) -- (1.1,-1.7);
 \filldraw (\x,0) circle (2pt) node[above] {\footnotesize $\label$};
 }
 \filldraw (1.1,-1.7) circle (2pt) ;
 
 \foreach \x in {0.5, 0.62,0.8} {
 \draw (1.1,-1.7-\x) circle (\x);
 }
 \draw[black, line width=1pt] (0.8,0.7) -- (2.3,-1.7);
 
 \end{tikzpicture}}\hspace{1em}&&\hspace{-2.5em}= 0.
 \eea
where a diagram without a cut should be understood as having a cut all the way to the right or, equivalently, to the left. Notice that this relation is valid for the loop integrand to all loop orders (but only for a single vertex). In other words, we can contract any number of pairs of fields in $\Hi$, since they are still associated to the same time in the largest time equation. 

%%%%%%%%%%%%%%%%%%%%%%%%%%%%%%%%%%%%%%%%%%%%%%%%%%%%%%%%%%%%%%%%%%%%%%%%%%%%

\paragraph{$V=2$ cut diagrams} For the exchange diagram we would find that the cut diagram representation of~\eqref{long2} contains five terms. However, if we specify a specific channel, namely a way to pairwise contract fields and the interaction Hamiltonians, the last term in~\eqref{long2} vanishes because an interaction Hamiltonian is contracted only with fields outside of its time ordering. Dropping this vanishing term we find the diagrammatic representation
\bea\label{exfin}
\hspace{-2em} \raisebox{-6ex}{%
\begin{tikzpicture}[scale=0.7, ball/.style = {circle, draw, align=center, anchor=north, inner sep=0}]
 \draw (-0.1,0) -- (4.3,0);
 \foreach \x/\label in {0.1/\hspace{1em}E_1\ldots,0.4/ , 1/\hspace{1em}E_m} {
 \draw (\x,0) -- (1.1,-1.5);
 \filldraw (\x,0) circle (2pt) node[above] {\fontsize{8}{10}\selectfont $\label$};
 }
 \foreach \x/\label in {3.2/E_{m+1}\ldots, 3.8/,4.1/\hspace{1em}E_n} {
 \draw (\x,0) -- (3.1,-1.5);
 \filldraw (\x,0) circle (2pt) node[above] {\fontsize{8}{10}\selectfont $\label$};
 }
 \filldraw (1.1,-1.5) circle (2pt);
 \filldraw (3.1,-1.5) circle (2pt);
 \draw (1.1,-1.5) to[out=80, in=100] (3.1,-1.5);
 \draw (1.1,-1.5) to[out=-80, in=-100] (3.1,-1.5);
 \draw (1.1,-1.5) to[out=30, in=150] (3.1,-1.5);
 \draw (1.1,-1.5) to[out=-30, in=-150] (3.1,-1.5);

 \node at (2.1,-1.5) {\fontsize{8}{10}\selectfont $...$};
 
\end{tikzpicture}}\hspace{-1em}-
\hspace{-1.5em}\raisebox{-6ex}{%
\begin{tikzpicture}[scale=0.7, ball/.style = {circle, draw, align=center, anchor=north, inner sep=0}]
 \draw (-0.1,0) -- (4.3,0);
 \foreach \x/\label in {0.1/\hspace{1em}E_1\ldots,0.4/ , 1/\hspace{1em}E_m} {
 \draw (\x,0) -- (1.1,-1.5);
 \filldraw (\x,0) circle (2pt) node[above] {\fontsize{8}{10}\selectfont $\label$};
 }
 \foreach \x/\label in {3.2/E_{m+1}\ldots, 3.8/,4.1/\hspace{1em}E_n} {
 \draw (\x,0) -- (3.1,-1.5);
 \filldraw (\x,0) circle (2pt) node[above] {\fontsize{8}{10}\selectfont $\label$};
 }
 \filldraw (1.1,-1.5) circle (2pt);
 \filldraw (3.1,-1.5) circle (2pt);
 \draw (1.1,-1.5) to[out=80, in=100] (3.1,-1.5);
 \draw (1.1,-1.5) to[out=-80, in=-100] (3.1,-1.5);
 \draw (1.1,-1.5) to[out=30, in=150] (3.1,-1.5);
 \draw (1.1,-1.5) to[out=-30, in=-150] (3.1,-1.5);

 \node at (2.1,-1.5) {\fontsize{8}{10}\selectfont $...$};

 \draw[black, line width=1pt] (2.0,0.2) -- (0.2,-1.6);
\end{tikzpicture}}\hspace{-1em}-
\hspace{-1.5em} \raisebox{-6ex}{%
\begin{tikzpicture}[scale=0.7, ball/.style = {circle, draw, align=center, anchor=north, inner sep=0}]
 \draw (-0.1,0) -- (4.3,0);
 \foreach \x/\label in {0.1/\hspace{1em}E_1\ldots,0.4/ , 1/\hspace{1em}E_m} {
 \draw (\x,0) -- (1.1,-1.5);
 \filldraw (\x,0) circle (2pt) node[above] {\fontsize{8}{10}\selectfont $\label$};
 }
 \foreach \x/\label in {3.2/E_{m+1}\ldots, 3.8/,4.1/\hspace{1em}E_n} {
 \draw (\x,0) -- (3.1,-1.5);
 \filldraw (\x,0) circle (2pt) node[above] {\fontsize{8}{10}\selectfont $\label$};
 }
 \filldraw (1.1,-1.5) circle (2pt);
 \filldraw (3.1,-1.5) circle (2pt);
 \draw (1.1,-1.5) to[out=80, in=100] (3.1,-1.5);
 \draw (1.1,-1.5) to[out=-80, in=-100] (3.1,-1.5);
 \draw (1.1,-1.5) to[out=30, in=150] (3.1,-1.5);
 \draw (1.1,-1.5) to[out=-30, in=-150] (3.1,-1.5);

 \node at (2.1,-1.5) {\fontsize{8}{10}\selectfont $...$};

 \draw[black, line width=1pt] (2.2,0.2) -- (4,-1.6);
\end{tikzpicture}}\hspace{-1em}+
\hspace{-1.5em} \raisebox{-6.6ex}{%
\begin{tikzpicture}[scale=0.7,ball/.style = {circle, draw, align=center, anchor=north, inner sep=0}]
 \draw (-0.1,0) -- (4.3,0);
 \foreach \x/\label in {0.1/\hspace{1em}E_1\ldots,0.4/ , 1/\hspace{1em}E_m} {
 \draw (\x,0) -- (1.1,-1.5);
 \filldraw (\x,0) circle (2pt) node[above] {\fontsize{8}{10}\selectfont $\label$};
 }
 \foreach \x/\label in {3.2/E_{m+1}\ldots, 3.8/,4.1/\hspace{1em}E_n} {
 \draw (\x,0) -- (3.1,-1.5);
 \filldraw (\x,0) circle (2pt) node[above] {\fontsize{8}{10}\selectfont $\label$};
 }
 \filldraw (1.1,-1.5) circle (2pt);
 \filldraw (3.1,-1.5) circle (2pt);
 \draw (1.1,-1.5) to[out=80, in=100] (3.1,-1.5);
 \draw (1.1,-1.5) to[out=-80, in=-100] (3.1,-1.5);
 \draw (1.1,-1.5) to[out=30, in=150] (3.1,-1.5);
 \draw (1.1,-1.5) to[out=-30, in=-150] (3.1,-1.5);

 \node at (1.7,-1.5) {\fontsize{8}{10}\selectfont $...$};
 
 \draw[black, line width=1pt] (2.1,0.2) -- (2.1,-2.4);
\end{tikzpicture}}\hspace{-1em}=0.
\eea
We will see in the next section how these identities can be transformed into cutting rules for correlators. 

%%%%%%%%%%%%%%%%%%%%%%%%%%%%%%%%%%%%%%%%%%%%%%%%%%%%%%%%%%%%%%%

\paragraph{$V>2$ cut diagrams} Here we briefly discuss the general properties of diagrams with more than two vertices. While it would be interesting to perform a serious combinatorial analysis of the problem, here we limit ourselves to some simple remarks. First of all, we notice that the number of terms appearing in the largest time equation grows fast and it's desirable to consider only the new constraints that arise at higher order. Moreover, the really useful power of these relations arises when a complicated diagram is reduced to the sum over products of simpler ones. To focus on this case, it is convenient to only discuss cases where all the external fields connected to a single vertex are treated as a single operator in the largest time equation. Then we have to deal with diagrams where each interaction vertex is attached to exactly one external line. Very generally we then have $V$ bulk vertices and $V$ external vertices. Given that a cut separates these $2V$ vertices into two groups, we know the largest time equation generates $2^{2V}$ terms. These are then related pairwise by complex conjugation as in~\eqref{even} and~\eqref{odd} leading to $2^{2V-1}$ terms. Of these terms, many vanish because of the general rule stated above~\eqref{above}. We were only able to find an upper bound on the number of vanishing diagrams by counting all cuts in which interaction vertices are not in the same time order as an external operator. These can be counted by summing the binomials $\binom{V}{k}$ over $1\leq k \leq V$, which gives $2^{V}-1$. This leads to the upper bound on the number $N$ of non-vanishing terms in the largest time equation (with one external leg per vertex)
\begin{align}
 N\leq 2^{2V-1}-2^V+1\,.
\end{align}
For example, for $V=1$ we find $N\leq 1$. In this case, the bound is saturated and we have a single term and we find the constraint
\bea
\raisebox{-3ex}{\begin{tikzpicture}
\draw (-0.5,0) -- (0.5,0);
\foreach \x in {0} {
 \draw (\x,0) -- (0,-1);
 \filldraw (\x,0) circle (2pt);
 \filldraw (0,-1) circle (2pt);}
 \end{tikzpicture}} = 0.
\eea
For $V=2$ we find $N\leq 5$. The actual number of non-vanishing terms is $N=4$. The one diagram that does vanish but is not accounted for in our bound is
\bea
\raisebox{-3em}{\begin{tikzpicture}
 \draw (0,0) -- (4,0);
 \draw (1,0) -- (1,-1.5);
 \draw (3,0) -- (3,-1.5);
 \filldraw (1,0) circle (2pt)  ;
 \filldraw (3,0) circle (2pt)  ;
 
 \filldraw (1,-1.5) circle (2pt);
 \filldraw (3,-1.5) circle (2pt);
  \draw (1,-1.5) -- (3,-1.5);
 \draw[black, line width=1pt] (0.1,0.3) -- (2.9,-1.8);
 \draw [black, line width=1pt](1.1,0.3) -- (3.28,-1.335);
 \draw [black, line width=1pt](2.9,-1.8) arc [start angle=45-180, end angle=68, radius=0.3];
\end{tikzpicture}}=0\,.
\eea
This is the fifth term in~\eqref{long2}\footnote{Depending on the channel it can be the fourth or fifth.}, and the external legs connected to each vertex have been represented graphically as a single line. It will be interesting to pursue this further in the future. \\

%%%%%%%%%%%%%%%%%%%%%%%%%%%%%%%%%%%%%%%%%%%%%%%%%%%%%%%%%%%%%%%%%%%%%%%%%%%%

\subsection{Cutting rules for correlators}

Finally, we would like to relate the propagator identities represented by cut diagrams to correlators with shifted kinematics. For technical reasons, we will restrict our discussion to fields of any mass in Minkowski but only massless and conformally coupled fields in de Sitter, but we include all derivative interactions. To do this, we first assume Hermitian analyticity, i.e. for all propagators, we assume 
\begin{align}\label{HA}
G(\eta,\eta',k)&=-G(\eta,\eta',-k)^*\,, & G(t,t',E)&=-G(t,t',-E)^*\,. 
 \end{align}
In the following, we will therefore restrict ourselves to these three cases, but include all derivative interactions, as they preserve hermitian analyticity.
This is true for all the fields we consider in this section. Furthermore, we write all expressions at the leading order of $\eta_0$ of the uncut correlator\footnote{This is necessary, because cut diagrams, will in some cases have subleading terms in $\eta_0$, that however cancel between terms.}. As we show in App.~\ref{app:D} this then allows us to relate the three types of cuts we consider in this section, to be related to the contact correlator, $B_n^c$, and exchange correlator $B_n^{ex}$ in the following way. For contact diagrams, we find
\bea\label{rules fin c}
 \hspace{-2em}\raisebox{-12ex}{\begin{tikzpicture}[ball/.style = {circle, draw, align=center, anchor=north, inner sep=0}]
 \draw (-1.2,0) -- (3.4,0);
 \foreach \x/\label in {-0.6/\hspace{1em}E_1\,\,\ldots,-0.3/ , 0.6/\hspace{-1em}E_m, 1.6/\hspace{1.5em}E_{m+1}\,\ldots, 2.5/,2.8/\hspace{1em}E_n} {
 \draw (\x,0) -- (1.1,-1.7);
 \filldraw (\x,0) circle (2pt) node[above] {\footnotesize $\label$};
 }
 \filldraw (1.1,-1.7) circle (2pt) ;
 
 \foreach \x in {0.5, 0.62,0.8} {
 \draw (1.1,-1.7-\x) circle (\x);
 }
 \draw[black, line width=1pt] (1.2,0.7) -- (-0.1,-1.7);
 
 \end{tikzpicture}}\hspace{1em}&&\hspace{-2.5em} = \frac{1}{2}\left[B^c_n(\{E_i\}_{i=1}^n)+(-1)^mB^{c}_n(\{-E_i\}_{i=1}^m,\{E_i\}_{i=m+1}^n))\right].
 \eea
 For exchanges, we consider cuts that run along internal lines, or cuts of external lines that all connect to the same vertex, as in~\eqref{exfin}. We then have
 \bea\label{rules fin e}
 \hspace{-2em} \raisebox{-7ex}{\begin{tikzpicture}[ball/.style = {circle, draw, align=center, anchor=north, inner sep=0}]
 \draw (-0.4,0) -- (4.6,0);
 \foreach \x/\label in {-0.1/\hspace{1em}E_1\,\,\ldots,0.2/ , 1/\hspace{-1em}E_m} {
 \draw (\x,0) -- (1.1,-1.5);
 \filldraw (\x,0) circle (2pt) node[above] {\footnotesize $\label$};
 }
 \foreach \x/\label in {3.2/\hspace{1.5em}E_{m+1}\,\ldots, 4/,4.3/\hspace{1em}E_n} {
 \draw (\x,0) -- (3.1,-1.5);
 \filldraw (\x,0) circle (2pt) node[above] {\footnotesize $\label$};
 }
 \filldraw (1.1,-1.5) circle (2pt) ;
 \filldraw (3.1,-1.5) circle (2pt) ;
 \draw (1.1,-1.5) to[out=80, in=100] (3.1,-1.5);
 \draw (1.1,-1.5) to[out=-80, in=-100] (3.1,-1.5);

 \draw (1.1,-1.5) to[out=30, in=150] (3.1,-1.5);
 \draw (1.1,-1.5) to[out=-30, in=-150] (3.1,-1.5);
 
 \node at (2.1,-0.8) {\footnotesize $y_1$};
 \node at (2.1,-1.1) {\footnotesize $y_2$};
 \node at (2.1,-1.5) {\footnotesize $...$};
 \node at (2.1,-1.9) {\footnotesize $y_{L}$};
 \node at (2.1,-2.2) {\footnotesize $y_{L+1}$};

 \draw[black, line width=1pt] (1.9,0.5) -- (-0.2,-1.6);
 
 \end{tikzpicture}}&&\hspace{-2.5em} =\frac{1}{2}\left[B^{ex}_n(\{E_i\})+(-1)^mB^{ex}_n(\{-E_i\}_{i=1}^m,\{E_i\}_{i=m+1}^n))\right],\\ \nonumber \label{rules fin e2}
 \hspace{-2em} \raisebox{-7ex}{\begin{tikzpicture}[ball/.style = {circle, draw, align=center, anchor=north, inner sep=0}]
 \draw (-0.4,0) -- (4.6,0);
 \foreach \x/\label in {-0.1/\hspace{1em}E_1\,\,\ldots,0.2/ , 1/\hspace{-1em}E_m} {
 \draw (\x,0) -- (1.1,-1.5);
 \filldraw (\x,0) circle (2pt) node[above] {\footnotesize $\label$};
 }
 \foreach \x/\label in {3.2/\hspace{1.5em}E_{m+1}\,\ldots, 4/,4.3/\hspace{1em}E_n} {
 \draw (\x,0) -- (3.1,-1.5);
 \filldraw (\x,0) circle (2pt) node[above] {\footnotesize $\label$};
 }
 \filldraw (1.1,-1.5) circle (2pt) ;
 \filldraw (3.1,-1.5) circle (2pt) ;
 \draw (1.1,-1.5) to[out=80, in=100] (3.1,-1.5);
 \draw (1.1,-1.5) to[out=-80, in=-100] (3.1,-1.5);

 \draw (1.1,-1.5) to[out=30, in=150] (3.1,-1.5);
 \draw (1.1,-1.5) to[out=-30, in=-150] (3.1,-1.5);
 
 \node at (1.7,-0.85) {\footnotesize $y_1$};
 \node at (1.7,-1.15) {\footnotesize $y_2$};
 \node at (1.7,-1.5) {\footnotesize $...$};
 \node at (1.7,-1.85) {\footnotesize $y_{L}$};
 \node at (1.7,-2.15) {\footnotesize $y_{L+1}$};
 
 \draw[black, line width=1pt] (2.1,0.2) -- (2.1,-2.4);
 
 \end{tikzpicture}}&&\hspace{-2.5em} =-\frac{B^{\text{c,cut}}_{m,L+1}(\{E_i\}_{i=1}^m,\{y_i\}_{i=1}^{L+1}) B^{\text{c,cut}}_{n-m,L+1}(\{E_i\}_{i=m+1}^n,\{y_i\}_{i=1}^{L+1})}{\prod_{i=1}^{L+1}{P(y_i)}},
 \eea
where $P(y)$, is the power spectrum of $\phi$ and we defined
\bea\label{rules fin d}
B^{\text{c,cut}}_{n,L}(\{E_i\}_{i=1}^n,\{y_i\}_{i=1}^{L})=\frac{1}{2}\left[B^{\text{c}}_{n+L}(\{E_i\}_{i=1}^n,\{y_i\}_{i=1}^{L})+(-1)^{L}B^{\text{c}}_{n+L}(\{E_i\}_{i=1}^n,\{-y_i\}_{i=1}^{L}))\right].
\eea

%%%%%%%%%%%%%%%%%%%%%%%%%%%%%%%%%%%%%%%%%%%%%%%%%%%%%%%%%%%%%%%%%%%%%%%%%%%%%%%%%%%%%%

\paragraph{1-vertex cutting rules}\label{cuttings} 

We can in principle get a multitude of cutting rules from~\eqref{CFTopt}, depending on how we identify operators with powers of $\phi$ and $\Hi$. We here show one example for contact diagrams. We use the relation~\eqref{rules fin c} between cut diagrams and analytically continued propagators inside the propagator identity~\eqref{contactfin}. The result can be manipulated as follows:
\bea\label{contactid}
&&B^{c}_n(\{E_i\}_{i=1}^n)-\frac{1}{2}\left[B^c_n(\{E_i\}_{i=1}^n)+(-1)^mB^{c}_n(\{-E_i\}_{i=1}^m,\{E_i\}_{i=m+1}^n))\right]\\ \nonumber
&&\qquad\qquad-\frac{1}{2}\left[B^c_n(\{E_i\}_{i=1}^n)+(-1)^{n-m}B^{c}_n(\{E_i\}_{i=1}^m,\{-E_i\}_{i=m+1}^n))\right]=0
\eea
We can then see that $B^{c}_n(\{E_i\}_{i=1}^n)$ cancels, and finally the we can invert the first $m$ energies, to obtain
\begin{tcolorbox}[colback=gray!10!white, colframe=white] 
\vspace{-1em}
\bea\label{contres}
B^{c}_n(\{E_i\}_{i=1}^n)+(-1)^n B^c_n(\{-E_i\}_{i=1}^n)=0.
\eea
\end{tcolorbox}

%%%%%%%%%%%%%%%%%%%%%%%%%%%%%%%%%%%%%%%%%%%%%%%%%%%%%%%%%%%%%%%%%%%

\paragraph{2-vertex cutting rules}\label{cuttings} 
For two insertions of an interaction Hamiltonian, the number of relations one can obtain from~\eqref{CFTopt} is in principle even larger. The rules we formulate here for the exchange focus on a particular channel, but the full rule then simply applies to all permutations.

If we insert $\mathcal{O}_1=\phi(t_0)^m$, $\mathcal{O}_2=\phi(t_0)^{n-m}$, $\mathcal{O}_3=\Hi^{(1)}(t)$, and $\mathcal{O}_4=\Hi^{(2)}(t')$ into~\eqref{CFTopt}, and denote the $s$ channel where the $\mathcal{O}_1$ are connected to the $\Hi^{(1)}$ and the remaining legs to $\Hi^{(2)}$, we obtain the rule for exchange diagrams at $L+1$-loops given in~\eqref{exfin}. Before inserting the relations found in~\eqref{rules fin e} into~\eqref{exfin}, note that using~\eqref{contres}, one can see that
\bea\label{cutiden}
B^{\text{c,cut}}_{n,L}(-\{E_i\}_{i=1}^n,\{y_i\}_{i=1}^{L})=(-1)^{n+1}B^{\text{c,cut}}_{n,L}(\{E_i\}_{i=1}^n,\{y_i\}_{i=1}^{L})
\eea
Now, again with~\eqref{rules fin e}, we can directly and generally relate to correlators with flipped energies. We denote the internal energies' dependence on the internal momenta, by $y_i=y_i(\vec{p_i})$. Then, the first three terms of~\eqref{exfin} combine, similarly to~\eqref{contactid}, to give
\bea
&&\frac{(-1)^{m+1}}{2}\left[B^{ex}_n(\{-E_i\}_{i=1}^m,\{E_i\}_{i=m+1}^n)+(-1)^{n}B^{ex}_n(\{E_i\}_{i=1}^m,\{-E_i\}_{i=m+1}^n)\right]=\\ \nonumber
&&\qquad=\int_{\vec{p}_1...\vec{p}_{L+1}}\frac{B^{\text{c,cut}}_{m,L+1}(\{E_i\}_{i=1}^m,\{y_i\}_{i=1}^{L+1}) B^{\text{c,cut}}_{n-m,L+1}(\{E_i\}_{i=m+1}^n,\{y_i\}_{i=1}^{L+1})}{\prod_{i=1}^{L+1}{P(y_i)}},
\eea
then flipping $\{E_i\}_{i=1}^m\rightarrow \{-E_i\}_{i=1}^m$ on both sides, and using~\eqref{cutiden} for the cut diagrams, we finally find 
\begin{tcolorbox}[colback=gray!10!white, colframe=white] 
\vspace{-1em}
\bea \label{finaleq}
&&\hspace{-2em}B^{\text{ex},s}_n(\{E_i\}_{i=1}^n)+(-1)^{n}B^{\text{ex},s}_n(\{-E_i\}_{i=1}^n)=\\ \nonumber
&&\qquad \qquad=2\int_{\vec{p}_1...\vec{p}_{L+1}}\hspace{-1em}\frac{B^{\text{c,cut}}_{m,L+1}(\{E_i\}_{i=1}^m,\{y_i\}_{i=1}^{L+1}) B^{\text{c,cut}}_{n-m,L+1}(\{E_i\}_{i=m+1}^n,\{y_i\}_{i=1}^{L+1})}{\prod_{i=1}^{L+1}P(y_i)}.
\eea
\end{tcolorbox}

For $L=0$ this simply reduces to a tree-level identity and the integrals can be ignored. This is a generalisation of the relation found in \cite{COT}, which we recover for $L=0, n=4$, and $m=2$.\\

We end this section with a number of comments. The cutting rules we obtained for the 1-vertex and 2-vertex diagrams here, by no means represent the full information one can obtain from~\eqref{CFTopt}. For instance in App.~\ref{app:C}, we derive the propagator identities for the 3-vertex case. Additionally, one could consider more general choices of operators in the largest time equation. However, the results we have presented in this work are not yet sufficient to relate these diagrams to correlators with shifted kinematics.

%%%%%%%%%%%%%%%%%%%%%%%%%%%%%%%%%%%%%%%%%%%%%%%%%%%%%%%%%%

\section{Scattering in de Sitter: a preview}\label{sec:Smatrix}

In this section\footnote{The results in this section were obtained in collaboration with Mang Hei Gordon Lee and more details will be presented in an upcoming paper.}, we present a brief discussion of scattering in de Sitter, which follows naturally from the in-out formalism we have introduced. The S-matrix we define is similar to an S-matrix that has recently been independently introduced by Melville and Pimentel in \cite{Melville:2023kgd}. We comment on the few differences and many similarities below. \\

Fig.~\ref{figextdS} suggests a natural way to define scattering in the (double) Poincar\'e patch of de Sitter: consider a state of $n$ free particles at the past null horizon, let them evolve in an interacting theory and project the resulting state on the tensor product of free particles at future null infinity. In the Schrodinger picture this looks very familiar:
\begin{align}
 S_{n,n'}=\bra{n',+\infty}\ket{n,-\infty}\,.
\end{align}
Following Wigner, a ``free particle" should correspond to an irreducible representation of the dS isometry group and is hence characterized by two quadratic Casimir operators, the conformal dimension $\Delta$ (related to the mass by~\eqref{Delta}) and spin $s$ (we use the notation of \cite{Hogervorst:2021uvp,Sun:2021thf,Penedones:2023uqc}), plus a set of eigenvalues for some conveniently chosen maximal abelian ideal (Cartan subalgebra). We choose to diagonalize the three (commuting) generators of spatial translations $P_i$ and denote single-particle states by $\ket{\Delta,\bfk,s,\sigma}$, with $\sigma$ the spin along some direction. These states are created by acting on the Bunch-Davies vacuum with creation operators, which in turn can be repackaged in the standard way into fields. We adopt a relativistic normalisation of the states. The in and out states of the S-matrix are then just tensor products of free particles,
\begin{align}
 \ket{n}=\bigotimes_a^n \ket{\Delta_a,\bfk_a,s_a,\sigma_a}\,.
\end{align}
To compute the S-matrix we work in perturbation theory in the interaction picture with a time evolution operator given by 
\begin{align}\label{Smat}
 S_{n,n'}&=\bra{n'}U(+\infty,-\infty)\ket{n}=\bra{n'}Te^{-i\int_{-\infty}^{+\infty}\Hi (\eta)d\eta}\ket{n}\,,
\end{align}
where again $\Hi $ is the interaction Hamiltonian written in terms of the interaction picture fields (free fields in the Heisenberg picture). This is the same in-out time-evolution operator\footnote{There is subtlety here. If we use the same $i\e$ prescription as for computing in-out correlators, namely~\eqref{inouteps}, we find a divergence because some incoming or outgoing particles have been inserted before or after all Hamiltonian interactions. In other words, since we are not projecting onto the vacuum of the free theory we should not use the $i\e$ rotation of the contour. Rather, we should adiabatically turn on interactions à la Gell-Mann and Low, namely with the shift $\Hi \to e^{-\e|\eta|}\Hi $ \cite{Gell-Mann:1951ooy}.} we defined in~\eqref{giodS} and hence we are naturally thinking of the ``extended" de Sitter spacetime in Fig.~\ref{figextdS}.
In the absence of IR divergences, this is a well-defined operator, as we discussed previously. The Feynman rules are pretty much the same as in Minkowski, except that we will use them in the time-momentum domain, as opposed to time-position or energy-momentum domains that are more familiar in Minkowski. The detailed prescription can be extracted from~\eqref{Smat}, but it is easier to notice that the S-matrix is related to the in-out correlators simply by replacing each off-shell external leg by an on-shell mode function evolving an incoming (outgoing) particle all the way to past (future) null infinity, as dictated by the LSZ projection. In our setup, this procedure works very similarly to Minkowski with the difference that one needs to use the dS mode function appropriate to the fields corresponding to the particle under consideration. This was nicely discussed in \cite{Melville:2023kgd} (see also \cite{Salcedo2022} for a similar projection of in-out correlators to the wavefunction). 

We define amplitudes $A$ by
\begin{align}\label{Adef}
 \bra{f}U(+\infty,-\infty)-1\ket{i}=i(2\pi)^4\delta^{(3)}(\ki-\ko)A_{if}\,,
\end{align}
for $\ket{i}$ and $\ket{f}$ some initial and final states respectively. Similarly to Minkowski, $A$ is proportional to the matrix element of $T=-i(S-1)$ but a crucial difference is that we strip off only the momentum-conserving Dirac delta and not the energy-conserving Dirac delta. The reason will be clear momentarily.\\

We are now in the position to compute the simplest scattering process of $n$ to $n'$ particles of mass $m^2=2H^2$, corresponding to a conformally coupled scalar, with the simple negative-frequency mode function \footnote{Note that we here drop the factor of $i$ for conformally coupled scalars from~\eqref{modefs}, for simplicity. This simply corresponds to a different choice of normalisation.}
\begin{align}
 f_{k}(\eta)=\eta \frac{H e^{-ik\eta}}{\sqrt{2k}}\,.
\end{align}
We choose a simple polynomial interaction $\Hi =\lambda\phi^{n+n'}/(n+n')!$ and restrict to $n+n'\geq 4$ to ensure the absence of IR divergences. The relativistic normalization of the states gives us
\begin{align}\label{contactA}
 \ket{\Delta,\bfk}=\sqrt{2|\bfk|} a^\dagger_\bfk\ket{0}\,.
\end{align}
Then to linear order in $\lambda$ the result is
\begin{align}\label{restreeamp}
 A_{nn'}=- \lambda \, (-iH \partial_{E_T})^{n+n'-4}\delta(E_T)\,,
\end{align}
where $E_T$ is the total energy accounting for the opposite sign of incoming and outgoing particles
\begin{align}
 E_T=- \sum_a^n |\bfk_a| + \sum_b^{n'} |\bfk_b|\,,
\end{align}
and we found the $(n+n'-4)$-th derivative of the Dirac delta. Notice that, at least in perturbation theory, the S-matrix enjoys crossing symmetry and so could simply choose all particles to be outgoing (this was also noticed in \cite{Melville:2023kgd}), so that $E_T$ is the usual positive sum of norms. \\

As a less trivial example, consider the exchange diagram mediating the elastic scattering of $3+r$ particles mediated by the interaction $\lambda \phi^{4+r}/(4+r)!$. This is IR finite as long as $r\geq 0$, hence the unusual definition of $r$. A direct calculation gives 
\begin{align}\label{scat}
 A_{3+r,3+r}&= \frac{\lambda^2 H^{2r}}{2} \sum_{l=0}^r b_l \frac{(\ksi-\Ei)^{1+r-l}+(\ksi+\Ei)^{1+r-l}}{2\ksi (-\Ei^2+\ksi^2)^{1+r-l}} \partial^{r+l}\delta(\Ei-\Eo) \,,
\end{align}
where
\begin{align}
 \Ei &= \sum_{a=1}^{3+r} |\bfk_a |\,, & \Eo&\equiv\sum_{a=4+r}^{6+2r} |\bfk_a |\,, \\
 \ki &=\sum_{a=1}^{3+r} \bfk_a & \ko & \equiv \sum_{a=4+r}^{6+2r} \bfk_a & b_l &\equiv \frac{r!}{l!} (-1)^{l} \,.
\end{align}
A useful check is that we should recover the tree-level single-exchange Minkowski amplitude for $r=0$, in which case the $\phi^4$ interaction is classically conformal. In this case, the sum disappears, the numerator cancels the factor $2\ksi$ at the denominator and we are left with the familiar $1/S$ with $S$ the Mandelstam variable $S=-\Ei^2+\ki^2$. The absence of a mass is what we would expect since a conformally coupled scalar is massless in Minkowski, where the Riemann tensor vanishes.

\paragraph{Unitarity and positivity} The amplitudes defined in~\eqref{Adef} satisfy the textbook generalized optical theorem 
\begin{align}
 A_{if}-A_{fi}^*=i\sum_X \int d\Pi_X\, (2\pi)^4 \delta^{(3)}(\ki-\bfk_X)A_{iX}A_{fX}^*\,,
\end{align}
where the sum is over all possible states and the only difference from Minkowski is that we have not removed any energy-conserving Dirac deltas, and so none needs to be added explicitly on the right-hand side. A powerful consequence of this theorem is that the right-hand side is manifestly positive for forward scattering and this constrains the imaginary part of $A_{ii}$ non-perturbatively. 

Let's check explicitly the optical theorem for the tree-level elastic scattering of four conformally coupled scalars in de Sitter with a polynomial interaction $\lambda \phi^5/(5!)$. We can copy the result from~\eqref{restreeamp} and the right-hand side is simply computed as 
\begin{align}
 \text{RHS}&=i \int \frac{dk_X^3}{(2\pi)^3} \frac{1}{2E_X}(2\pi)^4 \delta^{(3)}(\ki-\bfk_X)|A_{4,1}|^2\\
 &=2\pi i\lambda^2 H^2 \delta'(\Eo-\Ei)\frac{\delta'(\Ei-\ksi)}{2\Ei}\,. \label{RHS}
\end{align}
The left-hand side can be computed from~\eqref{scat} setting $r=1$. Two terms appear in the sum over $l$:
\begin{align} \text{LHS}&=\frac{\lambda^2 H^2}{2} 2i \Im \left[ -\frac{1} {(-\Ei^2+\ksi^2)} \delta''(\Ei-\Eo) + \frac{\ksi^2+\Ei^2}{\ksi (\Ei^2-\ksi^2)^2} \delta'(\Ei-\Eo)\right]
\end{align}
In the first term, corresponding to $l=1$, the Dirac delta needs to be integrated by parts. Hence the two terms can be combined and the numerator cancels the negative energy poles at $\Ei=-\ksi$ in the denominator, leaving only the positive-energy pole, at $\Ei= \ksi$, as expected for a physical and on-shell exchanged particle,
\begin{align}
 \text{LHS}&=i \frac{\lambda^2 H^2}{\ksi} \delta'(\Ei-\Eo) \Im \frac{1}{ (\Ei-\ksi+i\e)^2}\,.
\end{align}
One can now re-write $\Im(\Ei-\ksi+i\e)^{-2} = \partial_{\ksi}\Im(\Ei-\ksi+i\e)^{-1}$. Then, using the Sokhotski–Plemelj theorem, $\Im (x+i\e)^{-1}=-\pi \delta(x)$, we have $\partial_{\ksi}\Im(\Ei-\ksi+i\e)^{-1}=-\pi \delta'(\Ei-\ksi)$ to obtain precisely the same expression as the right-hand side in~\eqref{RHS}.

%%%%%%%%%%%%%%%%%%%%%%%%%%%%%%%%%%%%%%%%%%%%%%%%%%%%%%%%%%%%%%%%%%%%%%%%%%%%%%%%%%%%%%%%%%%%%%%%%%%%%%%%%%%

\section{Conclusions and outlook}\label{concl}

In this work, we have developed an in-out formalism that computes cosmological correlators. Our formalism is equivalent to the well-known in-in formalism for unitary, non-dissipative evolution and in the absence of IR divergences. The in-out formalism offers a welcome simplification of the Feynman rules and involves only the Feynman propagator. We have discussed a few applications of this formalism, such as the derivation of recursion relations for Minkowski correlators and cutting rules for de Sitter and Minkowski correlators.

Our results open many avenues for new exciting research:
\begin{itemize}
 \item Already from Fig.~\ref{figextdS} it is natural to define a scattering S-matrix in de Sitter by complete analogy with Minkowski. This S-matrix can be readily computed from the Feynman rules we have given for in-out correlators (with the usual amputation of external lines, which become on-shell mode functions). This S-matrix is very close to that recently introduced in \cite{Melville:2023kgd}, modulo some minor technical differences. The in-out S-matrix is interesting because it connects unitarity to positivity via the optical theorem, just like in Minkowski. We will present a thorough discussion in an upcoming paper \cite{upcoming}.
 \item Our formalism might be useful to make progress on understanding the renormalization of ultra-violet divergences in cosmological correlators (see \cite{Baidya:2017eho} for progress in this direction). Some of the outstanding issues include the apparent inequivalence of regularization procedures \cite{Weinberg:2005vy,Senatore:2009cf}, the characterization of counterterms and a systematic formulation of effective field theories in an expanding spacetime.
 \item The in-out formalism might simplify calculations and lead to new constraints in the cosmological bootstrap \cite{Baumann:2022jpr}, both for the de Sitter \cite{Arkani-Hamed:2018kmz} and the boostless case \cite{BBBB}.
 \item Our results might provide the key insight to understand the relation between the analytic structure of correlators and wavefunction coefficients. A study of the latter commenced in \cite{Arkani-Hamed:2017fdk} and was systematized and generalized recently in \cite{Salcedo2022}, where an identification and classification of singularities in Minkowski spacetime were presented. Later, in \cite{Agui-Salcedo:2023wlq}, it was proven that the branch cuts in the total energy appearing in the wavefunction always cancel in correlators (this fact had previously been noticed in a particular case in \cite{Lee:2023jby}). In \cite{Lee:2023kno}, it was shown that wavefunction singularities can be classified in "amplitude-like" singularities, which appear also in S-matrix elements, and "wavefunction-type" singularities that don't. There it was conjectured that only the amplitude-like singularities survive in correlators, while wavefunction-type singularities cancel out. The equivalence between in-in and in-out formalism seems to provide a rational for this phenomenon. 
 \item We have provided a formal argument for the equivalence of the in-in and in-out formalism, but we have presented explicit checks only at tree level. It would be important to make an explicit check at loop order. Moreover, our cutting rules for correlators might be useful to understand the peculiar behaviour of loop contributions to parity-odd correlators, recently computed in \cite{Lee:2023jby}. The surprising simplicity of the parity-odd loop correlators might be clarified by relating them to tree-level diagrams and leveraging the nice results derived in \cite{Stefanyszyn:2023qov}.
 \item As it is clear from our results, the number and complexity of the possible relation among correlators, such as the cutting rules, grow quickly for larger diagrams. It would be very interesting to see if a more abstract and probably geometric organising principle emerges, in analogy to the role that polytopes play in understanding wavefunction coefficients \cite{Arkani-Hamed:2017fdk,Arkani-Hamed:2018bjr,Benincasa:2019vqr,Benincasa:2019vqr,Albayrak:2023hie}. Perhaps the techniques developed in \cite{Anselmi:2016ane,Anselmi:2021hab} can help systematize our derivation of cutting rules.
\end{itemize}
Our main result, namely the derivation of an in-out formalism for cosmological correlators, is a technical one. However, it is not rarely the case that a different technical formulation of a problem leads to a new conceptual understanding or new unexplored connections. Posterity will judge.

%%%%%%%%%%%%%%%%%%%%%%%%%%%%%%%%%%%%%%%%%%%%%%%%%%%%%%%%%%%%%%%%%%%%%%%%%%%%%%%%%%%%%%%%%%%%%%%%%%%%%%%%%%%

\section*{Acknowledgements} 

We would like to thank Gordon Mang Hei Lee for many useful discussions and for collaboration on the results presented in Sec.~\ref{sec:Smatrix}. We are also thankful to Paolo Creminelli and Paolo Benincasa for useful discussions. E.P. has been supported in part by the research program VIDI with Project No. 680-47-535, which is (partly) financed by the Netherlands Organisation for Scientific Research (NWO). Y.D. acknowledges support
from the STFC. This work has been partially supported by STFC consolidated grant ST/T000694/1 and ST/X000664/1 and by the EPSRC New Horizon grant EP/V017268/1.

%%%%%%%%%%%%%%%%%%%%%%%%%%%%%%%%%%%%%%%%%%%%%%

\appendix

%%%%%%%%%%%%%%%%%%%%%%%%%%%%%%%%%%%%%%

\section{In-in equals in-out: the exchange diagram for general masses}\label{app:A}

After having established that for conformally coupled and massless scalars the in-in and in-out formalisms give the same result for an exchange diagram, we here establish this result for general masses. The derivation for the $rr$, $rl$ and $lr$ cases, hold for general masses, since they only rely on the equivalence for the contact diagram case, as shown in Sec.~\ref{checks}. Therefore, the only remaining part is the $ll$ case, which we discuss in the following. We start with the definitions of the in-in and in-out correlators for the exchange from Sec.~\ref{checks}, where we set couplings $\lambda_1 \lambda_2 = -1$ for simplicity. We have
\bea
\Gii^{ll}\hspace{-0.7em}&=&\hspace{-0.7em}\int_{-\infty(1+i\e)}^0\int_{-\infty(1+i\e)}^0 \frac{d\eta}{\eta^{4}} \frac{d\eta'}{\eta'^{4}}F_1 F_2\, \left[G^*_{F}(\eta,\eta';s)\prod_{a=1}^{n} G_{l}(\eta,\eta_{a};k_{a})\prod_{b=n+1}^{n+m} G_{l}(\eta',\eta_{b};k_{b})\right]\,,\nonumber \\ \nonumber
\Gio^{ll}\hspace{-0.7em}&=&\hspace{-0.7em} \int_{0}^{\infty(1-i\e)} \int_{0}^{\infty(1-i\e)} \frac{d\eta}{\eta^{4}}\frac{d\eta'}{\eta'^{4}}F_1 F_2\, \left[G_{F}(\eta,\eta';s)\prod_{a=1}^{n} G_{F}(\eta,\eta_{a};k_{a})\prod_{b=n+1}^{n+m} G_{F}(\eta',\eta_{b};k_{b})\right]\,. 
\eea

Let us first do some simplification on $\Gii^{ll}$. Note that from the contact diagram case, we know $\int_{-\infty(1+i\e)}^{+\infty(1-i\e)} \frac{d\eta'}{\eta'^{4}} F_1 \left[\prod_a G_l(\eta',\eta_a,k_a)\right]\rightarrow 0$. Therefore, for any intermediary $\eta$ we have
\bea
\int_{-\infty(1+i\e)}^{\eta} \frac{d\eta'}{\eta'^{4}} F_1 \left[\prod_a G_l(\eta',\eta_a,k_a)\right]=-\int_{\eta}^{+\infty(1-i\e)} \frac{d\eta'}{\eta'^{4}} F_1 \left[\prod_a G_l(\eta',\eta_a,k_a)\right].
\eea Let us expand the Feynman propagator in $\Gii^{ll}$, and then apply the identity above. We then have

\begin{align}
\Gii^{ll}
&=\int_{-\infty(1+i\e)}^{0} \frac{d\eta}{\eta^{4}} \int_{-\infty(1+i\e)}^{\eta} \frac{d\eta'}{\eta'^{4}} F_1 F_2\left[G_{l}(\eta',\eta;s)\prod_{a=1}^{n} G_{l}(\eta,\eta_{a};k_{a})\prod_{b=n+1}^{n+m} G_{l}(\eta',\eta_{b};k_{b})\right] \,, \nonumber \\ \nonumber
&\quad+ \int_{-\infty(1+i\e)}^{0} \frac{d\eta'}{\eta'^{4}} \int_{-\infty(1+i\e)}^{\eta'} \frac{d\eta}{\eta^{4}} F_1 F_2\left[G_{l}(\eta,\eta';s)\prod_{a=1}^{n} G_{l}(\eta,\eta_{a};k_{a})\prod_{b=n+1}^{n+m} G_{l}(\eta',\eta_{b};k_{b})\right] \,, \\ \nonumber
&=-\int_{-\infty(1+i\e)}^{0} \frac{d\eta}{\eta^{4}} \int_\eta^{\infty(1-i\e)} \frac{d\eta'}{\eta'^{4}} F_1 F_2\left[G_{l}(\eta',\eta;s)\prod_{a=1}^{n} G_{l}(\eta,\eta_{a};k_{a})\prod_{b=n+1}^{n+m} G_{l}(\eta',\eta_{b};k_{b})\right] \,, \\ \nonumber
&\quad-\int_{-\infty(1+i\e)}^{0} \frac{d\eta'}{\eta'^{4}} \int_{\eta'}^{\infty(1-i\e)} \frac{d\eta}{\eta^{4}} F_1 F_2\left[G_{l}(\eta,\eta';s)\prod_{a=1}^{n} G_{l}(\eta,\eta_{a};k_{a})\prod_{b=n+1}^{n+m} G_{l}(\eta',\eta_{b};k_{b})\right]\,. \end{align}
Now we can also expand $\Gio^{ll}$ and subtract $\Gii^{ll}$ from it, to see that they combine in a similar way as in the contact diagram case
\begin{align}
\Gio^{ll}-\Gii^{ll}&=\int_{-\infty(1+i\e)}^{\infty(1-i\e)}\frac{d\eta}{\eta^{4}} \int^{\infty(1-i\e)}_{\eta} \frac{d\eta'}{\eta'^{4}}F_1 F_2\, \left[G_{l}(\eta,\eta';s)\prod_{a=1}^{n} G_{l}(\eta,\eta_{a};k_{a})\prod_{b=n+1}^{n+m} G_{l}(\eta',\eta_{b};k_{b})\right]\,, \nonumber \\ \nonumber
&+\int_{-\infty(1+i\e)}^{\infty(1-i\e)}\frac{d\eta'}{\eta'^{4}} \int^{\infty(1-i\e)}_{\eta'} \frac{d\eta}{\eta^{4}}F_1 F_2\, \left[ G_{l}(\eta,\eta';s)\prod_{a=1}^{n} G_{l}(\eta,\eta_{a};k_{a})\prod_{b=n+1}^{n+m} G_{l}(\eta',\eta_{b};k_{b})\right]\,.
\end{align}

In the following, we will show that these terms vanish separately, and the derivation is the same for both terms. For concreteness, we focus on the second term. Furthermore, let us for now restrict $F_1$ and $F_2$ only to contain spatial derivatives. This simply leads to an overall factor, which we can drop. Furthermore, since we assume the integral is IR-finite, we simply need to show that the class of integrals 
\bea\label{intclass}
\Delta B:=\int_{-\infty(1+i\e)}^{\infty(1-i\e)}d\eta' \int^{\infty(1-i\e)}_{\eta'} d\eta \,\eta^{p_1}\eta'^{p_2}f_s(\eta)f_s(\eta')^*\prod_{a=1}^{n} f_{k_a}(\eta)\prod_{b=n+1}^{n+m} f_{k_b}(\eta'),
\eea
with $p_1,p_2\geq 0$, vanishes, where we dropped the overall factor of the external mode functions.

\begin{figure}[htb!]
 \centering
 \includegraphics[width=0.5 \textwidth]{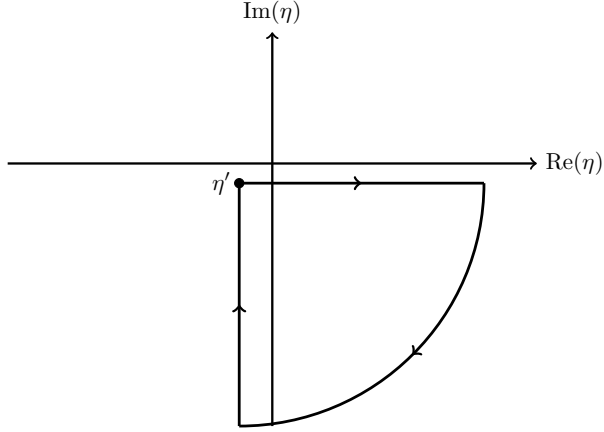}
 \caption{Choice of contour in the $\eta$ plane}
 \label{complex fig}
\end{figure}

Let us first focus on the inner integral over $\eta$. We note that $\eta'$ always has a small negative imaginary part, and therefore this integral is always in the lower half complex plane where $f_k(\eta)$ is analytic, and convergent at infinity. This means we can draw a contour along the $\eta'=const$ axis, up to $-i \infty$, and close the contour at infinity, to get a different representation of this integral. We show the contour in Fig~\ref{complex fig}
 . The contribution at infinity is zero, given the exponential convergence of $H_{\nu}^{(1)}$ in the lower half complex plane, and the inner integral can be written as
 \begin{align} \int^{\eta'}_{\eta'-i \infty} d\eta \eta^{p_1} f_{s}(\eta)\prod_{a=1}^n f_{k_a}(\eta)=i \int^{\infty}_{0} d\eta(\eta' - i \eta)^{p_1}f_{s}(\eta' - i \eta) \prod_{a=1}^n f_{k_a}(\eta' - i \eta) \nonumber\,,\end{align}
 where we substituted $\eta\rightarrow\eta' - i \eta$.
 
 Now let us get back to the main integral, where we can now change the order of integration. Furthermore, expanding out the the factor $(\eta' - i \eta)^{p_1}$, the integral again falls into the classes of integrals 
\begin{align}\label{key}
\Delta B =
-i \int^{\infty}_{0}d\eta\int_{-\infty(1+i\e)}^{\infty(1-i\e)}d\eta' \eta^{p_3}\eta'^{p_4}f_{s}(\eta' - i \eta)f_{s}(\eta')^* \prod_{a=1}^n f_{k_a}(\eta' - i \eta)\prod_{b=n+1}^{n+m} f_{k_b}(\eta')\, .
\end{align}
for some $p_3,p_4\geq0$. Now note that since $\eta>0$ and since $\eta'$ has a small negative imaginary part, the mode functions are always evaluated in the lower half complex plane. Therefore, the integrand in $\eta'$ is analytical over the whole region of integration. We, therefore, simply need to show that on the arc at negative infinity in the lower half complex plane, the integral vanishes.

Let us label $\sum_{a=1}^{n+m} k_a = k_T$ , then in the limit $\eta' \rightarrow -i \infty$ we have 
\bea
\eta^{p_3}\eta'^{p_4}f_{s}(\eta' - i \eta)f_{s}(\eta')^* \prod_{a=1}^n f_{k_a}(\eta' - i \eta)\prod_{b=n+1}^{n+m} f_{k_b}(\eta')\rightarrow \eta'^{p_3+1}\eta^{p_4+1} e^{-i k_T \eta'}e^{- (k_T+s) \eta},
\eea
where we see that the integral over $\eta'$ goes to zero, and that the integral over $\eta$ is finite. Therefore we conclude that the integral~\eqref{key} vanishes.
Finally, we note a couple of generalisations. The derivation for the term that comes from the Feynman propagator with $\eta \leftrightarrow \eta'$, is the same, simply with the consistent exchange $\eta \leftrightarrow \eta'$. Furthermore, in the presence of temporal derivatives, the derivation follows analogously, since time derivatives of the Hankel functions is a sum of Hankel functions with shifted mass. Therefore, since the formulas above only depend on the asymptotic behaviour and the analyticity of the mode functions, the resulting integrals fall under the same class of integrals as in~\eqref{intclass}, which vanishes.

%%%%%%%%%%%%%%%%%%%%%%%%%%%%%%%%%%%%%%
\section{More pole bagging in dS} \label{app:PoledS}
As started in~\eqref{ccgeneral}, we want to solve the integral
\bea\label{basecc}
B_n^{cc}&=& -(-i\eta_0)^{n} H^{2n-4}\left(\frac{ 1}{2 \pi i}\right)^{n-1} \int_{-\infty}^{\infty}\partial_{\omega_n}^{n-4}\left(\left(\prod_{i=1}^n \frac{d\omega_i}{\omega_i^2 -k_i^2+i \epsilon}\right)e^{i \omega_T\eta_0}\right)\delta(\omega_T). \eea

The derivatives can be fully done using Leibniz rule and then transferred to derivatives in $k_n$. We have
\bea
&&\partial_{\omega_n}^{k}\left(\left( \frac{1}{\omega_n^2 -k_n^2+i \epsilon}\right)e^{i \omega_T\eta_0}\right)=\sum_{l=0}^k{k\choose l}(i \eta_0)^{k-l}e^{i \omega_T\eta_0}\partial_{\omega_n}^{k-l}\left(\frac{1}{\omega_n^2 -k_n^2+i \epsilon}\right)\\ \nonumber
&&\qquad=-\sum_{l=0}^k{k\choose l}(i \eta_0)^{k-l}e^{i \omega_T\eta_0}\frac{1}{2 k_n}\partial_{k_n}^l\left(\frac{(-1)^l}{(\omega_n+k_n-i \epsilon)}-\frac{1}{(\omega_n-k_n+i \epsilon)}\right).
\eea
Having done all the derivatives, this is just a flat space correlator, however with the subtlety that we have some non-square terms in the denominator, such as $\omega_n\pm(k_n-i \epsilon)$. However, we can solve this by seeing that
\bea
\pm k_n B_n^{\text{flat}}&=&\left(\frac{1}{2 \pi i}\right)^{n-1} \int_{-\infty}^{\infty}\frac{d \omega_n}{\omega_n\pm(k_n-i \epsilon))}\left(\prod_{i=1}^{n-1} \frac{d\omega_i}{\omega_i^2 -k_i^2+i \epsilon}\right)e^{i \omega_T\eta_0}\delta(\omega_T). \\ \nonumber
\eea
Then putting this back into \ref{basecc}, we get 
\bea\label{flatder}
B_n^{cc}&=&(\eta_0 H)^{2n-4} \sum_{l=0}^{n-4}{n-4\choose l}(i \eta_0)^{-l} \left((-1)^l+1\right) \frac{1}{2 k_n}\partial_{k_n}^l(k_n B_n^{\text{flat}}). \eea

While it is nice to know, that we can always write conformally coupled scalars as derivatives of flat space ones, since we even know what the flat space answer is, we can fully write the conformally coupled solution. Shifting the index in the sum, and doing some simplification, we have
\bea\label{flatder}
B_n^{cc}&=& H^{2n-4}(-i\eta_0)^{n}k_T^{4-n}\frac{(n-4)!}{\prod_{i=1}^{n} 2k_i}\frac{1}{k_T} \sum_{l=0}^{n-4}\frac{(i \eta_0k_T)^l}{l!} \left((-1)^{n-l}+1\right). \eea
Interestingly, in the $n\rightarrow \infty$ limit, this is giving us an exponential. Therefore the final answer can be written as a Taylor expansion to a certain order in $\eta_0$ given by
\bea
B_{n}^{cc}&=&\frac{(n-4)! H^{n-4}(-H\eta_0)^{n}}{\left(\prod_{i=1}^{n}2 k_i\right)k_T^{n-3}}2\text{Re}\left[i^n e^{i \eta_0 k_T}\bigg|_{n-4}\right].
\eea
Then shifting $n\rightarrow n+4$ gives us the compact formula from the main text~\eqref{nicecc}.
%%%%%%%%%%%%%%%%%%%%%%%%%%%%%%%%%%%%%%

\section{Diagrammatic cutting rules for three vertices} \label{app:C}

In this appendix, we give details for the derivation of propagator identities for three-vertex diagrams. We do not show how to use hermitian analyticity to turn these relations into cutting rules for correlators with flipped energies. While our rules generalize to any number of external legs and loops, we focus on diagrams at tree level where all the external lines attached to a single vertex have been combined into a single one (this can always be done in Minkowksi and for conformally coupled scalars in de Sitter).

For three vertices we get multiple propagator identities, depending on the choice of operators. For the cases we study here, we always take two of the three external vertices to form \textit{together} one of the operators in the largest time equation. Then we consider a double exchange tree-level diagram with the following contractions:
\begin{table}[h!]
 \centering
 \begin{tabular}{ccccc}
 $\mathcal{O}_1=\phi(t_0)^{n_1}$ & & $\mathcal{O}_2 =\phi(t_0)^{n_2}$ & & $\mathcal{O}_3 =\phi(t_0)^{n_3}$ \\
 $\updownarrow$ & & $\updownarrow$ & & $\updownarrow$ \\
 $\mathcal{O}_4=\Hi^{(1)}(t)$ & $\leftrightarrow$ & $\mathcal{O}_5=\Hi^{(2)}(t')$ & $\leftrightarrow$ & $\mathcal{O}_6=\Hi^{(3)}(t'')$
 \end{tabular}
\end{table}
where $n_1+n_2+n_3=n$. Then we also consider a one-loop diagram with three vertices and the same contractions of operators as above with the addition of a contraction connecting $\mathcal{O}_4$ and $\mathcal{O}_6$ to make a loop. The presence of a loop will make a difference in the end result. Let us now proceed to derive the cutting rules simultaneously for both cases. 

As mentioned above, we get several different rules, depending on our choice of operators. We start by combining the left two vertices and take $\mathcal{O}_1=\phi(t_0)^{n_1+n_2}$ and the rest remains as is. To not have to go over the same parity argument as in the one and two vertex case again, we directly only write half the terms. We have
\bea\label{3vcut1}
0&\sim&\langle T [\phi(t_0)^n \Hi^{(1)}(t)\Hi^{(2)}(t')\Hi^{(3)}(t'')]\rangle\\ \nonumber
&-&\langle \phi(t_0)^{n_1+n_2}T [\phi(t_0)^{n_3} \Hi^{(1)}(t)\Hi^{(2)}(t')\Hi^{(3)}(t'')]\rangle \\ \nonumber
&+&\langle \bar{T} [\phi(t_0)^{n_1+n_2}\Hi^{(1)}(t)]T[\phi(t_0)^{n_3}\Hi^{(2)}(t')\Hi^{(3)}(t'')]\rangle\\ \nonumber
&+&\langle \bar{T} [\phi(t_0)^{n_1+n_2}\Hi^{(1)}(t)\Hi^{(2)}(t')\Hi^{(3)}(t'')]\phi(t_0)^{n_3}\rangle\\ \nonumber
&-&\langle \bar{T} [\phi(t_0)^{n_1+n_2} \Hi^{(1)}(t)\Hi^{(2)}(t')]T[\phi(t_0)^{n_3}\Hi^{(3)}(t'')]\rangle\\ \nonumber
&+&\langle \bar{T} [\phi(t_0)^{n_1+n_2} \Hi^{(2)}(t')]T[\phi(t_0)^{n_3}\Hi^{(1)}(t)\Hi^{(3)}(t'')],\rangle\\ \nonumber
\eea
Again the $\sim 0$ indicates that the identity is valid only after integrating over the time insertion of the Hamiltonian interactions. Note that for the double exchange the last term is zero, because $\Hi^{(1)}$ does not have a time ordering relative to $t_0$. Finally~\eqref{3vcut1} results in the following rules in terms of diagrams

\begin{equation}
0
=
\hspace{-0.8em}\resizebox{0.9\hsize}{!}{
\raisebox{-3ex}{\begin{tikzpicture}
 \draw (-0.1, 0) -- (2.2, 0);

 \foreach \x in {0.15, 1.05, 1.95} {
 \draw (\x, 0) -- (\x, -0.9);
 \filldraw (\x, 0) circle (1.5pt);
 \filldraw (\x, -0.9) circle (1.5pt);
 }

 \draw (0.15, -0.9) -- (1.05, -0.9);
 \draw (1.05, -0.9) -- (1.95, -0.9);
\end{tikzpicture}}
\hspace{-0.8em}
-
\hspace{-0.8em}
\raisebox{-3ex}{\begin{tikzpicture}
 \draw (-0.1, 0) -- (2.2, 0);

 \foreach \x in {0.15, 1.05, 1.95} {
 \draw (\x, 0) -- (\x, -0.9);
 \filldraw (\x, 0) circle (1.5pt);
 \filldraw (\x, -0.9) circle (1.5pt);
 }

 \draw (0.15, -0.9) -- (1.05, -0.9);
 \draw (1.05, -0.9) -- (1.95, -0.9);

 \coordinate (C) at (1.6,0.1);  
 \coordinate (D) at (-0.3,-0.9);  

 \draw[black, line width=1pt] (C) -- (D);

\end{tikzpicture}}
\hspace{-0.8em}
+
\hspace{-0.8em}
\raisebox{-3.9ex}{\begin{tikzpicture}
 \draw (-0.1, 0) -- (2.2, 0);

 \foreach \x in {0.15, 1.05, 1.95} {
 \draw (\x, 0) -- (\x, -0.9);
 \filldraw (\x, 0) circle (1.5pt);
 \filldraw (\x, -0.9) circle (1.5pt);
 }

 \draw (0.15, -0.9) -- (1.05, -0.9);
 \draw (1.05, -0.9) -- (1.95, -0.9);

 \coordinate (C) at (1.6,0.2);  
 \coordinate (D) at (0.6,-1.1);  

 \draw[black, line width=1pt] (C) -- (D);

\end{tikzpicture}}
\hspace{-0.8em}
+
\hspace{-0.8em}
\raisebox{-3ex}{\begin{tikzpicture}
 \draw (-0.1, 0) -- (2.2, 0);

 \foreach \x in {0.15, 1.05, 1.95} {
 \draw (\x, 0) -- (\x, -0.9);
 \filldraw (\x, 0) circle (1.5pt);
 \filldraw (\x, -0.9) circle (1.5pt);
 }

 \draw (0.15, -0.9) -- (1.05, -0.9);
 \draw (1.05, -0.9) -- (1.95, -0.9);

 \coordinate (A) at (2.35,-0.9);
 \coordinate (B) at (1.5,0.2);

 \draw[black, line width=1pt] (A) -- (B);
 
\end{tikzpicture}}
\hspace{-0.8em}
-
\hspace{-0.8em}
\raisebox{-3.9ex}{\begin{tikzpicture}
 \draw (-0.1, 0) -- (2.2, 0);

 \foreach \x in {0.15, 1.05, 1.95} {
 \draw (\x, 0) -- (\x, -0.9);
 \filldraw (\x, 0) circle (1.5pt);
 \filldraw (\x, -0.9) circle (1.5pt);
 }

 \draw (0.15, -0.9) -- (1.05, -0.9);
 \draw (1.05, -0.9) -- (1.95, -0.9);

 \coordinate (A) at (1.45,-1.1);
 \coordinate (B) at (1.45,0.2);

 \draw[black, line width=1pt] (A) -- (B);

\end{tikzpicture}}}
\end{equation}
and for the box-Loop
\begin{equation}
0=
\resizebox{0.9\hsize}{!}{
\raisebox{-3ex}{\begin{tikzpicture}
 \draw (0.0, 0) -- (1.4, 0);

 \foreach \x in {0.2, 1.2} {
 \draw (\x, 0) -- (\x, -1);
 \filldraw (\x, 0) circle (1.5pt);
 \filldraw (\x, -1) circle (1.5pt);
 }
 \filldraw (0.7, 0) circle (1.5pt);
 \filldraw (0.7, -0.5) circle (1.5pt);
 \draw (0.2, -1) -- (1.2, -1);
 \draw (0.7, -0.5) -- (0.7, 0);
 \draw (0.7, -0.5) -- (1.2, -1);
 \draw (0.7, -0.5) -- (0.2, -1);

\end{tikzpicture}}
-
\raisebox{-3ex}{\begin{tikzpicture}
 \draw (0.0, 0) -- (1.4, 0);

 \foreach \x in {0.2, 1.2} {
 \draw (\x, 0) -- (\x, -1);
 \filldraw (\x, 0) circle (1.5pt);
 \filldraw (\x, -1) circle (1.5pt);
 }
 \filldraw (0.7, 0) circle (1.5pt);
 \filldraw (0.7, -0.5) circle (1.5pt);
 \draw (0.2, -1) -- (1.2, -1);
 \draw (0.7, -0.5) -- (0.7, 0);
 \draw (0.7, -0.5) -- (1.2, -1);
 \draw (0.7, -0.5) -- (0.2, -1);
 
 \coordinate (A) at (0,-0.9);
 \coordinate (B) at (1.05,0.1);

 \draw[black, line width=1pt] (A) -- (B);

 \end{tikzpicture}}
+
\raisebox{-3.3ex}{\begin{tikzpicture}
 \draw (0.0, 0) -- (1.4, 0);

 \foreach \x in {0.2, 1.2} {
 \draw (\x, 0) -- (\x, -1);
 \filldraw (\x, 0) circle (1.5pt);
 \filldraw (\x, -1) circle (1.5pt);
 }
 \filldraw (0.7, 0) circle (1.5pt);
 \filldraw (0.7, -0.5) circle (1.5pt);
 \draw (0.2, -1) -- (1.2, -1);
 \draw (0.7, -0.5) -- (0.7, 0);
 \draw (0.7, -0.5) -- (1.2, -1);
 \draw (0.7, -0.5) -- (0.2, -1);
 
 \coordinate (A) at (0.5,-0.5);
 \coordinate (B) at (1.05,0.2);

 \draw[black, line width=1pt] (A) -- (B);

 \coordinate (AA) at (0.5,-1.1);
 \coordinate (BB) at (0.5,-0.5);

 \draw[black, line width=1pt] (AA) -- (BB);

\end{tikzpicture}}
+
\raisebox{-3ex}{\begin{tikzpicture}
 \draw (0.0, 0) -- (1.4, 0);

 \foreach \x in {0.2, 1.2} {
 \draw (\x, 0) -- (\x, -1);
 \filldraw (\x, 0) circle (1.5pt);
 \filldraw (\x, -1) circle (1.5pt);
 }
 \filldraw (0.7, 0) circle (1.5pt);
 \filldraw (0.7, -0.5) circle (1.5pt);
 \draw (0.2, -1) -- (1.2, -1);
 \draw (0.7, -0.5) -- (0.7, 0);
 \draw (0.7, -0.5) -- (1.2, -1);
 \draw (0.7, -0.5) -- (0.2, -1);

 \coordinate (CC) at (0.9,0.2);  
 \coordinate (DD) at (1.4,-0.9);  

 \draw[black, line width=1pt] (CC) -- (DD);

\end{tikzpicture}}
-
\raisebox{-3.3ex}{\begin{tikzpicture}
 \draw (0.0, 0) -- (1.4, 0);

 \foreach \x in {0.2, 1.2} {
 \draw (\x, 0) -- (\x, -1);
 \filldraw (\x, 0) circle (1.5pt);
 \filldraw (\x, -1) circle (1.5pt);
 }
 \filldraw (0.7, 0) circle (1.5pt);
 \filldraw (0.7, -0.5) circle (1.5pt);
 \draw (0.2, -1) -- (1.2, -1);
 \draw (0.7, -0.5) -- (0.7, 0);
 \draw (0.7, -0.5) -- (1.2, -1);
 \draw (0.7, -0.5) -- (0.2, -1);

 \coordinate (CC) at (0.95,0.2);  
 \coordinate (DD) at (0.95,-1.1);  

 \draw[black, line width=1pt] (CC) -- (DD);

\end{tikzpicture}}
+\raisebox{-3ex}{\begin{tikzpicture}
 \draw (0.0, 0) -- (1.4, 0);

 \foreach \x in {0.2, 1.2} {
 \draw (\x, 0) -- (\x, -1);
 \filldraw (\x, 0) circle (1.5pt);
 \filldraw (\x, -1) circle (1.5pt);
 }
 \filldraw (0.7, 0) circle (1.5pt);
 \filldraw (0.7, -0.5) circle (1.5pt);
 \draw (0.2, -1) -- (1.2, -1);
 \draw (0.7, -0.5) -- (0.7, 0);
 \draw (0.7, -0.5) -- (1.2, -1);
 \draw (0.7, -0.5) -- (0.2, -1);
 
 \coordinate (A) at (0.9,-0.8);

 \coordinate (D) at (-0.1,-0.8);  

 \draw[black, line width= 1pt] (D) -- (A);

 \coordinate (CC) at (0.9,0.2);  
 \coordinate (DD) at (0.9,-0.85);  

 \draw[black, line width=1pt] (CC) -- (DD);

\end{tikzpicture}}

}.
\end{equation}

On a diagrammatic level, if we change $n_1 \leftrightarrow n_3$ we simply get the same cutting rules, just mirrored along the middle axis. There are separate double-cutting rules if we consider the change $n_2 \leftrightarrow n_3$, which can be derived as well, but we omit writing them here explicitly.

%%%%%%%%%%%%%%%%%%%%%%%%%%%%%%%%%%%%%%%%%%%%%%%%%%%%%%%%%%%%%%%%%

\section{Relating cut diagrams to diagrams with flipped energies}\label{app:D}

In this appendix, we derive the relation of a cut diagram to a correlator with shifted kinematics. These are~\eqref{rules fin c},~\eqref{rules fin e} and~\eqref{rules fin e2}. We largely focus on conformally coupled and massless fields in de Sitter and in the end we comment on general masses in flat space. Notice that the derivations are largely the same. To streamline our notation we set the coupling constant $\lambda$ and $H$ to one since they can be easily reinstated. 

\paragraph{Conformally coupled and massless scalars in dS} The key ingredient to this derivation is that the mode functions enjoy the nice property
\bea\label{base}
f_k(\eta)=f_k(-\eta)^*,
\eea
which in particular also translates to all time derivatives
\bea
a(\eta)^n\partial_{\eta}^n f_k(\eta)=(a(-\eta)^n\partial_{-\eta}^n f_k(-\eta))^*.
\eea
Because time derivatives enjoy the same property as the mode function, henceforth we focus on interactions without time derivatives. As in the main text, we assume that all interactions are IR-finite. Using~\eqref{base} we find that the propagators obey
\bea
G^+(-\eta,-\eta',k)=G^+(\eta,\eta',k)^* ,\quad \text{and} \quad G_F(-\eta,-\eta',k)=G_F(\eta,\eta',k).
\eea
Finally, we note that our cutting rules are valid to leading order in $\eta_0 \to 0$, where $\eta_0$ is the time when operators are inserted. The order in $\eta_0$ in which we are interested is even for diagrams that are even under spatial parity, and odd for odd diagrams.

Let us start with deriving the result in~\eqref{rules fin c}. We consider a general scale invariant Hamiltonian interaction of the form\footnote{In order for the interaction Hamiltonian to be IR-finite, we eventually need additional factors of $\eta^{2n}$, coming from time derivatives. Crucially, however, this factor is invariant under $\eta\rightarrow-\eta$.}
\begin{align}
 \Hi =\int_\bfx \frac{1}{(n+2L)!} \,F(\partial_i) \eta^{n_d} \phi^{n+2L},
\end{align}
where $n_d$ is the number of spatial derivatives. In the following, we shorten the notation for $F(\partial_i)$, simply to $F$, since in Fourier space it is just a multiplicative factor. Now let us write out the cut diagram from~\eqref{rules fin c}, where we leave the integrals over the loop momenta implicit:
\bea
\int_{-\infty}^\infty \frac{d\eta}{\eta^4}\, F \eta^{n_d} \left(\prod_{i=1}^m G^+(\eta_0,\eta,k_i)\right)\left(\prod_{i=m+1}^n G_F(\eta_0,\eta,k_i)\right)\left(\prod_{i=1}^{L} G_{F}(\eta,\eta,y_i)\right).
\eea

Next, we transform $\eta\rightarrow-\eta$. For brevity of notation, we will not write out the loop propagators because, being invariant under time reversal, they are just spectators in the derivation. We have
\bea
&&\int_{-\infty}^\infty \frac{d\eta}{\eta^4}\, F \eta^{n_d} (-1)^{n_d}\left(\prod_{i=1}^m G^+(\eta_0,-\eta,k_i)\right)\left(\prod_{i=m+1}^n G_F(\eta_0,-\eta,k_i)\right) \\ \nonumber
&&=\int_{-\infty}^\infty \frac{d\eta}{\eta^4}\, F \eta^{n_d} (-1)^{n_d}\left(\prod_{i=1}^m G^+(-\eta_0,\eta,k_i)^*\right)\left(\prod_{i=m+1}^n G_F(-\eta_0,\eta,k_i)\right) \\ \nonumber
&&=\int_{-\infty}^\infty \frac{d\eta}{\eta^4}\, F \eta^{n_d} \left(\prod_{i=1}^m G^+(\eta_0,\eta,k_i)^*\right)\left(\prod_{i=m+1}^n G_F(\eta_0,\eta,k_i)\right), \\ \nonumber
\eea
where from the first to the second line we used the propagator identities under time reversal, and from the second to the last line we used that at the order in $\eta_0$ we are interested in, the correlator is even (odd) in $\eta_0$ exactly when we have an even (odd) number of spatial derivatives. We have therefore shown that to leading order we can use $G^+$ or $(G^+)^*$ for the cut legs interchangeably. If we now average these two contributions, we can write the cut diagram as
\bea
&&\frac{1}{2}\int_{-\infty}^\infty \frac{d\eta}{\eta^4}\, F \eta^{n_d}\left(\prod_{i=1}^m G^+(\eta_0,\eta,k_i)+\prod_{i=1}^m G^+(\eta_0,\eta,k_i)^*\right)\prod_{i=m+1}^n G_F(\eta_0,\eta,k_i) \\ \nonumber
&&=\frac{1}{2}\int_{-\infty}^\infty \frac{d\eta}{\eta^4}\, F \eta^{n_d}\left(\prod_{i=1}^m G_F(\eta_0,\eta,k_i)+\prod_{i=1}^m G_F(\eta_0,\eta,k_i)^*\right)\prod_{i=m+1}^n G_F(\eta_0,\eta,k_i) \nonumber \\ \nonumber
&&=\frac{1}{2}\int_{-\infty}^\infty \frac{d\eta}{\eta^4}\, F \eta^{n_d}\left(\prod_{i=1}^m G_F(\eta_0,\eta,k_i)+(-1)^m\prod_{i=1}^m G_F(\eta_0,\eta,-k_i)\right)\prod_{i=m+1}^n G_F(\eta_0,\eta,k_i)\nonumber \\ \nonumber
&&=\frac{1}{2}\left[B^c_n(\{k_i\}_{i=1}^n)+(-1)^mB^{c}_n(\{-k_i\}_{i=1}^m,\{k_i\}_{i=m+1}^n))\right],
\eea
where from the first to the second line we used the propagator identity that $G^+ +(G^+)^*=G_F+G_F^*$, which generalises to products, and from the third to the fourth line we used the Hermitian analyticity property. This concludes the derivation for the contact diagram case~\eqref{contactfin}. Next, we can do a very similar derivation for the exchange diagram. \\ 

Again, we do not explicitly write out any temporal derivatives, since any time derivative will not alter the time-reversal properties of the mode functions. We therefore consider the two Hamiltonian interactions
\begin{align}\label{Hamilex}
 \Hi =\int_\bfx \frac{1}{(m+L+1)!} \,F_1(\partial_i) \eta^{n_{d_1}}\phi^{m+L+1}+\frac{1}{(n-m+L+1)!} \,F_2(\partial_i) \eta^{n_{d_2}}\phi^{n-m+L+1}.
\end{align}
The spatial derivatives again only act as an external factor and we can omit writing them out explicitly. Next, we want to write the cut exchange diagram in~\eqref{exfin} in terms of propagators. Similarly to the contact case, we leave the loop integrals implicit, the cut diagram reads
\bea\nonumber
\int_{-\infty}^\infty \frac{d\eta}{\eta^4}\frac{d\eta'}{\eta'^4}\, F_1 F_2 \eta^{n_{d_1}}\eta'^{n_{d_2}} \prod_{i=1}^{m} G^+(\eta_0,\eta,k_i) \prod_{i=m+1}^{n}G_F(\eta_0,\eta',k_{i}) \prod_{i=1}^{L+1} G_F(\eta,\eta',y_{i}).
\eea 
Now let us flip both $\eta$ and $\eta'$ and directly apply the propagator identities,
\bea
&& \hspace{-1em}\int_{-\infty}^\infty \frac{d\eta}{\eta^4}\frac{d\eta'}{\eta'^4}\, F_1 F_2 (-1)^{n_{d_1}+n_{d_2}} \eta^{n_{d_1}}\eta'^{n_{d_2}} \prod_{i=1}^{m} G^+(-\eta_0,\eta,k_i)^* \prod_{i=m+1}^{n}G_F(-\eta_0,\eta',k_{i}) \prod_{i=1}^{L+1} G_F(\eta,\eta',y_{i})\nonumber \\ \nonumber
&&=\int_{-\infty}^\infty \frac{d\eta}{\eta^4}\frac{d\eta'}{\eta'^4}\, F_1 F_2 \eta^{n_{d_1}}\eta'^{n_{d_2}} \prod_{i=1}^{m} G^+(\eta_0,\eta,k_i)^* \prod_{i=m+1}^{n}G_F(\eta_0,\eta',k_{i}) \prod_{i=1}^{L+1} G_F(\eta,\eta',y_{i}),
\eea
where again we used that the exchange diagram at leading order in $\eta_0$ is even (odd) if the total number of derivatives is even (odd). Therefore using the same procedure of averaging the $G^+$ and $(G^+)^*$ terms, using propagator identities and Hermitian analyticity we get
\bea
&&\hspace{-1em}=\int_{-\infty}^\infty \frac{d\eta}{\eta^4}\frac{d\eta'}{\eta'^4}\, F_1 F_2 \eta^{n_{d_1}}\eta'^{n_{d_2}}\times\\ \nonumber
&&\qquad \times \left( \prod_{i=1}^{m} G^+(\eta_0,\eta,k_i)+ \prod_{i=1}^{m} G^+(\eta_0,\eta,k_i)^*\right) \prod_{i=m+1}^{n}G_F(\eta_0,\eta',k_i)\prod_{i=1}^{L+1} G_F(\eta,\eta',y_i)\\ \nonumber
&&\hspace{-1em}=\int_{-\infty}^\infty \frac{d\eta}{\eta^4}\frac{d\eta'}{\eta'^4}\, F_1 F_2 \eta^{n_{d_1}}\eta'^{n_{d_2}}\times\\ \nonumber
&&\qquad \times \left( \prod_{i=1}^{m} G_F(\eta_0,\eta,k_i)+ \prod_{i=1}^{m} G_F(\eta_0,\eta,k_i)^*\right) \prod_{i=m+1}^{n}G_F(\eta_0,\eta',k_i)\prod_{i=1}^{L+1} G_F(\eta,\eta',y_i)\\ \nonumber
&&\hspace{-1em}=\int_{-\infty}^\infty \frac{d\eta}{\eta^4}\frac{d\eta'}{\eta'^4}\, F_1 F_2 \eta^{n_{d_1}}\eta'^{n_{d_2}}\times\\ \nonumber
&&\qquad \times \left( \prod_{i=1}^{m} G_F(\eta_0,\eta,k_i)+(-1)^m \prod_{i=1}^{m} G_F(\eta_0,\eta,-k_i)\right) \prod_{i=m+1}^{n}G_F(\eta_0,\eta',k_i)\prod_{i=1}^{L+1} G_F(\eta,\eta',y_i)\\ \nonumber
&&=\frac{1}{2}\left[B^{ex}_n(\{k_i\}_{i=1}^n)+(-1)^mB^{ex}_n(\{-k_i\}_{i=1}^m,\{k_i\}_{i=m+1}^n))\right],
\eea
where these last steps work in the same way as for the contact diagram. \\

Finally let us come to the last remaining identity, namely the internal cut,~\eqref{rules fin d}. We again consider the interactions in~\eqref{Hamilex}. We note here that even if we have an interaction that involved different fields, the only propagator identity we will need is that for the mode functions $f^i$ and $f^j$ associated with the fields $\sigma_i$ and $\sigma_j$ we can write
\bea\label{intcut}
G^+_{i,j}(\eta,\eta',y) =\frac{ f_y^{i}(\eta)f_y(\eta_0)^*f_y(\eta_0) f_y^{j}(\eta')^*}{f_y(\eta_0)^*f_y(\eta_0)}=\frac{G^+_i(\eta_0,\eta,y)^*G^+_j(\eta_0,\eta',y)}{P(y,\eta_0)},\eea
where $P(y,\eta_0)$ is the power spectrum of $\phi$, and 
\begin{align}
 G^+_{i,j}(\eta,\eta',y)&=\langle\sigma_i (\eta)\sigma_j (\eta')\rangle\,, & G^+_i(\eta,\eta',y)&=\langle\sigma_i (\eta)\phi(\eta')\rangle\,. 
\end{align}
Let us return to the case where $\sigma_i=\phi$, since the derivation is the same. The diagram with $L$ cut loops reads
\bea
\int_{-\infty}^\infty \frac{d\eta}{\eta^4}\frac{d\eta'}{\eta'^4}\, F_1 F_2 \eta^{n_{d_1}}\eta'^{n_{d_2}} \prod_{i=1}^{m} G_F(\eta_0,\eta,k_i)^* \prod_{i=m+1}^{n}G_F(\eta_0,\eta',k_i)\prod_{i=1}^{L+1} G^+(\eta,\eta',y_i).
\eea 
Now using~\eqref{intcut} and putting it back into the cut diagram, we see that the integral over $\eta$ and $\eta'$ are now independent. Omitting an overall factor of $\left[ \prod_{i}^{L+1}P(y_i) \right]$ we get
\begin{align}
&\int_{-\infty}^\infty \frac{d\eta}{\eta^4} F_1 \eta^{n_{d_1}} \prod_{i=1}^{m} G_F(\eta_0,\eta,k_i)^*\prod_{i=1}^{L+1} G^+(\eta_0,\eta,y)^*\times\\ \nonumber
&\qquad \times \int_{-\infty}^\infty\frac{d\eta'}{\eta'^4}\,F_2 \eta'^{n_{d_2}} \prod_{i=m+1}^{n}G_F(\eta_0,\eta',k_i)\prod_{i=1}^{L+1} G^+(\eta_0,\eta',y)\\ \nonumber
&\qquad=(-1)^{m+L+1}%\frac{1}{\prod_{i,j,j\in L}P_{i,j}(y_i)}
\int_{-\infty}^\infty \frac{d\eta}{\eta^4} F_1 \eta^{n_{d_1}} \prod_{i=1}^{m} G_F(\eta_0,\eta,-k_i)\prod_{i=1}^{L+1} G^+(\eta_0,\eta,-y_i)\times\\ \nonumber
&\qquad\qquad \times \int_{-\infty}^\infty\frac{d\eta'}{\eta'^4}\,F_2 \eta'^{n_{d_2}} \prod_{i=m+1}^{n}G_F(\eta_0,\eta',k_i)\prod_{i=1}^{L+1} G^+(\eta_0,\eta',y_i) \\ \nonumber
&=-%\frac{1}{\prod_{i,j,j\in L}P_{i,j}(y_i)}
\frac{1}{2}\left[B^{c}_{m+L+1}(\{k_i\}_{i=1}^m,\{y_i\}_{i=1}^{L+1})+(-1)^{L+1}B^{c}_{m+L+1}(\{k_i\}_{i=1}^m,\{-y_i\}_{i=1}^{L+1})\right]\times \nonumber \\
&\qquad \times\frac{1}{2}\left[B^{c}_{n-m+L+1}(\{k_i\}_{i=m+1}^{n+m},\{y_i\}_{i=1}^{L+1})+(-1)^{L+1}B^{c}_{n-m+L+1}(\{k_i\}_{i=m+1}^n,\{-y_i\}_{i=1}^{L+1})\right],\nonumber
\end{align}
where we have used~\eqref{contres} to flip signs in the contact diagram. Finally, if we define
\bea
B^{\text{c,cut}}_{n,L}(\{E_i\}_{i=1}^n,\{y_i\}_{i=1}^{L})=\frac{1}{2}\left[B^{\text{c}}_{n+L}(\{E_i\}_{i=1}^n,\{y_i\}_{i=1}^{L})+(-1)^{L}B^{\text{c}}_{n+L}(\{E_i\}_{i=1}^n,\{-y_i\}_{i=1}^{L}))\right],
\eea
and reintroduce the power spectrum factors, we can write~\eqref{rules fin d} as
\begin{align}
-\frac{B^{\text{c,cut}}_{m,L+1}(\{E_i\}_{i=1}^m,\{y_i\}_{i=1}^{L+1}) B^{\text{c,cut}}_{n-m,L+1}(\{E_i\}_{i=m+1}^n,\{y_i\}_{i=1}^{L+1})}{\prod_{i=1}^{L+1}P(y_i)},
\end{align}
Here $B^{\text{c,cut}}_{m,L+1}$ refers of to one interaction Hamiltonian and $B^{\text{c,cut}}_{n-m,L+1}$ to the other.

\paragraph{Minkowski}
The proof for the same relations in flat space is roughly the same. We notice that the mode functions for any mass in flat space enjoy the same identity as in the dS case, namely
\bea\label{negt}
f_E(t)=\frac{e^{-iEt}}{\sqrt{2E}} \then f_E(t)=f_E(-t)^*\,.
\eea
This means, that if we do not consider temporal derivatives the proofs from above follow through immediately after we set $t_0=0$ without loss of generality.

Finally, when considering time derivatives in Minkowski, the rules derived above also hold. While derivatives of the mode functions do not satisfy~\eqref{negt} any longer, pairs of derivatives do. 
Since in flat space, diagrams with an odd number of time derivatives vanish, the results are true for any Hamiltonian.

%%%%%%%%%%%%%%%%%%%%%%%%%%%%%%%%%%%%%%%%%

%\bibliographystyle{plain} % or another style of your choice
%\bibliography{refs} % name of your .bib file without extension
\bibliographystyle{JHEP}
\small
\bibliography{refs}

\end{document}